\RequirePackage{fix-cm}
\documentclass[pdftex,twocolumn,epjc3]{svjour3}          
\RequirePackage[T1]{fontenc}
\smartqed  
\RequirePackage{graphicx}
\RequirePackage{mathptmx}      
\RequirePackage{flushend}
\RequirePackage[numbers,sort&compress]{natbib}
\RequirePackage[colorlinks,citecolor=blue,urlcolor=blue,linkcolor=blue]{hyperref}

\def\mevc{\ensuremath{\mathrm{MeV/}c}}
\def\gevc{\ensuremath{\mathrm{GeV/}c}}
\def\mev2c{\ensuremath{\mathrm{MeV/}c^2}}
\def\gev2c{\ensuremath{\mathrm{GeV/}c^2}}
\def\to{\ensuremath{\,\rightarrow\,}}
\def\pbar{\ensuremath{{\bar{p}}}}
\def\pbarp{\ensuremath{{\bar{p}p}}}
\def\PiPi{\ensuremath{{\pi\pi}}}
\def\omegapi0{\ensuremath{\omega\pi^0}}

\def\PiPiEta{\ensuremath{{\pi^0\pi^0\eta}}}
\def\PiEtaEta{\ensuremath{{\pi^0\eta\eta}}}
\def\KpKmPi0{\ensuremath{{K^+K^-\pi^0}}}
\def\pbarpToPi0Pi0Eta{\ensuremath{{\bar{p}p}\,\rightarrow\,\pi^0\pi^0\eta}}
\def\pbarpToPiEtaEta{\ensuremath{{\bar{p}p}\,\rightarrow\,\pi^0\eta\eta}}
\def\pbarpToKpKmPi0{\ensuremath{{\bar{p}p}\,\rightarrow\,K^+K^-\pi^0}}
\def\pbarpToKstarK{\ensuremath{{\bar{p}p}\,\rightarrow\,K^*(892) K}}

\def\pbarpToOmegaPi0{\ensuremath{{\bar{p}p}\,\rightarrow\,\omega\pi^0}}
\def\pbarpToPhiPi0{\ensuremath{{\bar{p}p}\,\rightarrow\,\phi(1020)\pi^0}}
\def\pbarpToKmKp{\ensuremath{{\bar{p}p}\,\rightarrow\,K^-K^+}}

\def\pbarpToKstarmKp{\ensuremath{{\bar{p}p}\,\rightarrow\,K^*(892)^- K^+}}
\def\OmegaToPi0Gamma{\ensuremath{\omega\,\rightarrow\,\pi^0\gamma}}
\def\OmegaToPipPimPi0{\ensuremath{\omega\,\rightarrow\,\pi^+\pi^-\pi^0}}
\def\PipPimPi0{\ensuremath{\pi^+\pi^-\pi^0}}

\usepackage{multirow}

\usepackage{adjustbox}

\journalname{Eur. Phys. J. C}
\begin{document}
\title{Coupled Channel Analysis of
\pbarpToPi0Pi0Eta,  \PiEtaEta\ and
  \KpKmPi0\ at 900 MeV/c and of \PiPi-Scattering Data}

\subtitle{The Crystal Barrel Collaboration}

\author{
  M.~Albrecht\thanksref{addr1} \and
  C.~Amsler\thanksref{addr4, e1} \and
  W.~D\"unnweber\thanksref{addr3} \and
  M.A.~Faessler\thanksref{addr3} \and 
  F.H.~Heinsius\thanksref{addr1} \and
  H.~Koch\thanksref{addr1} \and
  B.~Kopf\thanksref{addr1} \and
  U.~Kurilla\thanksref{addr1, e2} \and
  C.A.~Meyer\thanksref{addr2} \and
  K.~Peters\thanksref{addr1, e2} \and
  J.~Pychy\thanksref{addr1} \and 
  X.~Qin\thanksref{addr1} \and
  M.~Steinke\thanksref{addr1}\and
  U.~Wiedner\thanksref{addr1}
}

\thankstext{e1}{~Now at Stefan Meyer Institute, Austrian Academy of Sciences Vienna, 1090 Vienna, Austria}
\thankstext{e2}{~Now at GSI Helmholtzzentrum f\"ur Schwerionenforschung GmbH, 64291 Darmstadt, Germany}

\institute{%
~Ruhr-Universit\"at Bochum, 44801 Bochum, Germany\label{addr1}
\and
~Carnegie Mellon University, Pittsburgh, Pennsylvania 15213, USA\label{addr2}
\and
~Ludwig-Maximilians-Universit\"at M\"unchen, 80799 M\"unchen, Germany\label{addr3}
\and
~Physik-Institut der Universit\"at Z\"urich, CH{-}8057 Z\"urich, Switzerland\label{addr4}
}


\date{Received: date / Accepted: date}

\maketitle
\begin{abstract}
A partial wave analysis of antiproton-proton annihilation data in
flight at 900 \mevc\ into \PiPiEta , \PiEtaEta\ and \KpKmPi0\ is
presented. The data were taken at LEAR by the Crystal Barrel
experiment in 1996. The three channels have been coupled together with
$\pi\pi$-scattering isospin I=0 S- and D-wave as well as I=1 P-wave data
utilizing the K-matrix approach. Analyticity is treated using Chew-Mandelstam 
functions. In the fit all ingredients of the K-matrix, 
including resonance masses and widths, were treated as free parameters. In spite of the large number of parameters,
the fit results are in the ballpark of the values published by the Particle Data Group. 
In the channel  \PiPiEta\  a significant contribution of the spin exotic $I^G=1^-$
$J^{PC}=1^{-+}$ $\pi_1$-wave with a coupling to $\pi^0 \eta$ is observed. Furthermore the
contributions of $\phi(1020) \pi^0$ and
$K^*(892)^\pm K^\mp$ 
in the channel \KpKmPi0\ have been
studied in detail. 
The differential production cross section for the two reactions and
the spin-density-matrix elements for the $\phi(1020)$ and $K^*(892)^\pm$
have been extracted. No spin-align\-ment is
observed for both vector mesons. The spin density matrix elements have been also
determined for the spin exotic wave.

\keywords{\pbarp\ annihilation \and \PiPi\ scattering data \and coupled channel
  analysis \and K-matrix approximation \and
  Chew-Mandel\-stam function\and spin density matrix \and spin-exotic $\pi_1$}
\end{abstract}

\section{Introduction}
\label{intro_lab}
Two decades ago \pbarp\ annihilation data in flight from the Crystal Barrel experiment have
already been analyzed by combining different channels
\cite{Amsler:1995bf, Amsler:2002qq}. Such an approach provides good
means to face the challenges related to
the large number of possible initial \pbarp\ states and of
overlapping resonances with the same quantum numbers in the light
meson sector. The most important advantages compared to single channel
fits are the ability to provide additional constraints by sharing
common production amplitudes over different
channels and to describe the dynamical parts in a more sophisticated way
so that the conservation of unitarity and analyticity is better fulfilled.
During the
last years lots of efforts have been put into a better
understanding of the $\pi\pi$-scattering waves. By considering dispersion
relations and crossing symmetries the phase shifts and
inelasticities for energies below $\sqrt{s} \; < \; 1.425\;$\gev2c\ can be now
described very precisely \cite{GarciaMartin:2011cn} and have been
taken into account in this analysis. In addition over the last two
decades the
computing power has improved dramatically so that nowadays more
extensive coupled channel analyses can be performed on a reasonable
time scale. The reanalysis of
the Crystal Barrel data in combination with  $\pi\pi$-scattering data
therefore helps to better understand the production mechanism of light
meson states in the \pbarp\ annihilation process. \\
 The \pbarp\ data presented here have been measured by the Crystal Barrel
experiment at LEAR (Low Energy Antiproton Ring) in the year 1996. The analysis has been performed with
PAWIAN ({\bf PA}rtial {\bf W}ave {\bf I}nter\-active {\bf AN}alysis), a  powerful, user-friendly and
highly modular partial wave
analysis software package with the ability to support single and coupled 
channel fits with data obtained from different hadron spectroscopy 
experiments~\cite{Kopf:2014wwa}. The analysis of the channels 
\pbarpToPi0Pi0Eta,  \PiEtaEta\ and
  \KpKmPi0\ at a beam momentum of 900 \mevc\ coupled together with
  $\pi\pi$ scattering data could be considerably improved
  compared to previous analyses~\cite{Amsler:1995bf, Amsler:2002qq,Abele:1999en, Anisovich:1999jw} 
with emphasis on the following aspects:

\begin{itemize} 
\item The three channels
  are dominated by $f_0$ and $f_2$ resonances decaying to
  $\pi^0\pi^0$, $\eta \eta$ and $K^+ K^-$, respectively.  The goal therefore is 
  to determine the pole positions properly and to some extent
  the partial widths of these resonances utilizing the K-matrix
  technique with P-vector approach. The
  fundamental requirements of unitarity and
  analyticity are realized by making use of Chew-Mandel\-stam functions
  proposed by \cite{Wilson:2014cna,Tanabashi:2018oca}. The advantage of this description is that the search for resonances and the
  determination of their properties is not
  limited to the real axis of the complex energy plane where the data
  are located. Instead, it is possible to investigate the
  analytic structure over the full complex energy plane. 

\item The channel
  \KpKmPi0\ is of special interest. Only by coupling it to the other two channels
   it is feasible to disentangle the $f_{J}$ and $a_{J}$
  resonance contributions decaying to $K^+K^-$. This is possible because the
  \PiPiEta\ channel contains only the production amplitudes for
  the reactions \pbarp \to $a_{J}\pi^0$ and the \PiEtaEta\ channel
   only the amplitudes for \pbarp \to
  $f_{J}\pi^0$. 
  By sharing these production amplitudes and by constraining in addition the $\rho \pi^0$
  contribution with I=1 P-wave \PiPi-scattering data the decays proceeding via  
  $\phi(1020) \pi^0$ and $K^*(892)^\pm K^\mp$ can be well
  extracted. Based on the fitted amplitudes the spin density matrix (SDM)
  elements for the two vector mesons $\phi(1020)$ and $K^*(892)^\pm$ can be determined
  which provide the full information on the underlying production
  process. The comparison with the $\omega$ production in the channel
  \pbarpToOmegaPi0 \cite{Amsler:2014xta} delivers new information about the
  \pbarp\ production process of light mesons with strange-quark
  content. The reaction \pbarpToKpKmPi0\ has already been studied
  in detail for beam momenta between 1.0 and 2.5 \gevc~\cite{Baubillier:1976sb}. 
  Due to the limited number of events only SDM elements averaged over the 
  production angle have been determined.   
  With the data and the refined analysis 
  presented here it is even possible to extract the production angle dependence of the SDM elements with good accuracy.     

\item In \pbarp\ and $\bar{p} n$ annihilations the spin exotic wave $\pi_1$ was so far only visible in annihilations 
 at rest~\cite{Abele:1998gn, Abele:1999tf}. It was the aim of this refined analy\-sis to
 trace it also in \pbarp\ experiments in flight. Also here the
 extraction of the SDM elements might help to better understand the annihilation
 process for this kind of reaction.

\item The outcome of this analysis provides also new and very helpful insights for high
quality and high statistics experiments like PANDA
\cite{Lutz:2009ff}. One major physics topic of PANDA is the
spectroscopy of exotic and non-exotic states in the 
charmonium and open charm mass regions in \pbarp\ production and 
formation processes. In particular similarities between the \pbarp\
annihilation processes into the channels $\phi(1020)
\pi^0$, $K^*(892)^\pm
K^\mp$ and the channels  $J/\psi \pi^0$,
$D^* \bar{D}$ consisting of charm quarks can be expected.
\end{itemize}

\section{Crystal Barrel Experiment}
\label{CB_lab}
The Crystal Barrel detector has been designed with a cylindrical geometry along the beam axis.
A liquid hydrogen target cell with a length of 4.4\,cm and a diameter of 1.6\,cm was located in the center of the detector where
the \pbarp\ annihilation took place. Antiprotons that passed the target without annihilation were vetoed by a downstream scintillation detector. The target was surrounded by a silicon vertex detector, followed by a jet drift chamber which covered 90\,\% and 64\,\% of the full solid angle for the inner and outer layer, respectively. These devices together with 
a solenoid magnet providing a homogeneous 1.5\,T magnetic field parallel to the incident beam guaranteed a good vertex reconstruction, tracking and identification for charged particles. For the accurate measurement of the energy and flight direction of photons the detector was equipped with a barrel-shaped calorimeter consisting of 1380 CsI(Tl) crystals covering the full azimuth range of 360$^\circ$ and polar angles from 12$^\circ$ to 168$^\circ$. With this electromagnetic calorimeter, located between the jet drift chamber and the solenoid magnet, an energy resolution of $\sigma_E/E \approx$ 2.5\,\%  and an angular resolution of 1.2$^\circ$ in $\theta$ and $\phi$ each have been achieved for photons with an energy of 1\,GeV. A detailed description of the full detector can be found elsewhere~\cite{Aker:1992ny}.            

\section{Data Selection}
\label{datasel_lab}
The basic reconstruction and event selection was performed in analogy to older publications by the Crystal Barrel Collaboration (see e.g. \cite{Amsler:1995bf, Amsler:2002qq, Abele:1999en, Anisovich:1999jw}). In addition, neural networks were used to detect electromagnetic \cite{Degener:1995xq} and hadronic \cite{Berlich:1997qj} split-offs, that could be falsely registered as photon candidates in the electromagnetic calorimeter. For all three reactions considered here an exclusive reconstruction was performed. Thus, events are only accepted and subjected to further analysis if all final state particles have been detected.
\subsection{Selection Criteria}
\paragraph{$\pi^0\pi^0\eta:$}
This reaction results in a final state of six photons, thus the number of tracks is required to be zero. Furthermore, the number of photon candidates after application of the split-off detection must be exactly six. Since the final state photons can be combined in multiple ways to form the required $\pi^0$ and $\eta$ resonances, kinematic fits under the hypotheses $6\gamma$, $\pi^0\pi^0 \gamma\gamma$, $\pi^0\pi^0\eta$, $\pi^0\eta\eta$, $3\pi^0$, $\omega\omega$ and $3\eta$ are performed. For all hypotheses energy and momentum conservation are required, as well as additional constraints of the invariant two-photon mass to match the respective intermediate resonances $\pi^0$ or $\eta$. For the signal channel the fit must converge with a confidence level (CL) greater than 10\%, corresponding to a probability of $p>0.1$, while $p<0.001$ is required for most background hypotheses. For the $\omega\omega$ hypothesis a requirement of $p<0.01$ is used. A previous analysis of the $\pi^0\pi^0\eta$ final state based on the same data set has shown \cite{Amsler:2002qq}, that after the application of kinematic fits events originating from the reaction $\pbar p \to \omega (\to \gamma\pi^0) \pi^0\pi^0 \to 7\gamma$ represent the main source of remaining background. For these events in most cases one of the seven final state photons escapes the detection. To effectively suppress this type of background events, we make use of a sophisticated event-based method, which is described in \ref{qfactor_lab}.

\paragraph{$\pi^0\eta\eta:$}
Since also this reaction results in a six photon final state, the basic selection criteria are very similar to the ones applied in the $\pi^0\pi^0\eta$ case. Also here, multiple kinematic fits are performed to improve the quality of the selected data sample. All kinematic fits require conservation of energy and momentum and additional constraints to the $\pi^0$ or $\eta$ mass, where applicable. Again, the fit is required to converge for the signal channel with $p>0.1$. More stringent requirements are set to reduce background from the reactions $\pbar p \to \omega\omega$ ($p<0.01$) and $\pbar p \to 3\eta$ ($p<0.1$). Similar to the $\pi^0\pi^0\eta$ channel, it is expected that prominent background due to the reaction $\pbar p \to \omega \pi^0\pi^0 \to 7\gamma$ remains after the selection. This type of background can appear below both $\eta$ signals and is treated with the event-based background suppression described in \ref{qfactor_lab}.

\begin{figure}
   \includegraphics[width=0.24\textwidth]{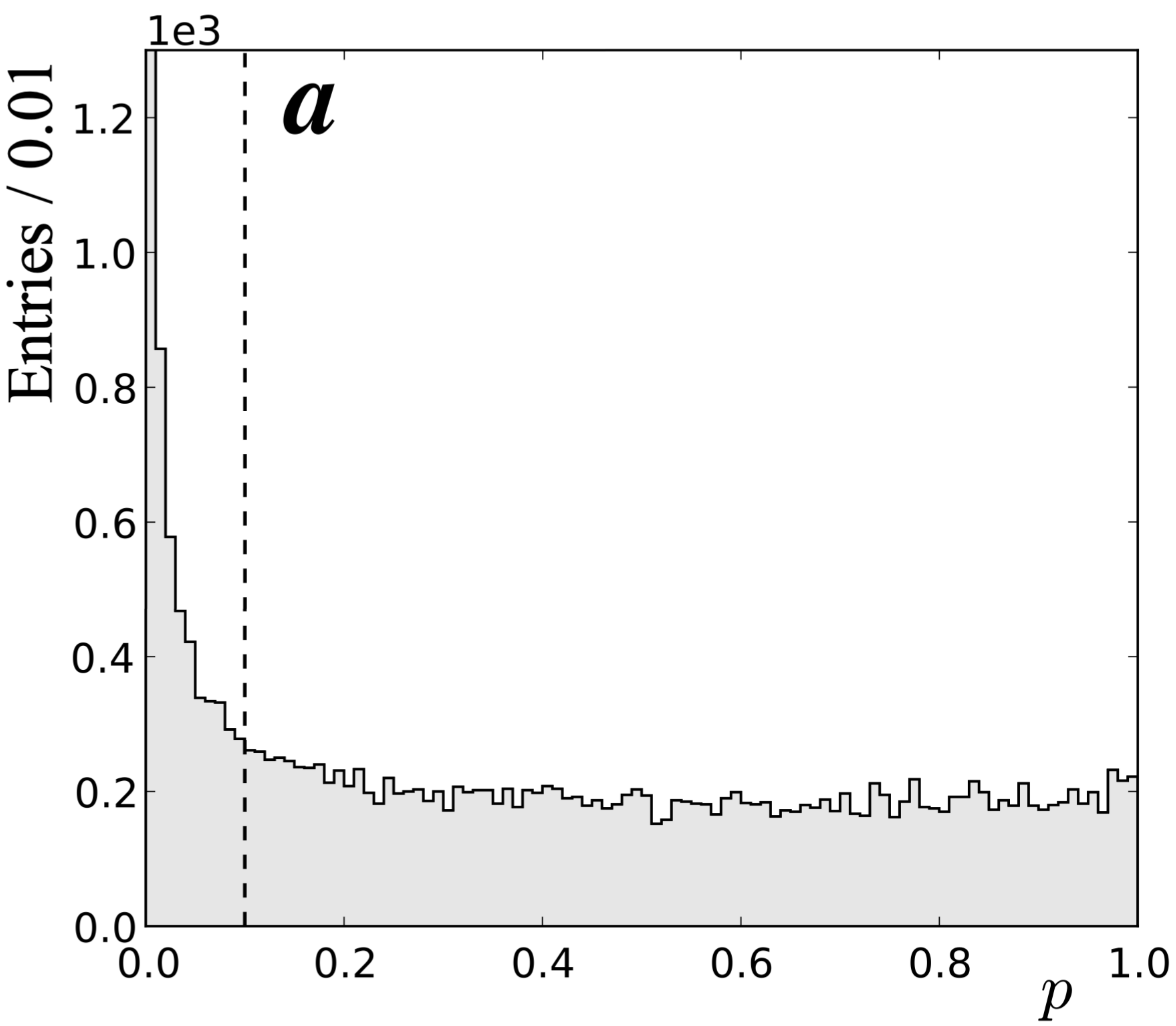} \includegraphics[width=0.23\textwidth]{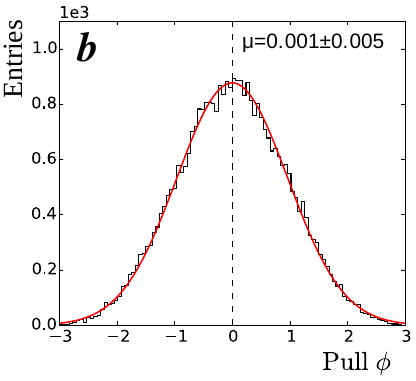}
   \includegraphics[width=0.23\textwidth]{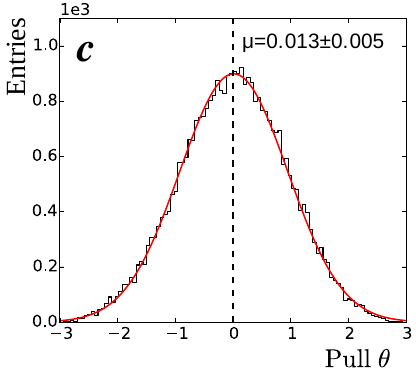} \includegraphics[width=0.23\textwidth]{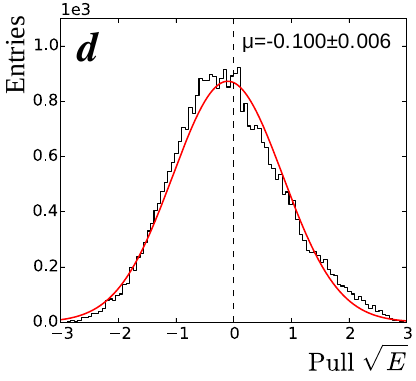} 
   \centering
   \caption{Confidence level and exemplary pull distributions obtained from the kinematic fit for the hypothesis $\pbar p\to K^+K^-\pi^0$. The confidence level (a) shows a flat distribution towards large $p$ values and an enhancement for low values, which is caused by background or misreconstructed events. The pull distributions for the angles $\phi$, $\theta$ (b,c) and the square root of the energy (d) for reconstructed photons show a Gaussian shape centered at zero with a width of approximately one. The mean values $\mu$ are extracted from the fit (red). All four distributions show the good quality of the selected data and indicate a well understood error matrix.}
   \label{fig:kkpi_CL_pull}
\end{figure}

\paragraph{$K^+ K^- \pi^0:$}
For this reaction, exactly two oppositely charged kaons are required, which must originate from a common vertex within the target cell. Furthermore, the number of accepted photon candidates after application of the split-off detection is required to be two. 
Since pions and kaons can not be separated easily by using the information on the differential energy loss from the drift chamber for momenta above 500\,\mevc\, the main background for the reaction under study is expected to be $\pbar p \to \pi^+\pi^- \pi^0$. To suppress background and to improve the quality of the data the events are subjected to kinematic fits under the hypotheses $\pbar p \to K^+ K^- \pi^0$, $K^+K^- \gamma\gamma$, $\pi^+\pi^-\pi^0$, $\pi^+\pi^-\eta$, $K^+K^-\eta$, $\pi^+\pi^-\eta'$ and $K^+ K^- \eta'$. For each hypothesis the conservation of momentum and energy are required (4 constraint fit) as well as an additional constraint to the invariant two-photon mass, which must match the mass of $\pi^0$, $\eta$ or $\eta'$, where applicable. To accept an event, it is required that the kinematic fit converges for the signal hypothesis with $p>0,1$, while for background suppression $p<0.01$ for the $\pi^+\pi^-\pi^0$-hypothesis and $p<10^{-5}$ for all other hypotheses is required. For the signal hypothesis, the distribution of the CL is found to be almost flat for $p>10\%$, while all pull distributions exhibit a Gaussian shape centered at $\mu=0$ with a width of $\sigma\approx1$. This indicates a high quality of the data and a properly adjusted error matrix. The CL distribution along with some pull distributions are exemplarily shown for this channel in Fig.\,\ref{fig:kkpi_CL_pull}. After application of these selection criteria the remaining background estimated from the sidebands of the $\pi^0$ signal in the distribution of the invariant two-photon mass is negligibly small. Thus, no further steps are taken to reduce remaining background and the finally selected 17529 events are used for the partial wave analysis.

\subsection{Additional Signal-Background Separation}
\begin{figure}
   \includegraphics[width=.24\textwidth]{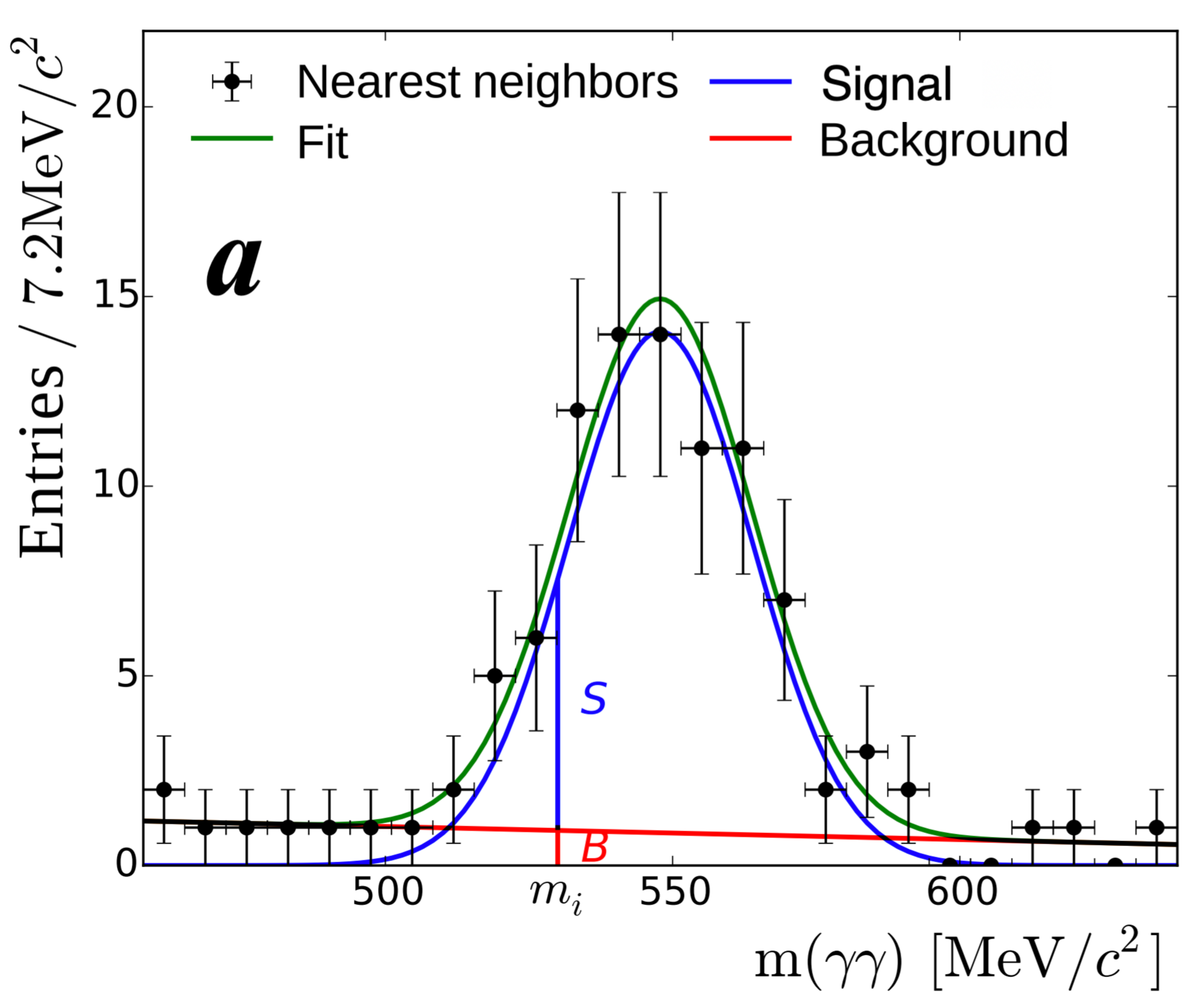}\includegraphics[width=.24\textwidth]{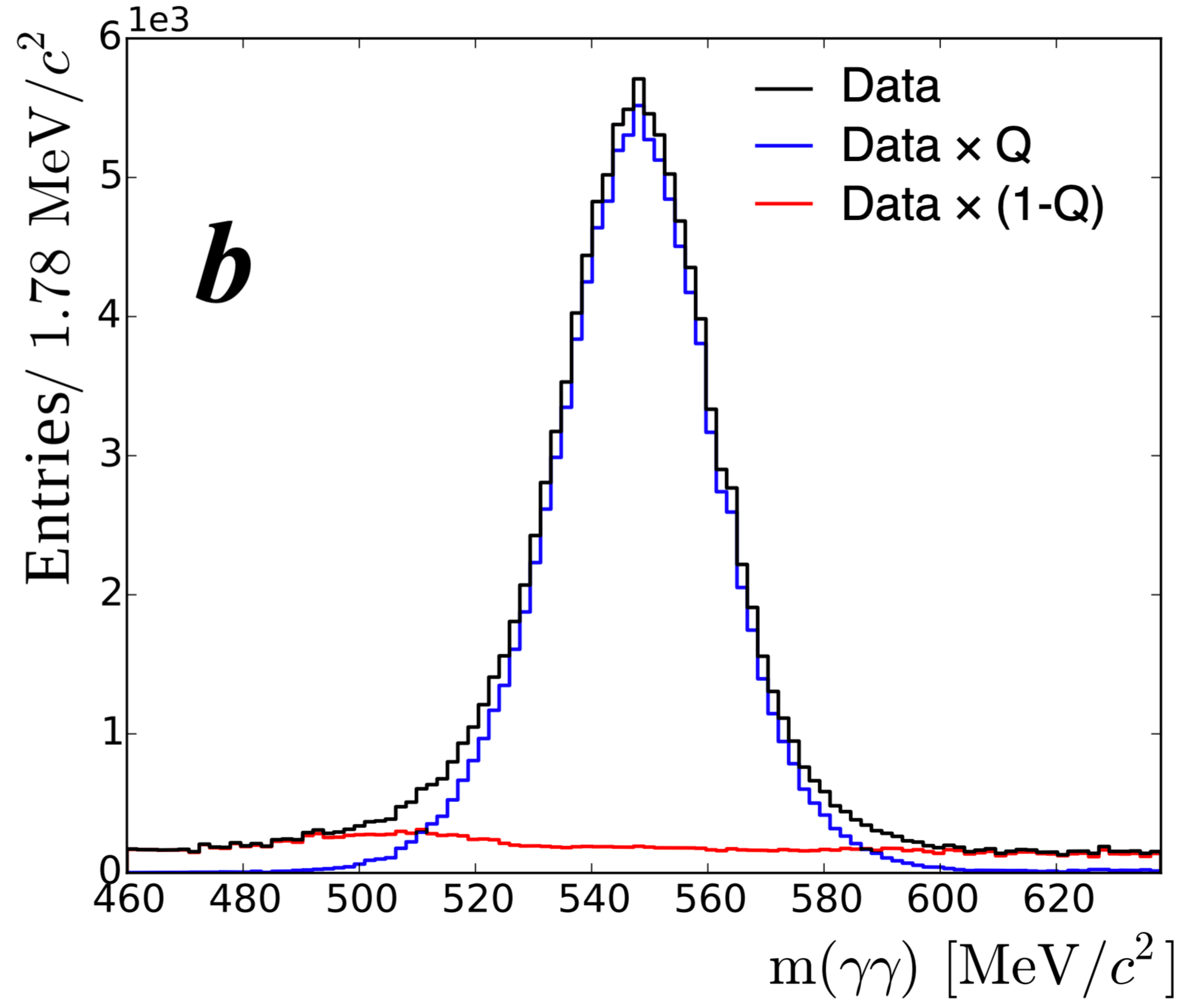}
   \caption{Performance of the $Q$-factor method for background below the $\eta$ signal appearing in the reaction $\pbar p \to \pi^0\pi^0\eta$. (a) shows the invariant two-photon mass for 100 nearest neighbors of a selected $\pi^0\pi^0\eta$ event. The green curve shows the total fit result. The blue and red shapes represent the signal and background contributions, respectively. The vertical line denotes the position of the actual event. (b) shows the invariant two-photon mass for all events after application of the $Q$-factor method for the identified signal (blue) and background (red) components.}
   \label{fig:williams}
\end{figure}

The remaining background contribution for the reactions resulting in a final state consisting of six photons can in principle stem from various sources and is mostly irreducible using simple one-dimensional selection criteria. To identify and suppress this type of background, a signal weight factor $Q$ is assigned to each event. In contrast to other methods, as e.g. the side-band subtraction method for binned data, the procedure employed here is an event based method. No details about the sources of all non-interfering background contributions must be known a priory. This method is described in detail in \cite{Williams:2008sh} and was successfully applied in earlier analyses by the CLAS collaboration \cite{Williams:2009ab,Williams:2009aa} as well as in a recent re-analysis of Crystal Barrel data \cite{Amsler:2014xta}. In principle the method relies on the fact, that background events cannot reproduce the narrow mass shape of resonances as e.g. $\pi^0$ or $\eta$ in the corresponding invariant mass spectrum.
To apply the method the nearest neighbors in the phase space for each event must be identified. Therefore, a metric containing relevant kinematic variables must be defined. As described above, the main background for the reaction $\pbar p \to \pi^0\pi^0\eta$ stems from processes resulting into a $7\gamma$ final state, where no $\eta$ meson is involved in the decay. Therefore, the method is applied to the invariant two-photon mass in the $\eta$ signal region. The metric was chosen to consist of four kinematic variables, namely the polar production angle of the $\eta$ meson in the center-of-mass frame, the polar and azimuthal decay angles of one of the $\pi^0$'s in the $\pi^0\pi^0$ helicity frame as well as the polar decay angle of the $\eta$ meson in the $\pi^0\eta$ helicity frame. Fig.\,\ref{fig:williams} (a) shows the distribution of the invariant two-photon mass for the 100 nearest neighbors of a selected event identified with the metric described above. The $Q$-factor for the event under consideration is then defined as the signal-to-background fraction at the position of the event, obtained from an unbinned fit to this distribution containing a description of the signal (described by a Gaussian function) and a linear background, as visualized in Fig.\,\ref{fig:williams} (a). 
The same event weight method has been applied for the reaction $\pbar p \to \pi^0\eta\eta$. Also here, the dominant background contribution stems from reactions, which do not involve an $\eta$ meson. In contrast to the reaction described above, the $Q$-factor method now has to be applied to the two-dimensional invariant two-photon distribution to simultaneously consider background below both $\eta$ resonances. The metric again consists of four variables, namely the polar production angle of the $\pi^0$ in the center-of-mass frame, the polar and azimuthal decay angles of one of the two $\eta$ mesons in the $\eta\eta$ helicity frame and the polar decay angle of that $\eta$ meson in the $\pi^0\eta$ helicity frame. In this case a number of 200 nearest neighbors is identified for each event and the signal is described by a two-dimensional Gaussian function, while the background is parameterized with a linear function independently for both invariant two-photon masses. The $Q$-factor is then calculated in the same way as for the one dimensional case. \\ 
    The performance of the $Q$-factor method is evaluated using dedicated Monte Carlo samples which are generated using proper amplitude models for signal and background contributions. The background can be clearly identified and after application of the $Q$-factor method we obtain clean data samples which are used as input for the partial wave analysis. 

\label{qfactor_lab}

\subsection{Overview of the Selected $\pbar p$ Data Samples}

\begin{figure}
   \includegraphics[width=0.48\textwidth]{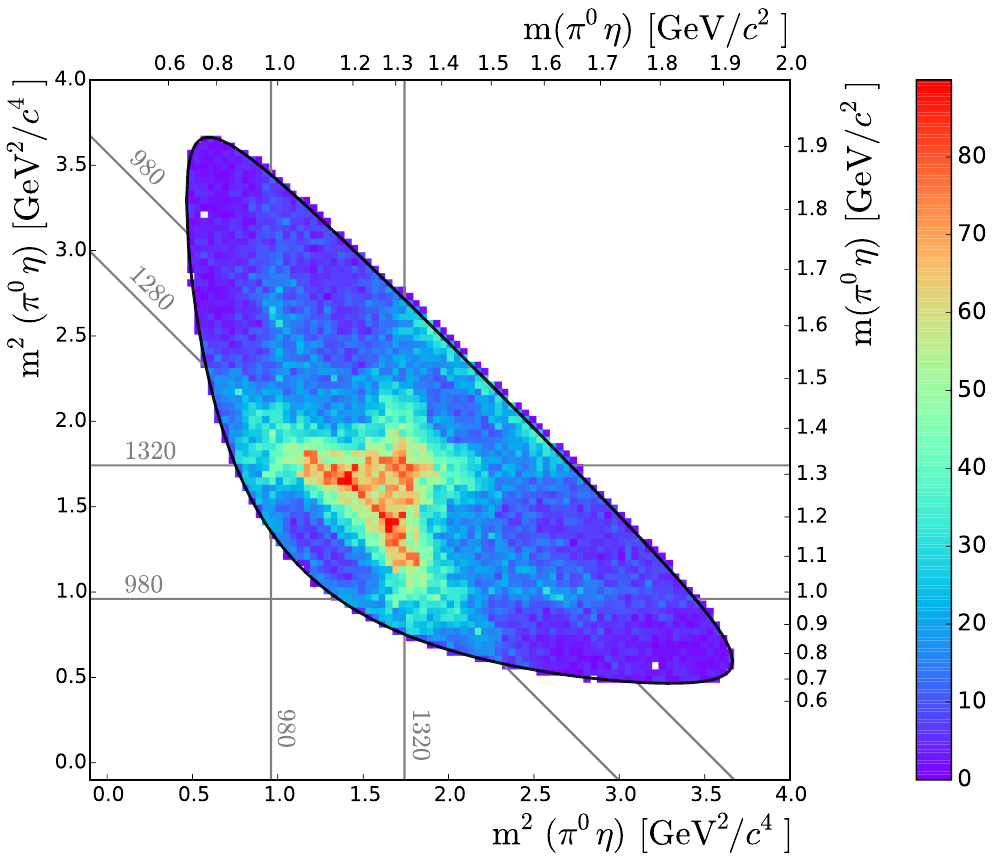}
   \centering
   \caption{Dalitz plot for the $\pi^0 \pi^0 \eta$ data after application of the background suppression,  not corrected for acceptance. Some resonance masses of interest are marked with thin black lines and gray labels (two entries per event).}
   \label{fig:pipieta_dalitz}
\end{figure}

\begin{figure}
   \includegraphics[width=0.49\textwidth]{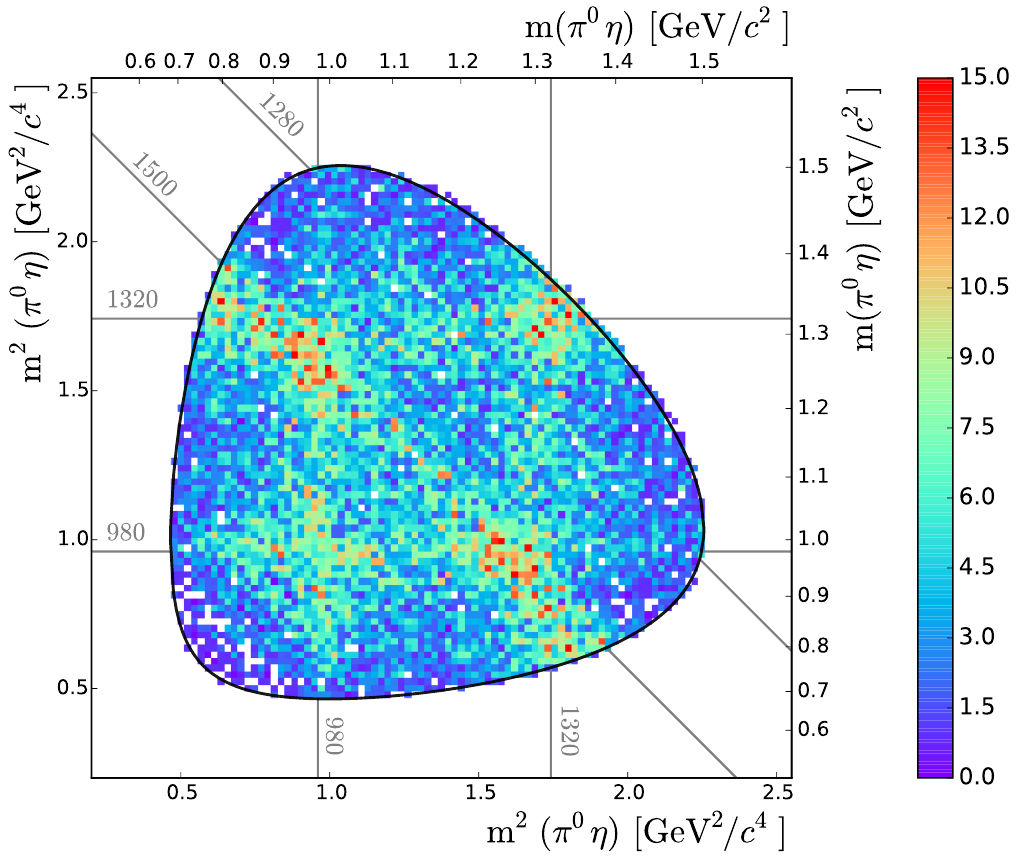}
   \centering
   \caption{Dalitz plot for the $\pi^0 \eta \eta$ data after application of the background suppression, not corrected for acceptance. Some resonance masses of interest are marked with thin black lines and gray labels (two entries per event).}
   \label{fig:pi0etaeta_dalitz}
\end{figure}

\begin{figure}
   \includegraphics[width=0.49\textwidth]{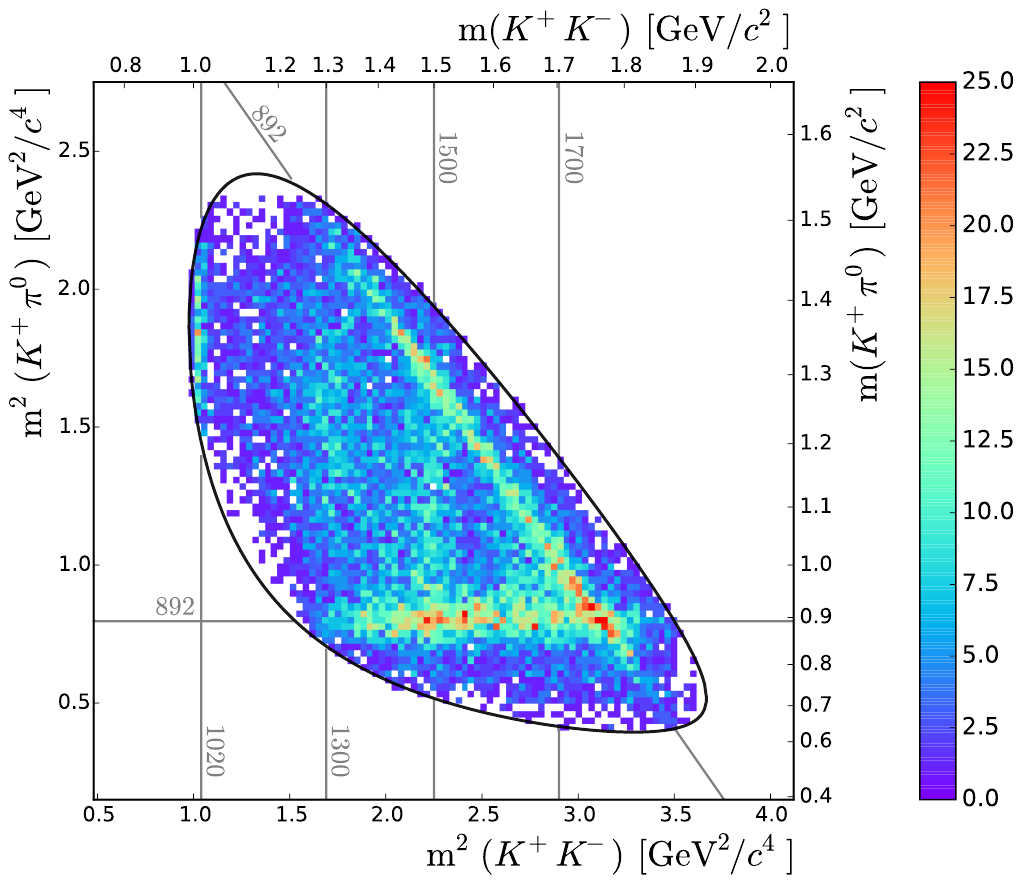}
   \centering
   \caption{Dalitz plot for the $K^+ K^- \pi^0$ data after application of the background suppression, not corrected for acceptance. Some resonance masses of interest are marked with thin black lines and gray labels.}
   \label{fig:kkpi_dalitz}
\end{figure}

After application of the $Q$-factor method, a sample of 90408 signal events for the $\pi^0\pi^0\eta$ channel is obtained. 
Figure \ref{fig:pipieta_dalitz} shows the Dalitz plot of the selected and $Q$-weighted $\pi^0\pi^0\eta$ events. Prominent structures possibly originating from contributions of the $a_2(1320)$ as well as the $f_2(1270)$ meson are clearly visible. Furthermore, possible structures around 1\,GeV$/c^2$ can be seen, which may originate from contributions of the $a_0(980)$ and $f_0(980)$ states.\\ 
Figure \ref{fig:pi0etaeta_dalitz} shows the corresponding plot for the $\pi^0\eta\eta$ channel, for which in total 10533 weighted events were retained. Also here, a strong structure at the mass of the
$a_2(1320)$ meson is observed, as well as a rather strong signal of the $a_0(980)$ in comparison with the $\pi^0\pi^0\eta$ channel. A strong contribution around 1.5\,\gev2c\ is obvious, originating from an $f_2^{'}(1525)$, an $f_0(1500)$ or both. Figure \ref{fig:kkpi_dalitz} shows the Dalitz-plot for all 17529 selected events for the $K^+K^-\pi^0$ channel. It is dominated by bands corresponding to contributions of the $K^*(892)^\pm$ meson decaying into $K^\pm\pi^0$. At the edge of the phase space a structure corresponding to the $\phi(1020)$ meson is visible, as well as structures around 1.3, 1.5 and 1.7\,\gev2c. The respective invariant mass plots and the efficiency corrected decay angular distributions are shown in Fig.~\ref{fig:ResultPiPiEta}, \ref{fig:ResultPiEtaEta} and \ref{fig:ResultKKPi}.
Table \ref{tab:Data} summarizes the size of the selected samples and the corresponding number of signal events.
\begin{table}[h]
   \caption{Number of selected events for all three reactions}

   \label{tab:Data}      
   \footnotesize
   \centering
   \begin{tabular}{l r r} 
      \hline\noalign{\smallskip} 
      Reaction      & Total number & Signal events \\
      $\pbar p \to$ & of events    & $\sum Q$ \\
      \noalign{\smallskip}
      \hline
      \noalign{\smallskip} 
      $\pi^0 \pi^0\eta$ & 97372 & 90408 \\
      $\pi^0 \eta\eta$ & 11905 & 10533 \\
      $K^+K^-\pi^0$ & 17529 & 17529 \\
      \\
   \end{tabular}
\end{table}

\subsection{Overview of the Scattering Data}
\label{overview_scattdata}
Most of the $f_0$, $f_2$ and $\rho$ resonances in the light meson
sector are characterized by the coupling to several decay
channels. Therefore different sets of scattering data are
included which results in an adequate
consideration of unitarity. The individual sets of scattering data
used in the coupled channel fit are as follows:
For the reaction $\pi\pi \rightarrow \pi\pi$,\ $I=0$\ $S$- and $D$-wave ($S0$- and $D0$-wave, associated with 
$f_0$ and $f_2$ resonances)
  and $I=1$\ $P$-wave ($P1$-wave, associated with $\rho$ resonances) are taken into account from
  \cite{GarciaMartin:2011cn} for the energy region from the $\pi\pi$
  threshold up to $\sqrt{s}\;=\;1.425\,$\gev2c. These model
  independent descriptions for the phase shift and the
  inelasticity  are based on
  dispersion relations and crossing symmetries. Since the errors provided in \cite{GarciaMartin:2011cn}
  are very small and only based on the
  uncertainties of the underlying theory, systematic errors of
  0.001 for the inelasticity and of $0.01^\circ$ for the phase motion have been added for
  the combined fit with the \pbarp\ data. The remaining energy range
  between  $\sqrt{s} \; > \; 1.425\,$\gev2c\ and $\sqrt{s} \; <\;1.9\,$\gev2c\ is covered by the CERN-Munich
  data~\cite{Hyams:1973zf} where the solutions have been chosen which are labeled with (-$\;$-$\;$-) and (-$\;$+$\;$-) 
  in \cite{Ochs:2013gi} and \cite{Hyams:1975mc}, respectively. 
  The modulo squares of the T-matrix for the S0- and D0-wave of the scattering
  process $\pi\pi \rightarrow KK$ are taken
  from~\cite{Longacre:1986fh} and for  $\pi\pi \rightarrow \eta\eta$
  from~\cite{Binon:1983ny}. The data from~\cite{Binon:1984ip} are used
  for the S0-wave scattering process $\pi\pi \rightarrow
  \eta\eta^\prime$.  
  A summary of all  scattering data is shown in Fig.~\ref{fig:ResultsScattering}.

\section{Partial Wave Analysis}
\label{pwa_lab}
\subsection{Amplitudes}

\begin{figure*}[ht]
   \centering
   \includegraphics[width=\textwidth]{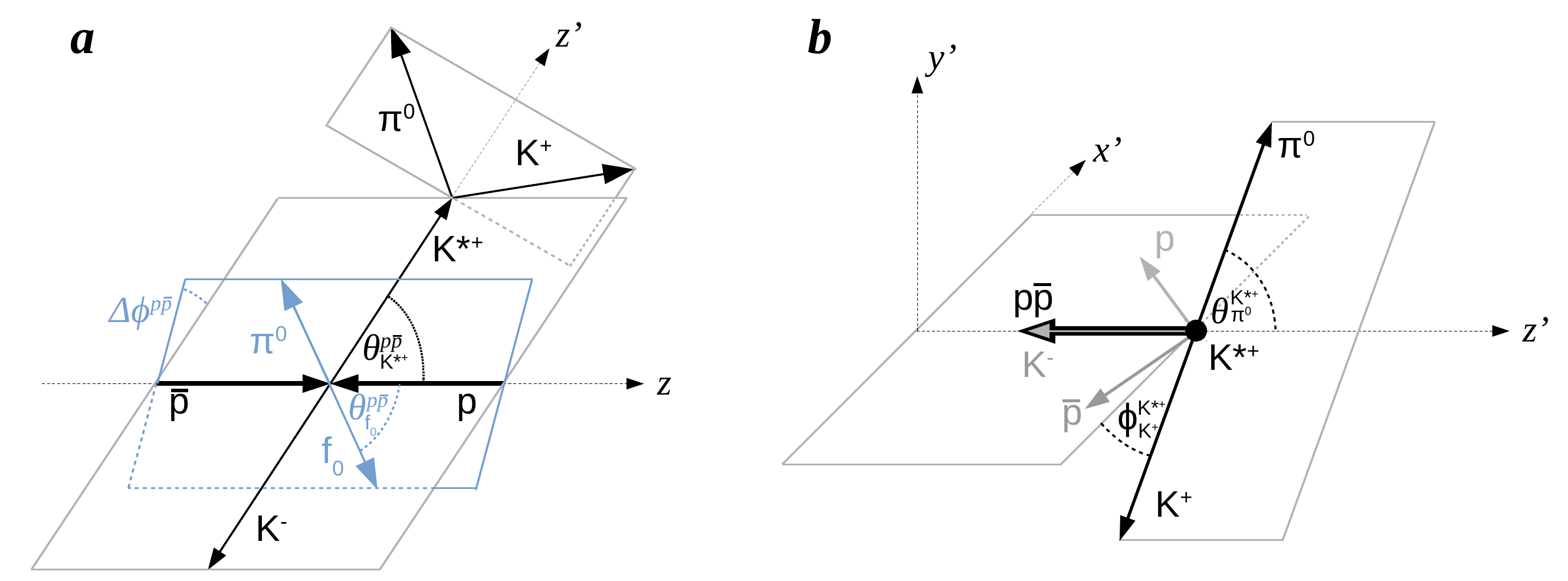}
   \caption{Graphical representation of the most important kinematic variables for the production (a) and decay (b) reference frames. As an example, a scenario corresponding to the $\pbarp\to K^+ K^-\pi^0$ channel was chosen. The $z$-axis in (a) is  defined by the flight direction of the antiproton, while the directions of the $x-$ and $y$-axes are arbitrary and only defined in the laboratory frame by convention ($y$-axis pointing up in vertical direction). The $\pbarp$ system then decays to $K^{*+} K^-$, with a further decay of the $K^{*+}$ into $K^+\pi^0$ (black), or e.g. into $f_0\pi^0$ with the $f_0$ decaying into $K^+ K^-$ (blue). The angle $\Delta\phi^\pbarp$ is the angle between the planes of these two decay branches. According to eqn.~(\ref{eq:XProdAmp}) $K^{*+}$ stands for $X$. In the helicity frame of the $K^{*+}$ depicted in (b), the $z'$ axis is defined by the opposite direction of the $\pbarp$ system or equivalently of the $K^-$ particle. The $x'-z'$ plane is given by the production plane, spanned by the flight directions of the proton and antiproton. Accordingly, the $y'$ axis is perpendicular to this production plane. $K^{*+}$ corresponds to $X$, $\pi^0$ to $s_1$ and $K^+$ to $s_2$ which are given in eqn.~(\ref{eq:XDecAmp}) .}
   \label{fig:drawingProdDecay}
\end{figure*}

\emph{Description of the $\pbarp$ annihilation amplitudes:} The amplitudes for the individual \pbarp\ channels are defined in a
similar way as explained in~\cite{Amsler:2014xta}.
The various contributing initial \pbarp\ states are expressed as an expansion
in terms of $I^G\;J^{PC}$ states. It has turned out
from semi-classical calculations~\cite{Mundigl:1991jp} as well as from
data analyses, as described in~\cite{Amsler:2014xta} for example, that
only states up to a maximal orbital momentum $L^{max}_\pbarp $ of the \pbarp\ system contribute.
The complete
reaction chain starting from the
\pbarp\ annihilation system down to the decay into the final state
particles is taken into account. The underlying description is based
on the helicity formalism. The considered
angles, illustrated in Fig.~\ref{fig:drawingProdDecay}, are the polar and azimuthal angle of the production of
the intermediate resonance $X$ in the \pbarp\ rest frame with respect to the
direction of the \pbar\ beam ($\theta^{\pbarp}_X, \phi^{\pbarp}_X$) and the azimuthal and polar angle
of the $X$ decay in its helicity system, in which the y-axis is
defined to be parallel to the normal vector of the production plane ($\theta^X_{s_1},\phi^X_{s_1}$). 
$X$ stands either for an isolated single resonance or for a partial wave containing several resonances with defined spin/isospin quantum numbers.\\
 The differential
cross section is described in terms of the transition amplitude depending on
the helicities of the involved particles, which is subdivided into the \pbarp\
initial state amplitude $A^{\pbarp \rightarrow J^{PC}}_{\lambda_p \;
  \lambda_\pbar}$,
the production amplitude   $B^{
  J^{PC} \rightarrow X s_r}_{\lambda_X \lambda_{s_r}}$ and the decay amplitude
$C^{
  X \rightarrow s_1 s_2}_{\lambda_{s_1} \lambda_{s_2}}$
of the intermediate resonance $X$. The differential cross section for each of the three $\pbarp$ annihilation reactions is
given by:
\begin{eqnarray}
  \label{eq:crossSection1} 
  \frac{d\sigma}{d\tau} \propto w  =  \sum_{\lambda_p \lambda_\pbar} \Big|
 \sum_{J^{PC}}  A^{\pbarp \rightarrow J^{PC}}_{\lambda_p \; \lambda_\pbar}  
 \Big( \sum_{X} \big( \sum_{\lambda_X} B^{J^{PC} \rightarrow X s_r}_{\lambda_X \lambda_{s_r}} \cdot
C^{
  X \rightarrow s_1 s_2}_{\lambda_{s_1} \lambda_{s_2}}  \big) \Big)
  \Big|^2, \nonumber \\
\end{eqnarray} where $\tau$ stands for the phase space, $s_r$ describes the recoil particle of X and $s_1$, $s_2$ are the decay
particles of $X$. $\sum_{X}$ runs over all waves and resonances in all possible sub-channels, like $(K^-\pi^0)K^+$, 
$(K^+K^-)\pi^0$. This expression is equivalent to summing incoherently over the \pbarp\
triplet states ($S_\pbarp = 1$) and the singlet state  ($S_\pbarp =
0$)~\cite{Koch:2012PandaNote}. The initial state amplitude $ A^{\pbarp
  \rightarrow J^{PC}}_{\lambda_p \; \lambda_\pbar}$ is expandable in
LS-states:
\begin{eqnarray}
  \label{eq:pbarAmp}
  A^{\pbarp \rightarrow J^{PC}}_{\lambda_p \; \lambda_\pbar} &=& \sum_{I}
                                                  i^{J^{PC}}(I^{J^{PC}}_\pbarp)\left( \sum_{L_{\pbarp }, S_{\pbarp }}\right.
  \langle 1/2 , \lambda_p ,  1/2 , -\lambda_\pbar | S_\pbarp ,  \lambda_\pbarp  \rangle
   \nonumber \\
  && \left. \cdot\langle
     L_{\pbarp }, 0 , S_\pbarp,
     \lambda_\pbarp | J^{PC},  \lambda_\pbarp  \rangle \cdot \alpha^{\pbarp
     \rightarrow J^{PC}}_{L_{\pbarp }S_\pbarp} \right), 
\end{eqnarray} where $ i^{J^{PC}}(I^{J^{PC}}_\pbarp)$ represents the isospin
contributions $I^\pbarp$=0 and $I^\pbarp $=1 for the relevant
initial \pbarp\ state. The expansion into
the LS-scheme is taken into account by the sum over the orbital
momenta $L_\pbarp$ and the spins $S_\pbarp$ of the \pbarp\ system
with the Clebsch-Gordan coefficients for the coupling of the antiproton and proton spins to $S_\pbarp$ and for the coupling of $L_\pbarp $ and $S_\pbarp$ to $J^{PC} $. The complex fit
parameter   $\alpha^{\pbarp \rightarrow J^{PC}}_{L_{\pbarp }S_\pbarp}$ is proportional to the partial wave amplitude $T^{\pbarp \rightarrow J^{PC}}_{L_{\pbarp }S_\pbarp}$ and
includes some additional constant prefactors which are not explicitly specified here.

The production amplitude reads:
\begin{eqnarray}
  \label{eq:XProdAmp}
  B^{ J^{PC}\rightarrow X s_r}_{\lambda_X \lambda_{s_r}}
  &=& D^{J^{PC} *}_{\lambda_{\pbarp} \, \lambda_X} (\theta^{\pbarp}_X, \phi^{\pbarp}_X) \cdot
          \langle I_X, I_{Xz} , I_{s_r}, I_{s_r z}| I^{J^{PC}}_{\pbarp }, 0 \rangle
                                                                      \nonumber \\
 && \Big( \sum_{L_{X s_r}\;S_{Xs_r}} B^{L_{X s_r}}(\sqrt{s}, m_X, m_{s_r})                                                                
     \nonumber \\ &&  \langle J_X, \lambda_X, J_{s_r}, \lambda_{s_r}
    |  S_{X s_r} , \lambda_{X s_r} \rangle  \nonumber \\
&&  \langle L_{X s_r}, 0, S_{X s_r}, \lambda_{X s_r} | J^{PC}
   \lambda_{X s_r} \rangle
   \cdot \alpha^{J^{PC} \rightarrow X s_r}_{L_{X s_r} S_{X s_r}} \Big),
   \nonumber \\
\end{eqnarray}
where $ D^{J^{PC} *}_{\lambda_{\pbarp}\,\lambda_X}$ denotes the complex conjugate Wigner-D function 
for the decay of the \pbarp\ system to $X$ and $s_r$. It is worth noting that also the imaginary 
part of the
Wigner-D function must be used here. The absolute azimuth angle  $\phi^{\pbarp}_X$ is not unambiguously defined 
for unpolarized \pbarp\ measurements. However, the difference of this angle between two different 
particle
subsystems is accessible and thus considered by this complex function. The first 
Clebsch-Gordan coefficient describes the isospin
coupling followed by the loop over all possible $L_{X s_r} \, S_{X s_r}$ combinations.  $B^{L_{X s_r}}$ 
describes the orbital momentum
dependent Blatt-Weisskopf production barrier factor. X stands for the set of isospin and spin quantum numbers of the selected sub-channel. 
The complex fit parameter
$\alpha^{J^{PC} \rightarrow X s_r}_{L_{X s_r} S_{X s_r}}$ includes again some additional and not  explicitly
specified constant prefactors.

The decay amplitude is given by:
 \begin{eqnarray}
  \label{eq:XDecAmp}
  C^{X \rightarrow s_1 s_2}_{\lambda_{s_1} \lambda_{s_2}} &=&
                                                               D^{J_X*}_{\lambda_X
                                                              0}(\theta^X_{s_1},\phi^X_{s_1})
                                                              \cdot
                                                               \langle
                                                                      I_{s_1},
                                                                      I_{s_1
                                                              z}
                                                                      ,
                                                                      I_{s_2},
                                                                      I_{
                                                                      s_2
                                                                      z}|
                                                                      I_X, I_{Xz}  \rangle
                                                                      \nonumber \\
 && \cdot F_X^{L_{s_1 s_2}}(m_X, m_{s_1}, m_{s_2}) \cdot
                                                               \alpha^{X
                                                               \rightarrow
                                                               s_1
                                                              s_2}_{L_{s_1
                                                              s_2} , 0},
 \end{eqnarray}
 here  the complex conjugate Wigner D-function  $D^{J_X*}_{\lambda_X \, 0}$
depends also on the azimuthal angle $\phi^X_{s_1}$. $m_X$ is the energy of the two-body sub-channel. 
Since both final state particles exhibit a total spin of 0 only one LS-combina\-tion remains with $L_{s_1 s_2}=J_X$ and $S_{s_1 s_2}=0$, 
which is included in the parameter $\alpha^{X \rightarrow s_1s_2}_{L_{s_1s_2} , 0}$. The term 
 $F_X^{L_{s_1 s_2}}(m_X, m_{s_1},
 m_{s_2})$ represents the dynamics of the partial wave which is described either by the
 K-matrix formalism or by the Breit-Wigner parametrization as discussed in
 more detail in the following section~\ref{dynamics}. Since the
 complex fit parameters  $\alpha^{\pbarp\rightarrow J^{PC}}_{L_{\pbarp }S_\pbarp }$,
 $\alpha^{J^{PC} \rightarrow X s_r}_{L_{X s_r} S_{X s_r}}$,
 and $\alpha^{X \rightarrow s_1 s_2}_{L_{s_1 s_2} , 0}$ are not 
 independent of each other a certain subset of them
 is fixed so that exactly one
  solution is provided for the fit procedure. For a single channel scenario it must be ensured that each
 product of these fit parameters exhibits one independent complex parameter. Additional constraints such as fixing
 one global phase in the coherent terms according to eq.~(\ref{eq:crossSection1}) or the treatment of shared 
 parameters for the coupling of different channels lead to further restrictions in the choice of the set of free fit 
 parameters.\\
 For the channels $\PiPiEta$ and $\PiEtaEta$
 only one isospin component is allowed which is $I = 0$ for
 $\PiPiEta$ and $I = 1$ for $\PiEtaEta$. The channel $\KpKmPi0$, however, contains excited
 ka\-ons such as $K^*(892)^\pm$ and probably also the $(K^\pm\pi^0)_S$ wave. The reaction chains (sub-channels) 
containing these intermediate resonances with strange
 quark content must be treat\-ed slight\-ly
 differently compared to the equations above. These
 channels do not exhibit a well defined C and G parity and can originate 
  from both isospin components $I_\pbarp = 0$  and $I_\pbarp
 = 1$ of the \pbarp\
 system.  Therefore it is necessary to expand the two
 particle systems 
 $K^*(892)^\pm K^\mp$  and $(K\pi)_S^\pm K^\mp$ to
 $C$ and $G$ eigenstates for the relevant $I^G \;
 J^{PC}$ \pbarp\ state and the
 isospin coupling in eq.~(\ref{eq:pbarAmp} - \ref{eq:XDecAmp}) must be
 replaced by appropriate prefactors for considering C-and G- symmetry.\\ 
 \emph{Description of the $\pi\pi$  scattering
    amplitudes:} As mentioned in~\ref{overview_scattdata}, scattering data from the model independent calculations 
from~\cite{GarciaMartin:2011cn} are used for the reaction $\pi\pi \rightarrow \pi\pi$ below $\sqrt{s}\;=\;1.425\,$\gev2c. The remaining 
$\pi\pi$-scattering amplitudes are derived in the
  usual way from $\pi N$ -scattering measurements \cite{Anisovich:2002ij}. For small 4-momentum transfers t they are nearly independent from t and 
the s-dependence of the $\pi\pi$ scattering can be extracted in the form of scattering matrices $T_{ij}$. i stands for the initial and j for the 
final channel, such as $\pi\pi$, $K\bar{K}$,
$\eta\eta$ or $\eta\eta^\prime$. The T-matrix can be para\-metrized in terms of K-matrix elements (see \ref{dynamics}). The scattering data 
for elastic reactions are provided in terms of phase shifts and
inelasticities, while for inelastic channels, like \linebreak
$\pi\pi \rightarrow
K\bar{K}$ and  $\pi\pi \rightarrow \eta \eta$, the moduli squared of the
T-matrix, e.g. $(2J+1) \; \rho_{\pi\pi} \; |T|^2 \; \rho_{K\bar{K}}$, are taken (see Fig. \ref{fig:ResultsScattering}).
 
 \subsection{Dynamics}
 \label{dynamics}
It is well known that Breit-Wigner parameterizations are only
adequate for descriptions of relatively narrow and isolated
resonances located far away from any thresholds and with a strong coupling to
not more than one channel~\cite{Tanabashi:2018oca}. However, many light mesons
with the same quantum numbers are broad, overlapping
with each other and decaying into several different channels. Thus
more sophisticated descriptions are needed for those non-trivial
dynamics. The analysis pre\-sent\-ed here makes use of the K-matrix
formalism for the 2-body scattering processes and of the P-vector approach for the $\pbarp$-channels ~\cite{Aitchison:1972ay, Chung:1995dx}.
The P-vector approach is assumed to provide an effective description not only for the non-trivial production mechanism, but also for the effect of rescattering.\\
\emph{Description of the scattering processes:} The S-matrix of a 2-body scattering process can be written as
\begin{equation}
S =I + 2i \,\rho^{1/2} \; T \, \rho^{1/2}
,\end{equation}
 with $I$ being the identity and $\rho$ the phase space diagonal matrix. $T$ is the Lorentz-invariant 
transition matrix element. For reasons of simplicity a possible dependence on the orbital momentum L is neglected here.\\
 $T$ can be written in terms of the K-matrix:
\begin{equation}
T(s) = (I \, +  \,  K(s) \; C(s))^{-1} \; K(s),
\end{equation}
where $s$ is the squared energy of the two-body system and $C(s)$ is the Chew-Mandelstam matrix which is
 a diagonal matrix by definition. For the fits
presented here the elements of this matrix are calculated by the functions as defined
in~\cite{Wilson:2014cna} and are connected
to the phase space elements by:
\begin{equation}
 \label{equ:CewMadelstamAnsRho}
Im \; C_{ii}(s) \, = \, - \rho_{ii}(s),
\end{equation}
where i represents one specific channel. In case of the decay into two stable particles with masses $m_1$ and $m_2$ 
and with s being real and 
above the threshold the phase space element reads:
\begin{equation}
 \label{equ:RhoDefault}
\rho_{ii}(s, m_1, m_2) \, = \, \sqrt{ \Bigl(1-\frac{(m_1+m_2)^2}{s} \Bigr) \cdot \Bigl(1-\frac{(m_1-m_2)^2}{s}\Bigr) }
\end{equation}
and is normalized such that $\rho_{ii}(s, m_1, m_2) \rightarrow 1$ as $s \rightarrow \infty$. 
The K-matrix itself is based on the description
in~\cite{Pennington:2007se}:
\begin{samepage}
\begin{eqnarray}
  \label{equ:kmatri}
   K_{ij}(s)  =   &\frac{s-s_0}{s_{norm}}& \cdot \sum_{\alpha} B^L_{\alpha_i}(q_i,q_{\alpha_i}) \cdot \nonumber \\
         & & \left( \frac{g^{bare}_{\alpha_i} \; g^{bare}_{\alpha_j}}{ {m^{bare}_\alpha}^2-s} + \sum_n  \tilde{c}_{nij} \cdot s^n \right) 
             \cdot B^L_{\alpha_j}(q_j,q_{\alpha_j})    
\end{eqnarray}
\end{samepage}
where i and j stand for the reaction channels. The bare
  parameters $g^{bare}_{\alpha_i}$ and  $m^{bare}_\alpha$ represent the coupling
  strength to the channel i and the mass of the resonance $\alpha$ in
  the K-matrix representation, respectively.
$B^L_{\alpha_i}(q_i,q_{\alpha_i})$ denote the Blatt-Weisskopf barrier
factors with the breakup momentum $q_i$ and the
resonance breakup momentum $q_{\alpha_i}$ depending on the orbital decay angular
momentum $L$ in the channel $i$.
The
$s$ dependent polynomial terms of the order $n$ together with the
parameters $\tilde{c}_{nij}$
describe 
background contributions, which are allowed to be added to the K-matrix without
violating unitarity. $(s-s_0)/s_{norm}$ represents the Adler
zero term where $s_0$ is the Adler zero
position for the elastic scattering amplitude.  Based on ChPT $s_0$ is set to $m_{\pi^0}^2/2$ for the
$(\pi\pi)_S$-wave with isospin I=0~\cite{Rupp:2004rf}. For the parameterization of the
$(K\pi)_S$ wave with isospin I=1/2 $s_0$=0.23~\gev2c\ with $s_{norm}= m_K^2 + m_\pi^2$ is used
~\cite{Pennington:2007se}.\\
 \emph{Description of the three-body $\pbarp$ annihilation
    channels:} Here the dynamics is described in the P-vector
approach (see eq.~(\ref{eq:XDecAmp})) by \cite{Aitchison:1972ay}: 
\begin{equation}
\label{eq:Fvector}
F^p_l \; = \; \sum_j (I \;  +  \; K(s) \; C(s))_{lj}^{-1}
\cdot P^p_j,
\end{equation}
where $s$ is the energy squared of the respective two-body sub-channel. $p$ stands for the production of the wave or resonance. $P^p_j$ represents one element of the P-vector 
taking into account the production process of the sub-channel X. $\sum_j$ runs over all channels relevant for the partial wave under consideration. $l$ is one of the two-body channels relevant for the respective annihilation channel. The F-vector is equivalent to the T-matrix of the two-body scattering process. 
The P-vector has to
exhibit the same pole structures as the K-matrix and is
defined as:
\begin{equation}
\label{eq:Pvector}
 P^p_i =  \sum_{\alpha} \left( \frac{ \beta^p_{\alpha} \; g^{bare}_{\alpha_i}}{{m^{bare}_\alpha}^2
  -s} +\sum_n   c^p_{ni} \cdot s^n\right) B^L_{\alpha_i}(q_i,q_{\alpha_i}),
\end{equation}
\noindent where $\beta^p_{\alpha}$ is the complex parameter representing the strength of the production process. The remaining terms describe an
eventual energy dependent background term. In the
  coupled channel fit the minor dependency on the different $\bar{p}p$ annihilation processes
  is neglected and the same effective
  production background terms have been used for the different $J^{PC}$ initial $\bar{p}p$ states.
The masses and decay couplings are the same as in the scattering case,
only the
production strengths are parameters to be fitted to the annihilation data. 
For the specific case of the $a_0$ wave represented by a K-matrix 
consisting of the resonances $a_0(980)$ and $a_0(1450)$ and the channels 
$\pi^0\eta$ and $K^+K^-$ with constant background terms produced via $J^{PC} = 1^{++}$ in the reaction
$\bar{p}p \rightarrow a_0 \pi^0 \rightarrow K^+ K^- \pi^0$ eq.~(\ref{eq:Fvector}) and eq.~(\ref{eq:Pvector}) 
read:
\begin{equation}
\label{eq:FvectorExample}
F^{1^{++}}_{(K^+ K^-)} \; = \; \sum_{j=(\pi^0\eta)}^{(K^+K^-)} (I \; + \; K(s) \; C(s))_{{(K^+ K^-)} \; j}^{-1} \; \;
P^{1^{++}}_j
\end{equation}
with
\begin{equation}
\label{eq:PvectorExample}
 P^{1^{++}}_{(K^+ K^-)} =  \sum_{\alpha= a_0(980)}^{a_0(1450)}
\left( \frac{ \beta^{1^{++}}_{\alpha} \; g^{bare}_{\alpha_{K^+K^-}}}{{m^{bare}_\alpha}^2 -s}
+ \, c_{0 \; (K^+ K^-)} \right)
\end{equation}
and
\begin{equation}
 P^{1^{++}}_{(\pi^0\eta)} =  \sum_{\alpha= a_0(980)}^{a_0(1450)}
\left( \frac{ \beta^{1^{++}}_{\alpha} \; g^{bare}_{\alpha_{\pi^0\eta}}}{{m^{bare}_\alpha}^2 -s}
+ \, c_{0 \; (\pi^0\eta)} \right).
\end{equation}
The Blatt-Weisskopf barrier factors are not needed here because only the orbital
momentum L=0 is contributing.

\subsection{Fits to Data}
For the coupled channel fit a minimization function is used that considers all individual channels properly. The
\pbarp\ data, provided by the full information of each event located
in a multidimensional phase space volume, are
treated slightly differently from the scattering data which
are given by one-dimensional diagrams assigned with errors.   
For the \pbarp\ channels an unbinned maximum likelihood
minimization procedure is used. Input for
this method are the selected data with the event weights $Q_i$ as well as
phase-space distributed Monte Carlo events. For properly taking into account the detector 
resolution and acceptance the GEANT3 transport code has been used.
To consider the correct reconstruction efficiency the Monte Carlo
events had to undergo the same reconstruction and selection criteria as applied for data events
and described in section \ref{datasel_lab}. 
The extended likelihood
function $\mathcal{L}$ for each individual channel k is defined as:
\begin{eqnarray}
\label{equ:likelihood}
\mathcal{L}_k \propto n_{data}!\cdot\exp\Big(-\frac{(n_{data}-\overline{n})^2}{2n_{data}}\Big)\cdot\prod_{i=1}^{n_{data}} \frac{w(\vec{\tau_i}, \vec{\alpha}) \, \epsilon(\vec{\tau_i})}
{\int w(\vec{\tau}, \vec{\alpha}) \, \epsilon(\vec{\tau}) \, \mathrm{d}\tau} \nonumber & \\
\end{eqnarray}
where $n_{data}$ denotes the number of data events in the channel k, 
 $\vec{\tau}$ the phase-space coordinates, $\vec{\alpha}$ the complex fit parameter, 
$\epsilon(\vec{\tau})$ the acceptance and reconstruction efficiency at the position $\vec{\tau}$ and
$\overline{n}=n_{data} \cdot \int w(\vec{\tau}, \vec{\alpha}) \, \epsilon(\vec{\tau}) \, \mathrm{d}\tau / 
\int \epsilon(\tau)\,\mathrm{d}\tau$.
$w(\vec{\tau}, \vec{\alpha})$ is the weight as given in
eq.~(\ref{eq:crossSection1}). $\alpha$ is the set of
  parameters, like coupling constants, production strengths and bare resonance parameters.
By logarithmizing eq.~(\ref{equ:likelihood}), approximating the integrals 
with Monte Carlo events and introducing the weight $Q_i$ for each event, the 
function to be minimized is then given by:
\begin{eqnarray}
\label{equ:likelihoodAprox}
 -\ln\,\mathcal{L}_k &\approx& -\sum_{i=1}^{n_{data}} Q_i \cdot \ln w(\vec{\tau_i}, \vec{\alpha}) \nonumber \\
       & &   +\Big(\sum_{i=1}^{n_{data}} Q_i\Big) \, \cdot \, \ln\Big( \frac{\sum_{j=1}^{n_{MC}} w(\vec{\tau_j}, \vec{\alpha})}{n_{MC}} \Big) \nonumber \\
       & & + \frac{1}{2} \, \cdot \, \Big(\sum_{i=1}^{n_{data}} Q_i\Big) \, \cdot \, \Big(\frac{\sum_{j=1}^{n_{MC}} w(\vec{\tau_j}, \vec{\alpha})}{n_{MC}}-1\Big)^2 , 
\end{eqnarray}
where $n_{MC}$ represents the number of selected Monte Carlo events for the channel k.\\
 The scattering data are provided by one dimensional data points
with errors for each scattering channel $k^\prime$ and here $\chi^2$ functions are introduced for
the minimization procedure. Due to the fact that the relation between $\chi^2$
and   $-\ln\,\mathcal{L}$ can be approximated by $\chi^2 =
 -2 \cdot \ln\,\mathcal{L}$  the total negative likelihood function to be
minimized is finally defined as:
\begin{eqnarray}
-\ln\,\mathcal{L}_{total} = \sum_{\pbarp\ channel \; k}
  -\ln\,\mathcal{L}_{k} + \sum_{scatt. \; channel\; k^\prime} 0.5 \cdot \chi^2_{k^\prime}
\end{eqnarray}  

\subsection{Choice of the Best Hypothesis}
The analysis was performed in two steps. In the first step, which was not 
performed with the full machinery as described before, the parameter space was 
investigated in order to fix reasonable start values for the final analysis.\\
Phase 1: The starting point was the performance of single channel fits for the three
\pbarp\ reactions individually. Several different hypotheses have been
tested in order to get a first glance on the potentially contributing
resonances. Based on
the outcome from~\cite{Amsler:2014xta} 
the maximal orbital momentum of the initial \pbarp\ system has been
chosen to be $L^{max}_\pbarp \,=\, 4$ for all fits. Also the production amplitudes have been 
limited to $L^{max}_{X\,s_r} \,=\, 5$. The dynamics of the scalar S wave has been realized by
the K-matrix parameterization with fixed parameters from~\cite{Anisovich:2002ij}. The $(K\pi)_S$
wave with isospin $I=1/2$ contributes in \pbarpToKpKmPi0. It
has been expressed for the single as well as for the coupled channel fits by the K-matrix
parameterization from the FOCUS
experiment~\cite{Pennington:2007se} containing only the two channels $K\pi$ and $K\eta^\prime$ and the
resonance pole $K^{*\pm}_0(1430)$. All remaining resonances
have been taken into account by 
Breit-Wigner approximations. In order to achieve meaningful results all
masses and widths have been fixed to PDG values. Only the narrow
$\phi(1020)$ resonance has been treated slightly differently. To
take into account the
detector resolution  a Voigtian
function, a convolution of a Breit-Wigner and a Gaussian
parameterization, has been used. The total width of $\phi(1020)$
has been fixed to the very precisely known value of 4.2~MeV$/c^2$~\cite{Tanabashi:2018oca} in order
to avoid unphysical correlations with the fit parameter representing the
description of  the detector resolution.\\
Phase 2: Based on the outcome
of these fits with simplified and rudimentary descriptions, more
sophisticated coupled channel fits taking into account the relevant scattering data
have been started. Apart from few isolated resonances the Breit-Wigner
and the scalar S wave parameterizations have been replaced by
K-matrix descriptions as specified in~\ref{dynamics}. The masses, coupling strengths and 
relevant background
terms have been released for all contributing
resonances. Only the narrow $\phi(1020)$ resonance, the $K^*(892)^\pm$ and the fixed parametrization for the 
$(K\pi)_S$-wave have been treated in the same way as done before for the single channel fits. \\
The best fit hypothesis had the following ingredients:
\begin{itemize}
   \item $f_0$-wave ($I^GJ^{PC}=0^+0^{++}$): 5 K-matrix poles ($f_0(500)$, $f_0(980)$, $f_0(1370)$, $f_0(1500)$, $f_0(1710)$) with 5 channels 
($\pi\pi$, $K\bar{K}$, $\eta\eta$, $\eta \eta^\prime$, $2\pi \; 2\pi$). $2\pi \; 2\pi$ is treated as an effective channel with $m_1 = m_2 = m_\pi+m_\pi$ according to eq.~(\ref{equ:RhoDefault}) covering all channels with the decay into four pions.\\
 Similar descriptions for the $f_0$-wave have been used in two other previous analyses. A complementary K-matrix approach
with the same decay channels and the same number of poles has been chosen in~\cite{Anisovich:2002ij}. The advantage was that a richer set of data samples was considered. On the other hand the treatment of analyticity was more rudimentary since Chew-Mandelstam functions have not been taken into account. In~\cite{Ropertz:2018stk} instead
high-accuracy dispersive representations have been applied based on a three channel description only.    
\item $f_2$-wave ($I^GJ^{PC}=0^+2^{++}$): 4 K-matrix poles ($f_2(1270)$, $f^\prime_2(1525)$, $f_2(1810)$, $f_2(1950)$) with 4 channels 
($\pi\pi$, $2\pi \; 2\pi$, $K\bar{K}$, $\eta\eta$).\\
The decay mode to $\eta\eta^\prime$ was not used here, since the relevant resonances are expected to couple only weakly 
to this channel. Also here $2\pi$ $2\pi$ is used as an effective decay mode. 
\item $\rho$-wave ($I^GJ^{PC}=1^+1^{--}$): 2 K-matrix poles ($\rho(770)$, $\rho(1700)$) with 3 channels ($\pi\pi$, $K\bar{K}$ and $2\pi \; 2\pi$). 
This wave is only relevant for the \pbarp\ annihilation channel to \KpKmPi0.  
\item $a_0$-wave ($I^GJ^{PC}=1^-0^{++}$): 2 K-matrix poles ($a_0(980)$, $a_0(1450)$) with 2 channels ($\pi^0 \eta$, $K\bar{K}$).\\
Only these two channels were directly measurable via the $\pbar p$-data. No information from $\pi\pi$-scattering data 
could be used.
An effective channel for covering all decay modes into $2\pi\;2\pi$ is not introduced here. Fits with additional free
parameters related to such an effective channel lead to unreasonable results since no scattering data and thus no 
constraints on inelasticities can be used. 
\item $a_2$-wave ($I^GJ^{PC}=1^-2^{++}$): 2 K-matrix poles ($a_2(1320)$, $a_2(1700)$) with 2 channels ($\pi^0 \eta$, $K\bar{K}$).\\
For the choice of the decay channels the same arguments as before hold.
\item $(K\pi)_S$-wave ($I\ J^{P}=1/2\ 0^{+}$): Fixed K-matrix parameterization from the FOCUS experiment~\cite{Pennington:2007se} 
with one pole ($K^*_0(1430)$) and 2 channels ($K\pi, K\eta^\prime$).
\item $\pi_1$-wave ($I^GJ^{PC}=1^-1^{-+}$): 1 K-matrix pole with two channels ($\pi\eta, \pi \eta^\prime$). This description 
is motivated by the recent analysis of COMPASS data~\cite{Rodas:2018owy}, where the observed rapid phase 
      shifts of the $1^{-+}$ wave in $\pi \eta$ and in $\pi \eta^{\prime}$ can be 
explained by the presence of only one pole.       
\item $\phi(1020), \, K^*(892)^\pm$\\
These isolated resonances occurring in the channel\linebreak 
$\KpKmPi0$ are parametrized by Breit-Wigner functions.
 \end{itemize} 
 In addition, constant background terms for the K-matrix as well as for the P-vector were needed for the
three waves $f_0$, $f_2$ and $\rho$ in order to get consistent good
results for the simultaneous description of the \pbarp\ and scattering
data. All background terms of the effective $2\pi \; 2\pi$ channel for
the $f_0$-, $f_2$- and $\rho$-wave have been fixed to zero.
 For the dynamics of the $a_0$- and $a_2$-wave no background terms
 have been considered and the $\pi_1$-wave is described by a constant background term
only for the $\pi \eta$ channel.\\
The K-matrix parameters for the $f_0$-, $f_2$- and $\rho$-wave obtained by the best fit are
summarized in the supplemental material. The parameterizations for the $a_0$, $a_2$ and $\pi_1$-waves are not
listed there since the descriptions are simplified and limited to two channels each.\\ 
 For the selection of the best fit hypothesis the Bayesian information criterion (BIC) and the
Akaike information criterion (AIC)~\cite{Schwarz:1978AnnStat, Burnham:2002Springer} have been used, which are given by
\begin{eqnarray}
  \label{eq:BIC}
\mathrm{BIC} = -\ln\,\mathcal{L} + \mathrm{ndf} \cdot ln(N)
\end{eqnarray}
and
\begin{eqnarray}
  \label{eq:AIC}
\mathrm{AIC} = -\ln\,\mathcal{L} + 2 \cdot \mathrm{ndf},
\end{eqnarray}
with ndf being the number of free fit parameters and $N$ the number of data events. The best fit is characterized by 
a minimal value. The penalty term related to the number of free parameters is larger in BIC than in AIC. Therefore the exclusive
use of  AIC generally tends to overfitting, while the more stringent criterion BIC instead favors solutions with less 
parameters and tends to underfitting. A very significant result is achieved when both
criteria prefer the same hypothesis. In Tab.~\ref{tab:FitLHs} the BIC- and AIC-values for the best fit and 
for alternative fits are summarized. The best hypothesis is selected by having the lowest AIC+BIC 
value. A better BIC value is achieved for 
the hypothesis with only three $f_2$ poles. However, the worsening based on the $\Delta$AIC value 
is considerably larger. Slightly better AIC-values are achieved for
the hypothesis by adding the $\rho_3(1690)$ resonance and for the ones
by adding the $\phi(1680)$  and a contribution of the $\pi_1$ to the $\pi\eta\eta$ channel, respectively. 
But due to the additional large number of free parameters the BIC criteria gets dramatically worse compared to the best hypothesis. Alternatively to the $\pi_1$ contribution with one pole also this spin-exotic wave containing two
poles was tested but discarded because of worse $\Delta$BIC- and
$\Delta$AIC values. Also the description of the $\pi_1$-wave with a relativistic 
Breit-Wigner function led to a  significantly worse fit
result. 
In addition fits have been performed by adding one more pole for the $f_0$-, $f_2$-, $a_0$- and $a_2$-waves 
which are not explicitly listed in Tab.~\ref{tab:FitLHs}.
 These fits do not yield significant improvements. In most cases
the additional poles move far away from the real axis into the complex energy plane and mimick the slowly varying behavior of the K-matrix background terms.

\begin{table}[htb]
\caption{Likelihood values and the  number of free 
  parameters for fits with the best and alternative hypotheses.
  $\Delta$NLL, $\Delta$ndf, $\Delta$BIC and $\Delta$AIC are the differences of the
obtained negative log likelihood values, the number of free parameters
as well as the BIC and AIC values between the alternative and the best
hypothesis.
The fits marked with (*) are taken into account for the
estimation of the systematic uncertainties. Exemplarily, the results
using the scattering data for the solution (-$\;$-$\;$-) from
\cite{Ochs:2013gi} are shown.}

\label{tab:FitLHs}      
\footnotesize
\centering
\adjustbox{max width=\linewidth}{
\begin{tabular}{l  r r r r r r} 
  \hline\noalign{\smallskip}
 hypothesis   & NLL & ndf & $\Delta$NLL & $\Delta$ndf & $\Delta$BIC & $\Delta$AIC\\
\noalign{\smallskip}
  \hline \noalign{\smallskip}
  best hypothesis & -43198 & 595 & 0  & 0 & 0  & 0  
  \\
  \\ \noalign{\smallskip}
  w/ 3 $f_2$ poles  (*) & -43072 & 562 & 126 & 33 & -135 & 185
     \\  \noalign{\smallskip}
  w/  $\pi_1 \; \eta$ (*) & -43239 & 615 & -41 & 20 & 150 & -43
  \\ \noalign{\smallskip}     
  w/  $\phi(1680) \; \pi^0$ (*) & -43223 & 619 & -25 & 24 & 230 & -3
  \\ \noalign{\smallskip}                                                  
  w/ $\rho_3(1690) \; \pi^0$ (*) & -43267 & 633 & -69 & 38 & 305 & -64
    \\
  \\  \noalign{\smallskip}      
  $\pi_1$ w/ Breit-Wigner  & -43151 & 592 & 47 & -3 & 57 & 87 
    \\ \noalign{\smallskip}
   w/  2$\,\pi_1$ poles & \multirow{2}{*}{-43212} & \multirow{2}{*}{612} & \multirow{2}{*}{-14} & \multirow{2}{*}{17} & \multirow{2}{*}{171} & \multirow{2}{*}{12} 
 \\ 
  in  $\pi^0 \pi^0 \eta$  &  & & &  & &
 \\ \noalign{\smallskip}
  w/  3$\,\rho$ poles  & -43206 & 613 & -8 & 18 & 194 & 19
 \\ \noalign{\smallskip}
  w/ 1 $a_2$ pole  & -43074 & 578 & 124 & 17 & 49 & 214
    \\ \noalign{\smallskip}
  w/o  $\pi_1 \; \pi^0$  & -43030 & 570 & 168 & 25 & 44 & 286
    \\  \noalign{\smallskip} 
  w/ 4 $f_0$ poles & -42957 & 559 & 241 & 36 & 61 & 410
      \\  \noalign{\smallskip}
  w/ 1 $a_0$ pole & -42955 & 572 & 243 & -23 & 351 & 439
   \\
  \noalign{\smallskip}
  \hline
\end{tabular}
}
\end{table}

\subsection{Comparison of Data and Fit}

\subsubsection{Results for \pbarp\ Data}
\begin{figure}[ht]
\hspace{-4mm}\includegraphics[width=0.5\textwidth]{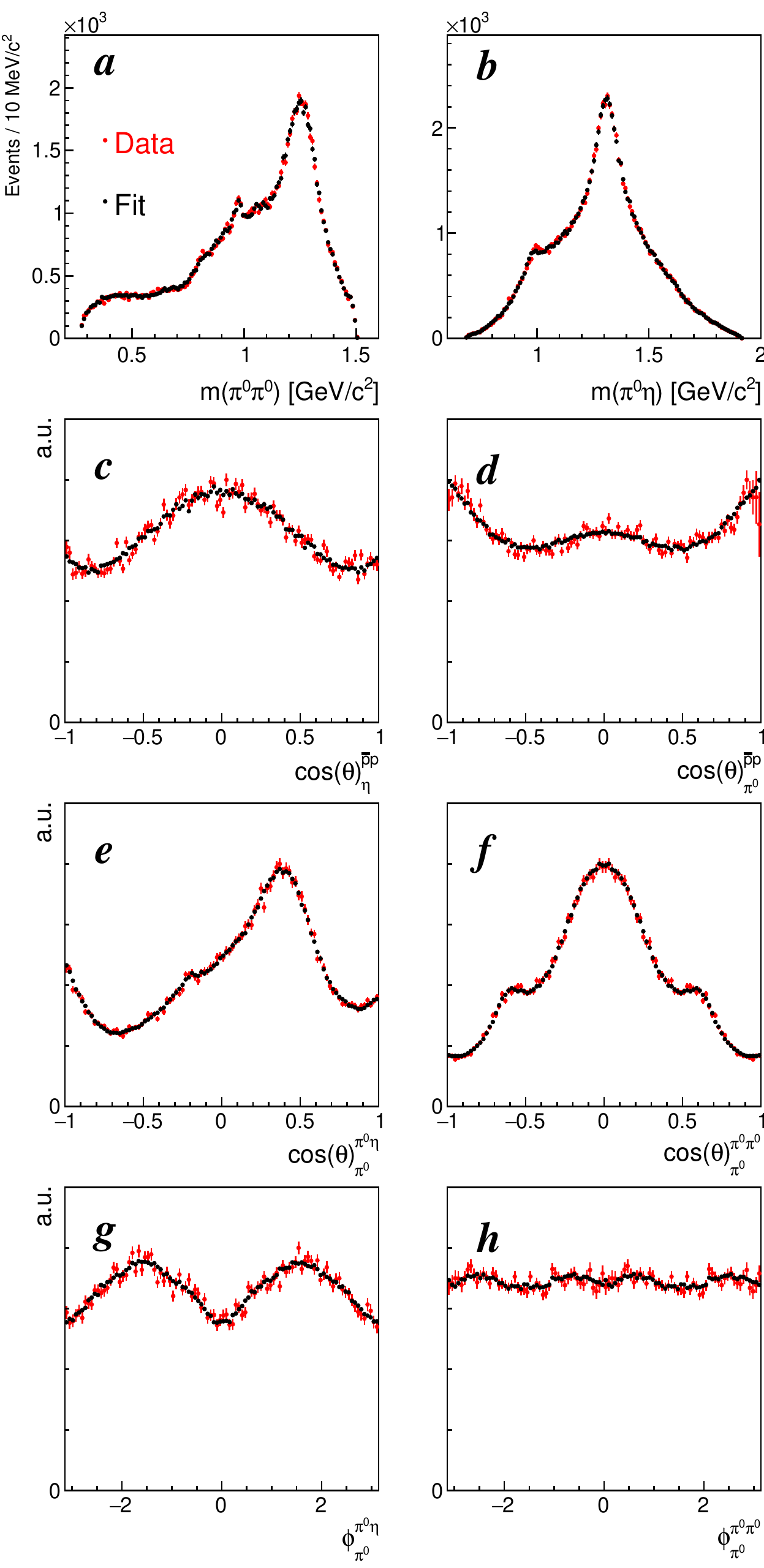}
\centering
\caption{Not acceptance corrected invariant $\pi^0\pi^0$ and $\pi^0\eta$ mass distributions
  (a and b), acceptance corrected decay angular distributions for the
  production (c and d) and for the decay (e-h) of the reaction
  \pbarpToPi0Pi0Eta\ . The data are marked with red and the
 fit result is illustrated by black dots with error bars.}
\label{fig:ResultPiPiEta}       
\end{figure}

\begin{figure}[ht]
\hspace{-4mm}\includegraphics[width=0.5\textwidth]{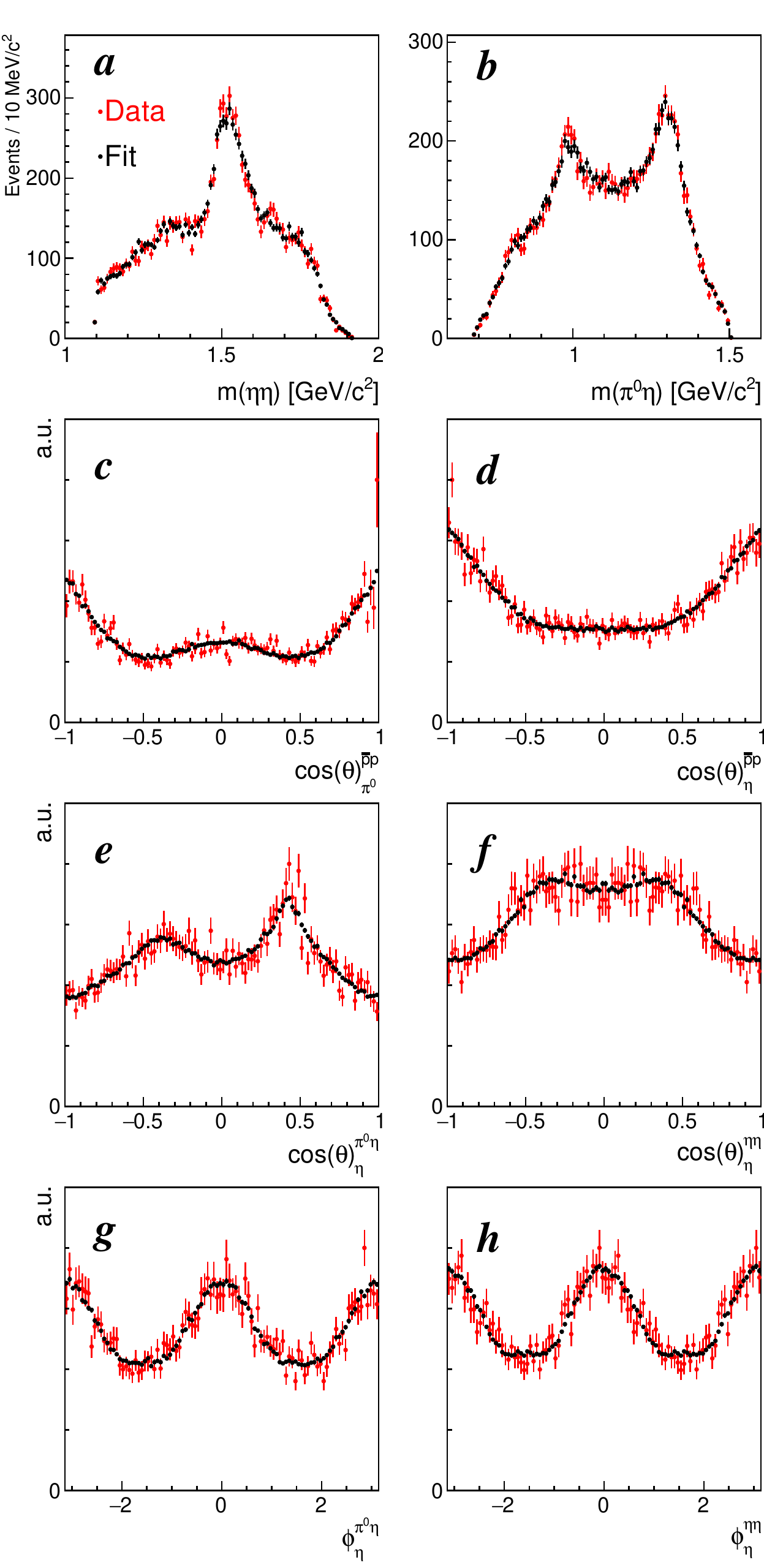}
\centering
\caption{Not acceptance corrected invariant $\eta\eta$ and $\pi^0\eta$ mass distributions
  (a and b), acceptance corrected decay angular distributions for the
  production (c and d) and for the decay (e-h) of the reaction
  \pbarpToPiEtaEta\ . The data are marked with red and the
 fit result is illustrated by black dots with error bars.}
\label{fig:ResultPiEtaEta}       
\end{figure}

 \begin{figure}[ht]
\hspace{-4mm}\includegraphics[width=0.5\textwidth]{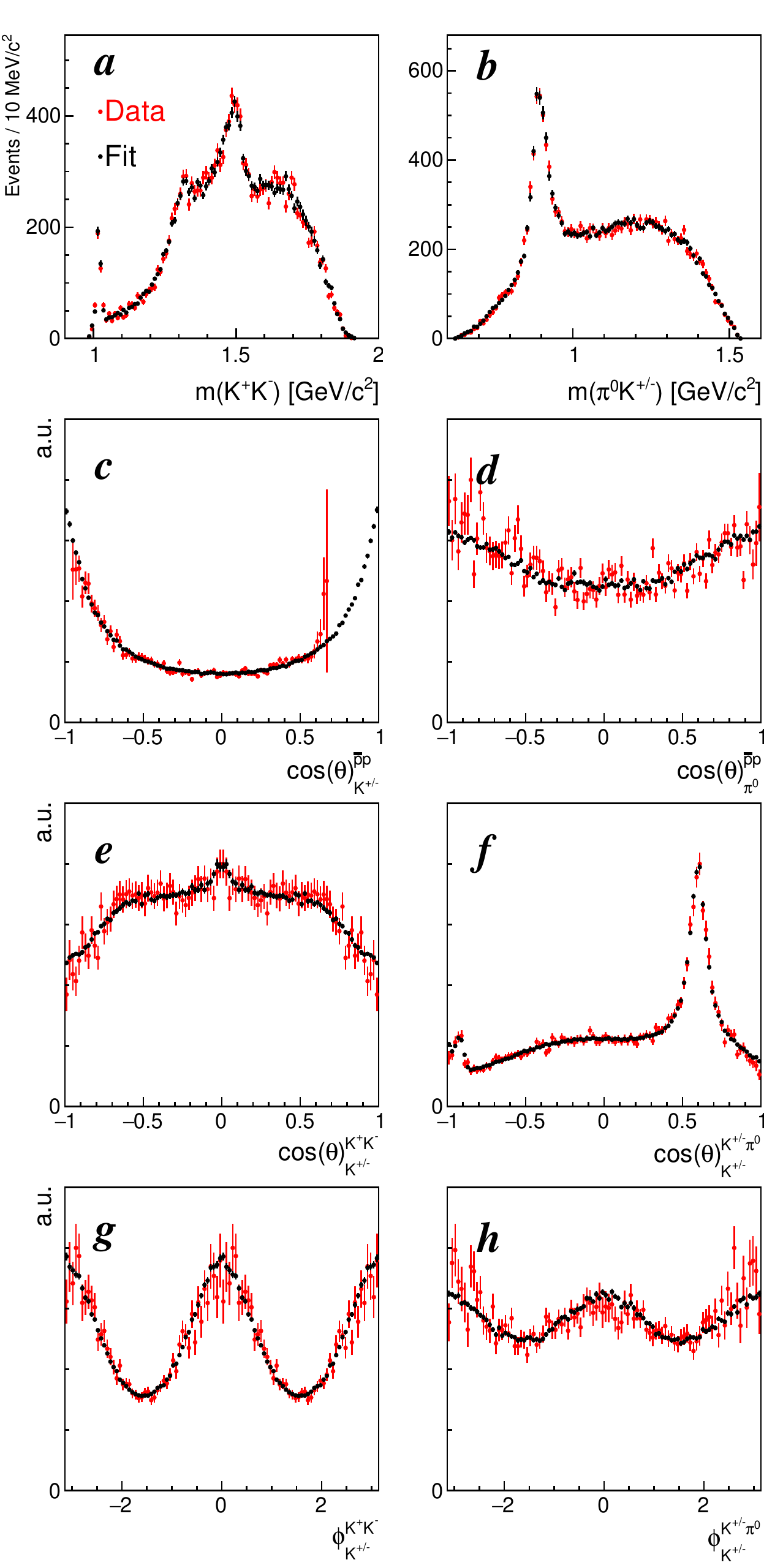}
\centering
\caption{Not acceptance corrected invariant $K^+K^-$ and $K^\pm\pi^0$ mass distributions
  (a and b), acceptance corrected decay angular distributions for the
  production (c and d) and for the decay (e-h) of the reaction
  \pbarpToKpKmPi0\ . The data are marked with red and the
 fit result is illustrated by black dots with error bars.}
\label{fig:ResultKKPi}       
\end{figure}

\begin{figure}[htb]
\vspace{2.5mm}
\begin{tabular}{cc}
\hspace{-5mm}\includegraphics[width=0.26\textwidth]{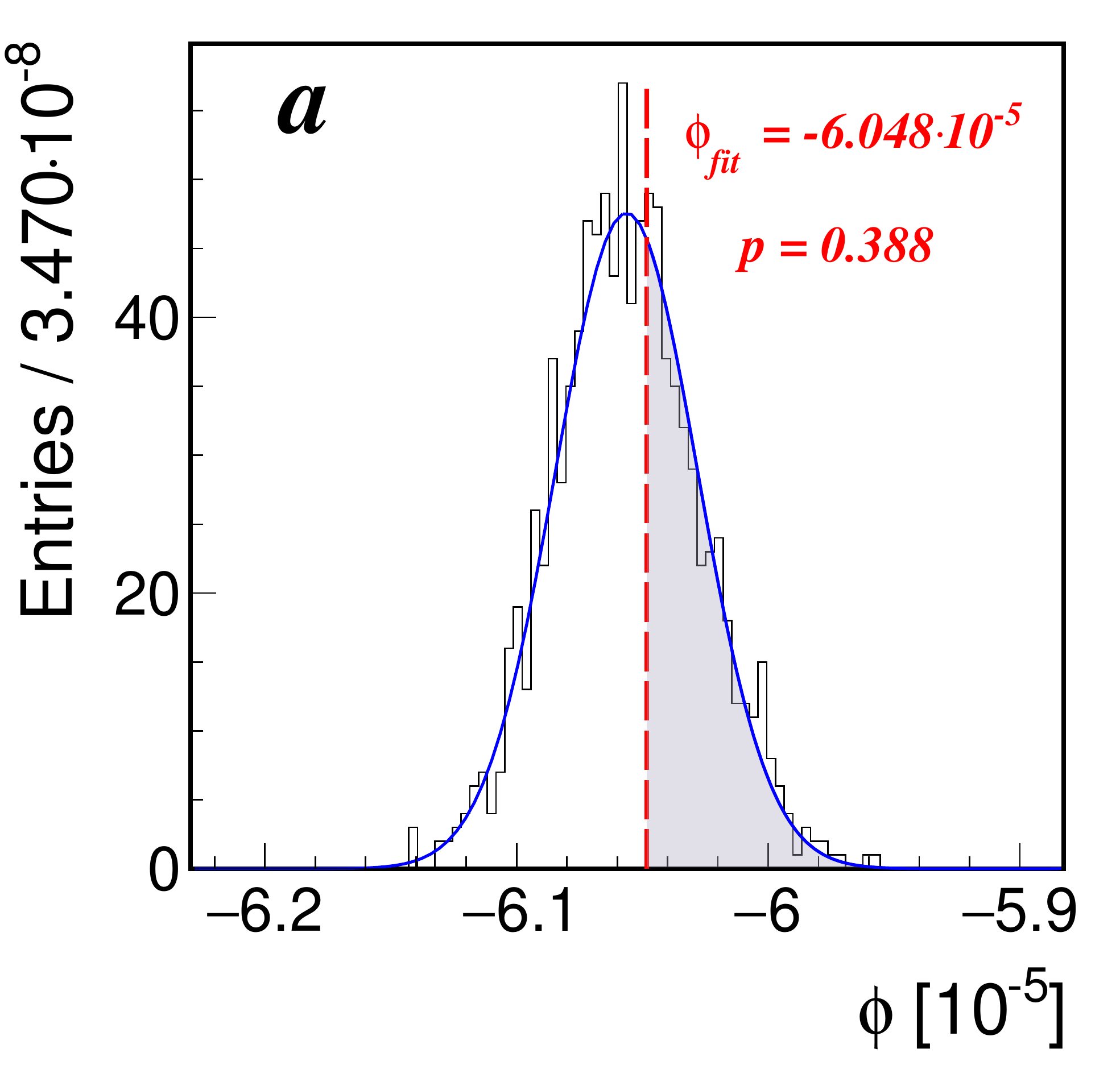} &
\hspace{-5mm}\includegraphics[width=0.26\textwidth]{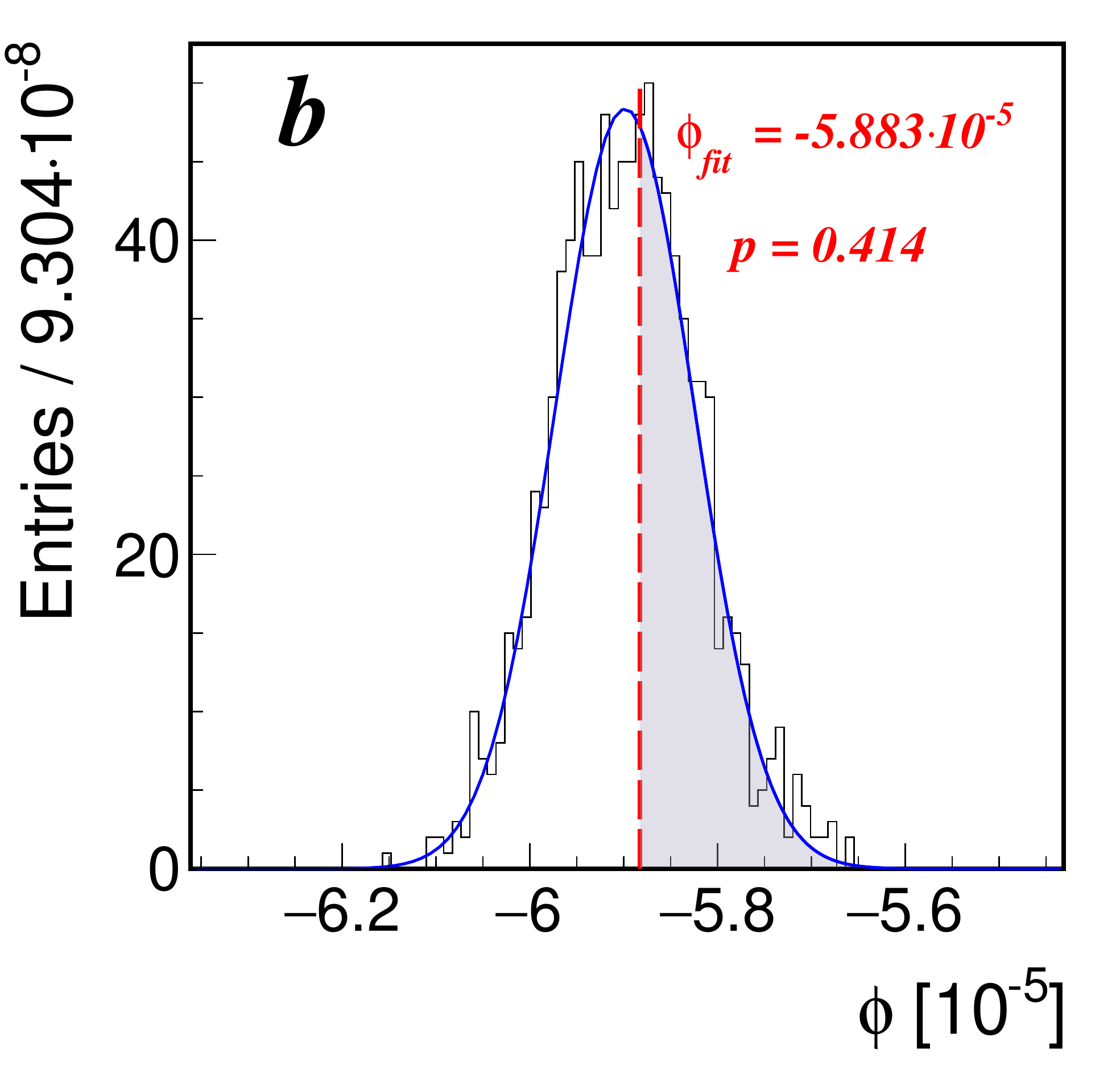} \\
\hspace{-5mm}\includegraphics[width=0.26\textwidth]{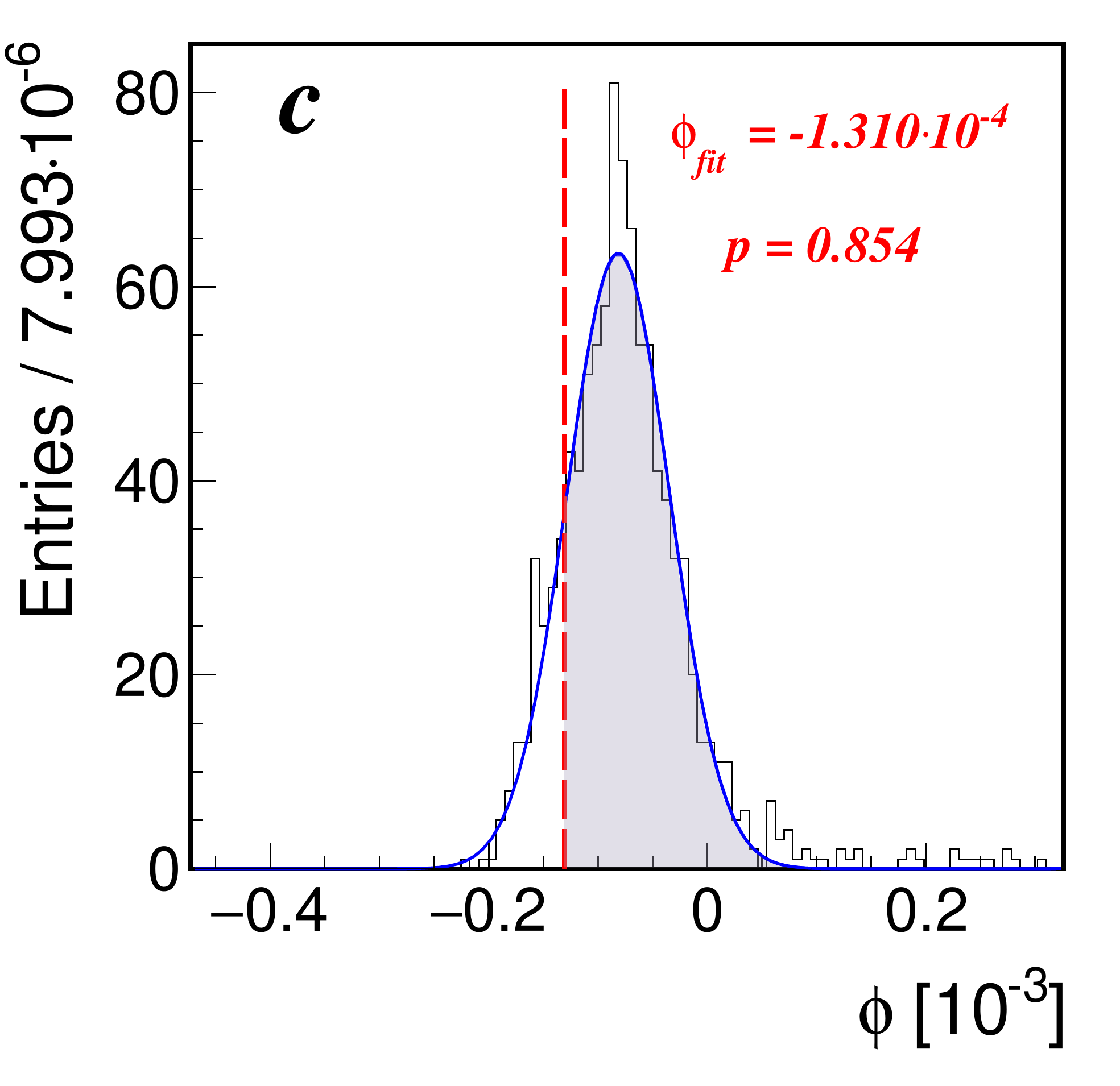} &\\
\end{tabular}
\vspace{-1mm}
\caption{Goodness of fit. Statistical energy $\phi$ distribution and corresponding 
p-value for $\bar{p}p\to$ (a) $\PiPiEta$, (b) $\PiEtaEta$ and (c)
$\KpKmPi0$. The red dashed line marks the position of the nominal statistical energy
$\phi_{fit}$ and the gray area illustrates the number of scenarios with energies above $\phi_{fit}$.}
\label{fig:GOF}       
\end{figure}

The fitted mass and decay angular distributions
for the reactions \pbarpToPi0Pi0Eta\ , \PiEtaEta\ and \KpKmPi0\
are compared with the data in Fig.~\ref{fig:ResultPiPiEta}, \ref{fig:ResultPiEtaEta}
and \ref{fig:ResultKKPi}, respectively. A good description of the data
can be clearly seen for all projections. In Fig.~\ref{fig:ResultKKPi} (c) an acceptance hole is clearly visible in the production
angular distribution of the $K^\pm\pi^0$ system for 
$cos \theta^\pbarp _{K^\pm \pi^0}$  between 0.7 and 1. The loss of acceptance is caused by the limited coverage of the
jet drift chamber in the very forward direction.
However, this range can be fairly covered by the extrapolation of the
fit result.\\  
 A goodness of fit test of the obtained probability density function to the data has been 
performed by utilizing a multivariate analysis based on the concept of statistical
energy~\cite{Aslan:2004}. The underlying principle is the comparison of the event density 
distribution in the phasespace volume between the reconstructed data sample and 
a set of Monte Carlo samples which are generated with the obtained fit parameters
and different random seeds each.
This powerful binning free approach makes use of the distance between the events 
in the multidimensional phase-space volume. The p-value is calculated from the number 
of the comparative scenarios with energies above the nominal statistical energy
$\phi_{fit}$  divided by the number of all scenarios. The general idea behind and all details 
of the energy test are explained extensively in~\cite{Aslan:2004}. As shown in
Fig.~\ref{fig:GOF}, p-values of 
0.388, 0.414 and 0.854 demonstrate the good description of 
the data for all three channels \PiPiEta\ , \PiEtaEta\ and \KpKmPi0\ ,
respectively.
Furthermore this outcome
reveals that the model selection based on the BIC and AIC values is a
good choice for the extraction of the best fit hypothesis.\\ 
 The reliability of the fit procedure has also been tested by an input-output check. 
Based on the parameter file obtained by the best fit to the data (called reference fit), Monte Carlo samples for 
the individual channels have been generated. The reliability has 
been checked by performing fits where the start parameters were 
randomized within 15~$\sigma$ for the amplitude and 10~$\sigma$ for the resonance parameters. 
After the fitting procedure the obtained physical quantities are compared with the outcome of 
the reference fit. The agreement is good and the deviations of all relevant quantities, i.e.
  differential cross sections, masses, widths, partial widths 
and spin density matrix elements, are less than $3\sigma$ of the statistical uncertainties.

\subsubsection{Results for Scattering Data}

\begin{figure*}[htb]
\begin{tabular}{ccccc}
\hspace{-3mm}\includegraphics[width=.25\textwidth]{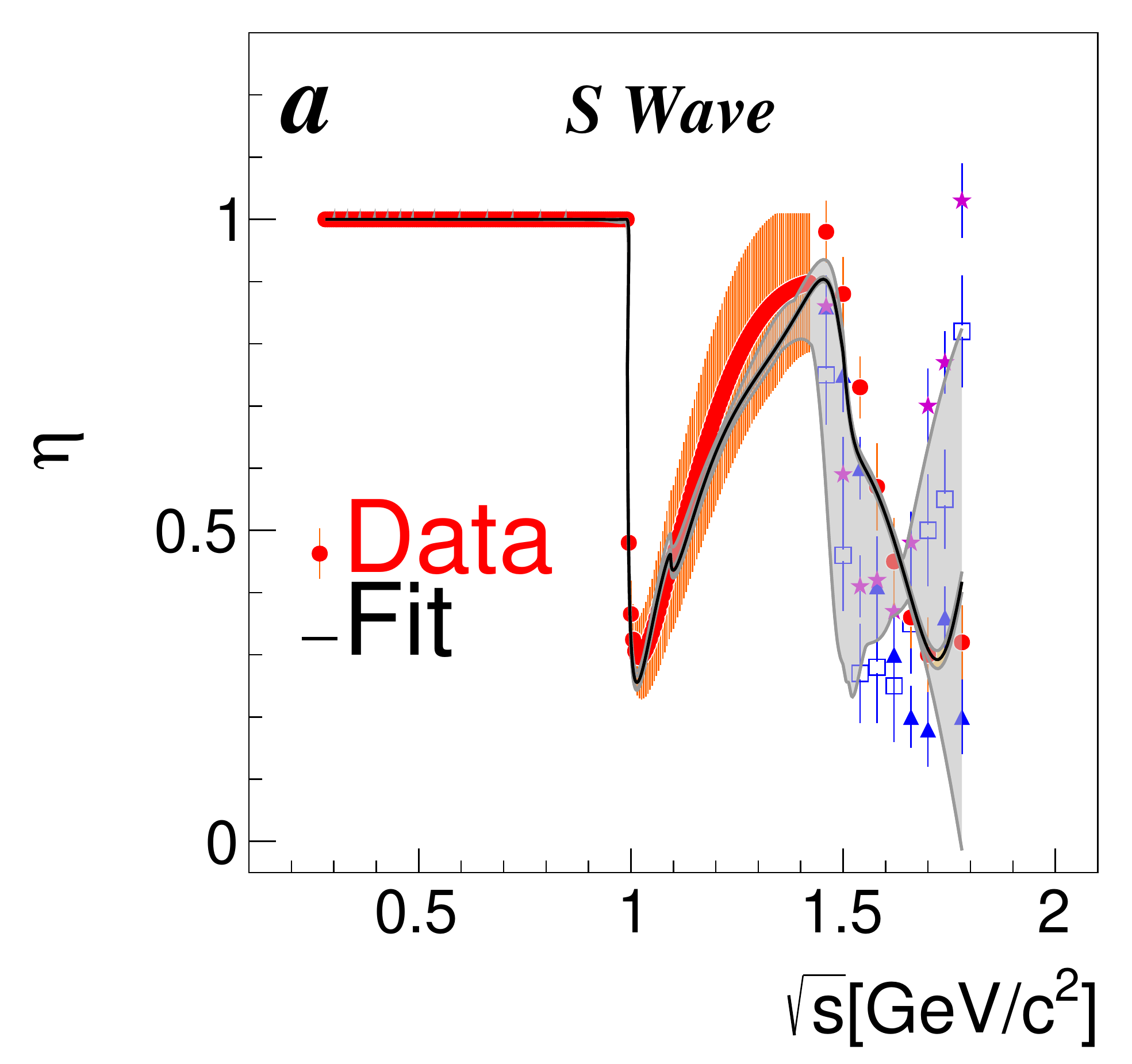} &
\hspace{-3mm}\includegraphics[width=.25\textwidth]{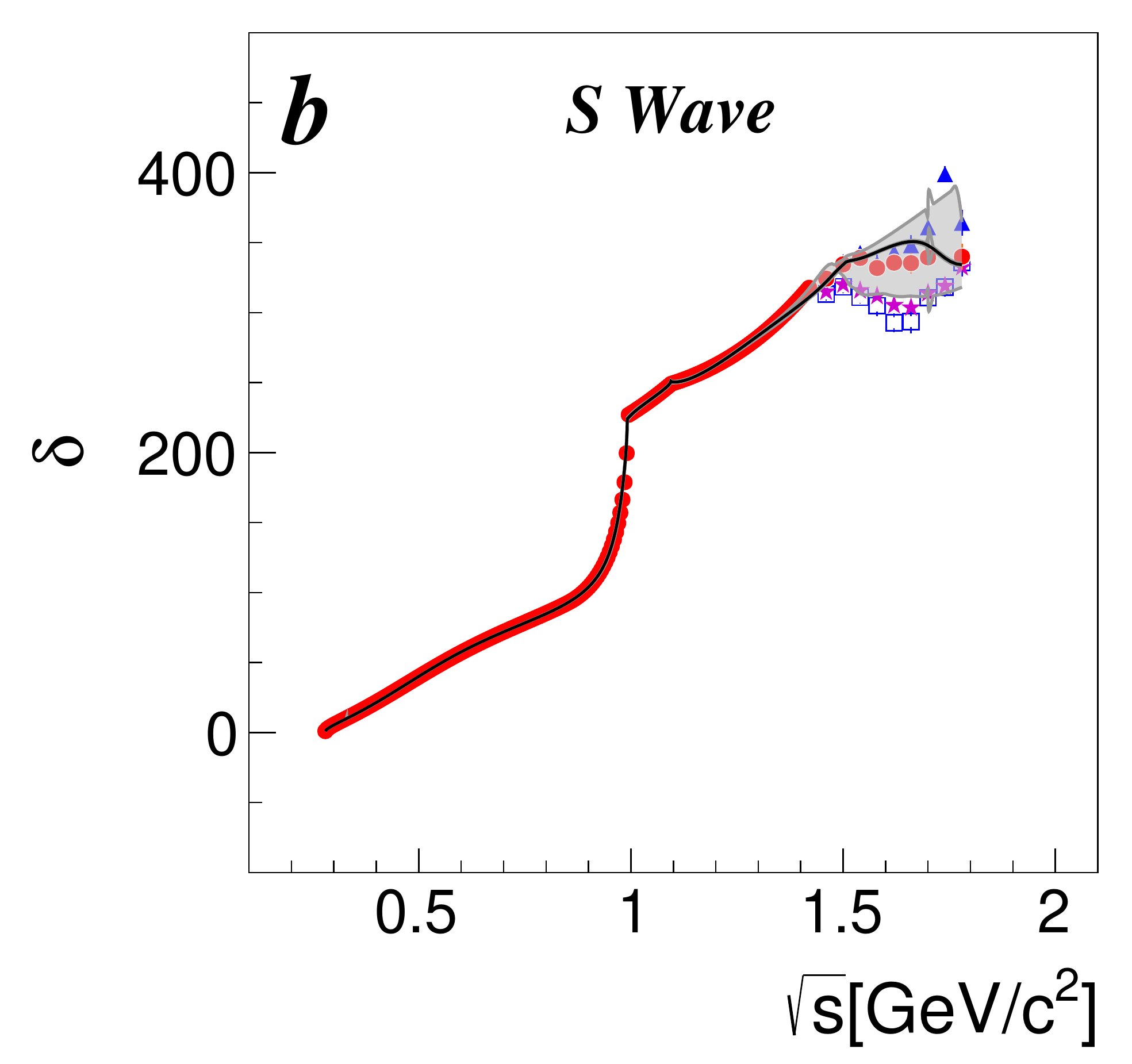}      &
\hspace{-3mm}\includegraphics[width=.25\textwidth]{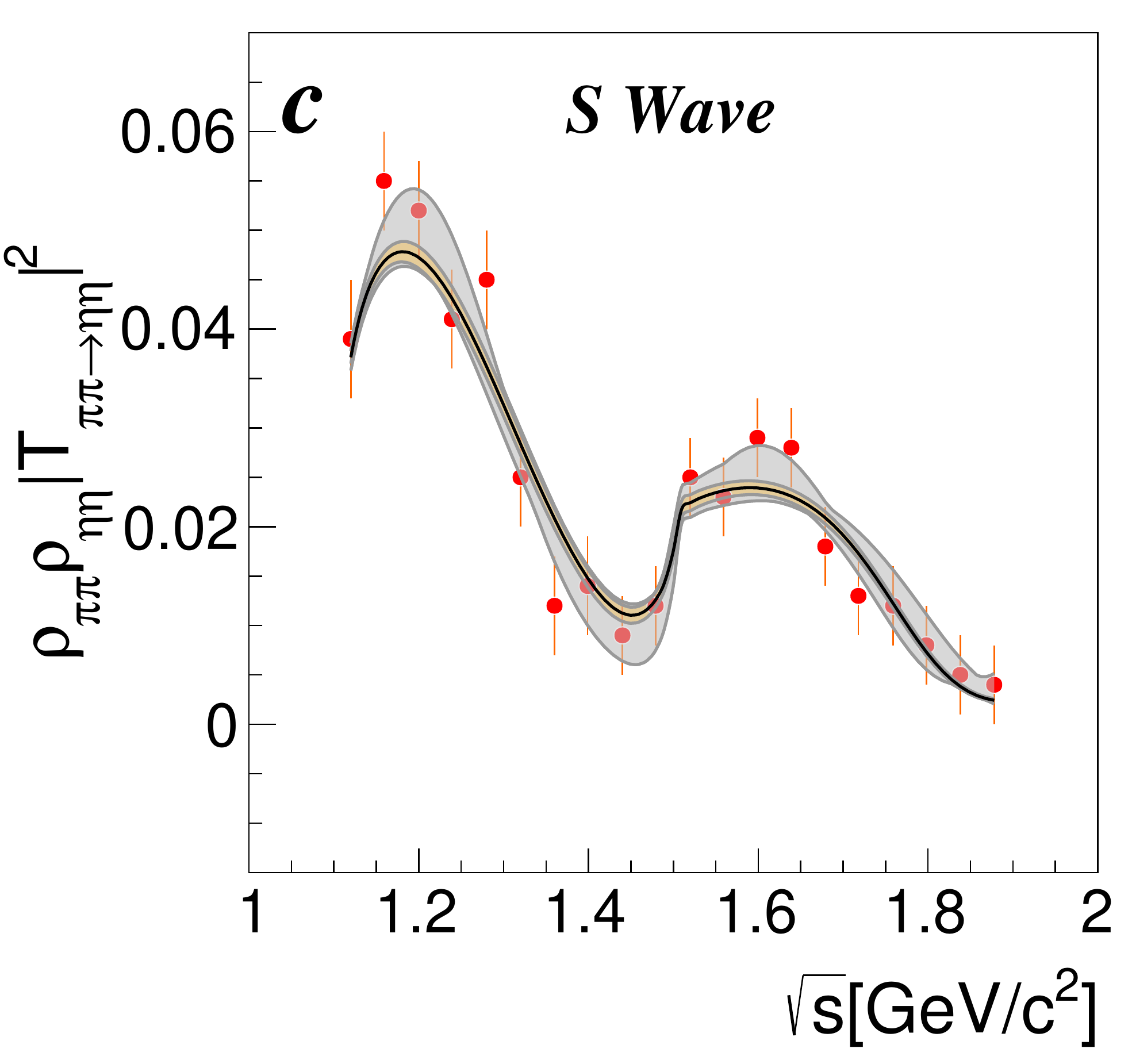} &
\hspace{-3mm}\includegraphics[width=.25\textwidth]{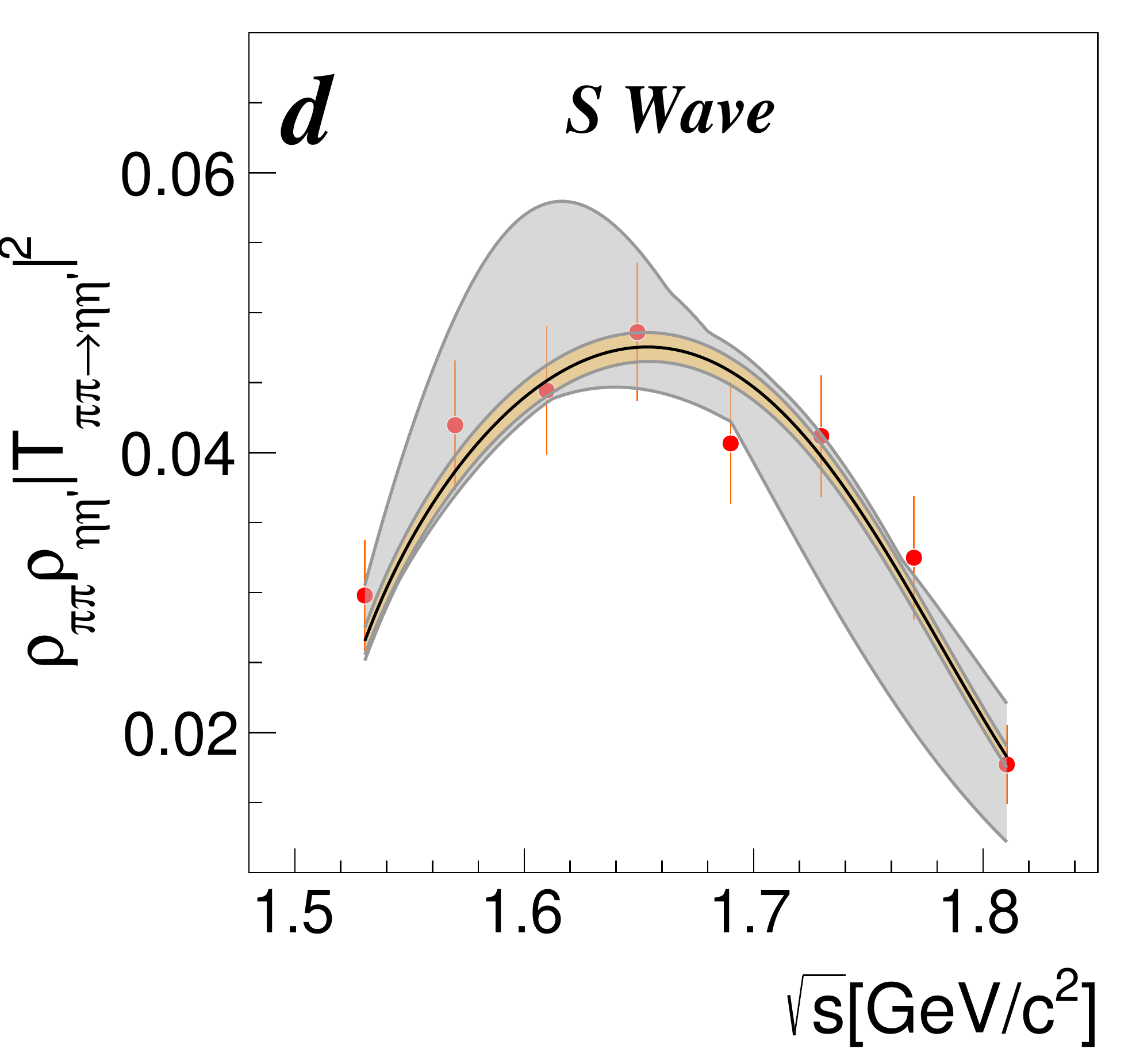} \\
\hspace{-3mm}\includegraphics[width=.25\textwidth]{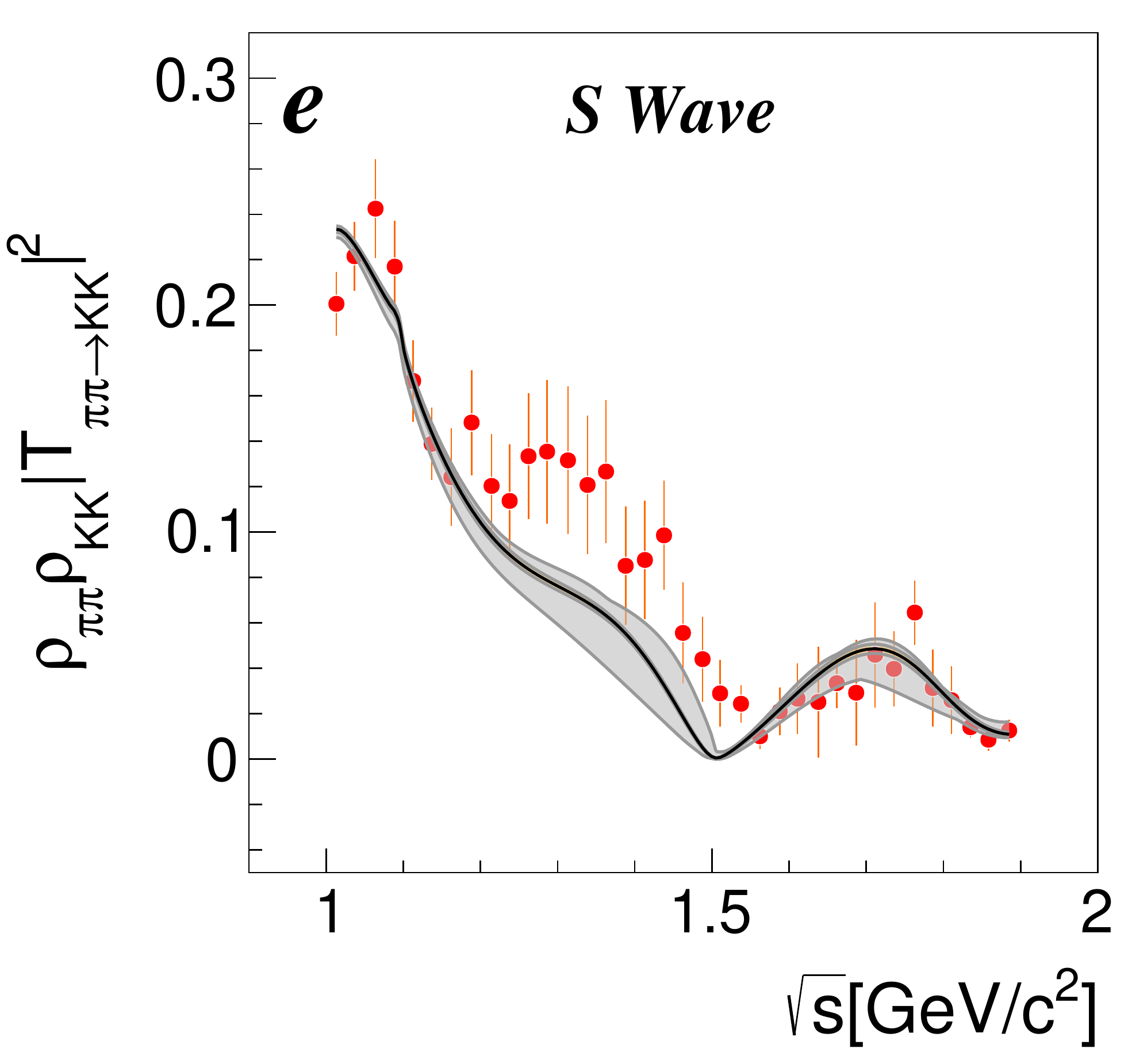}     & 
\hspace{-3mm}\includegraphics[width=.25\textwidth]{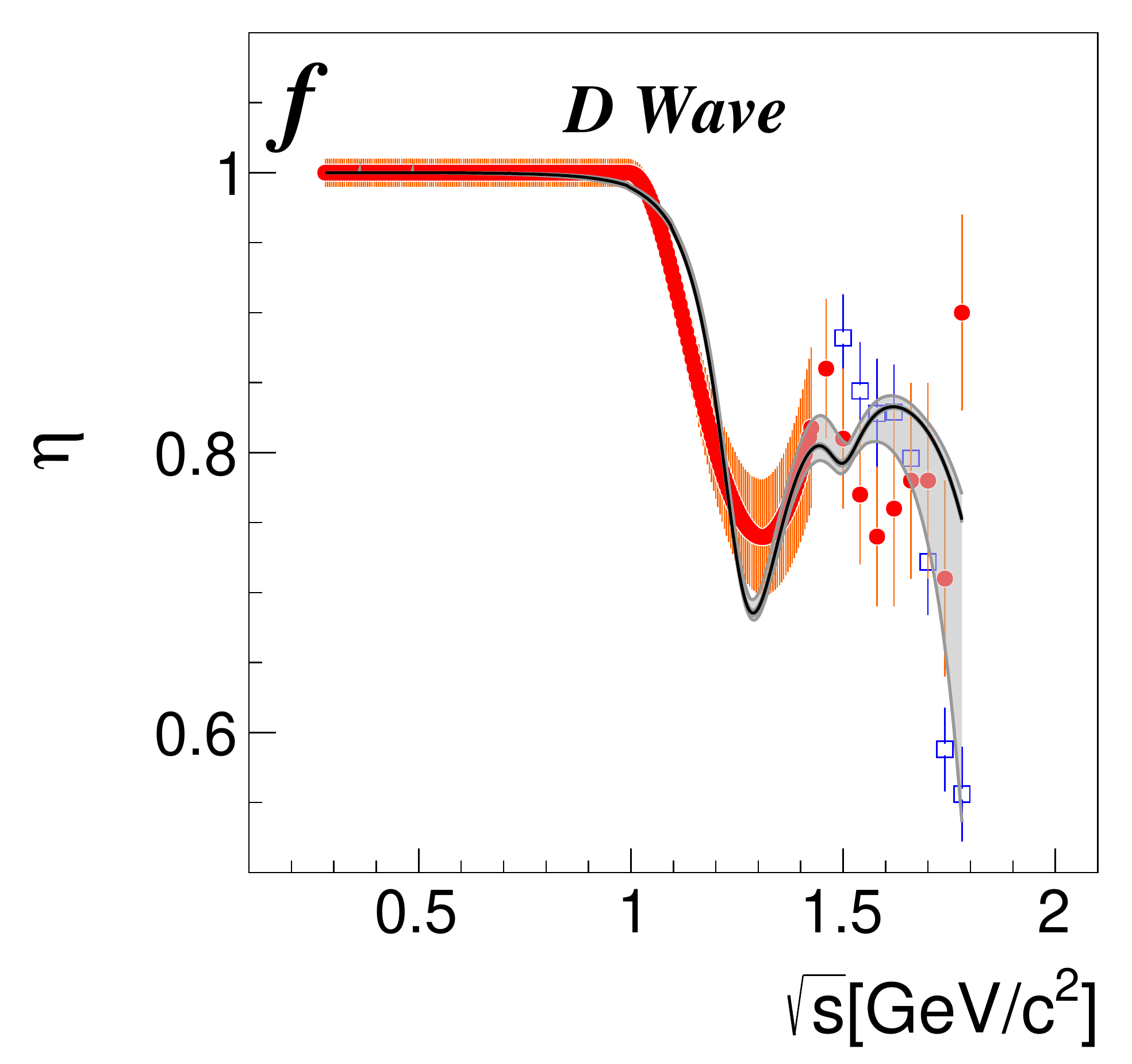}      &
\hspace{-3mm}\includegraphics[width=.25\textwidth]{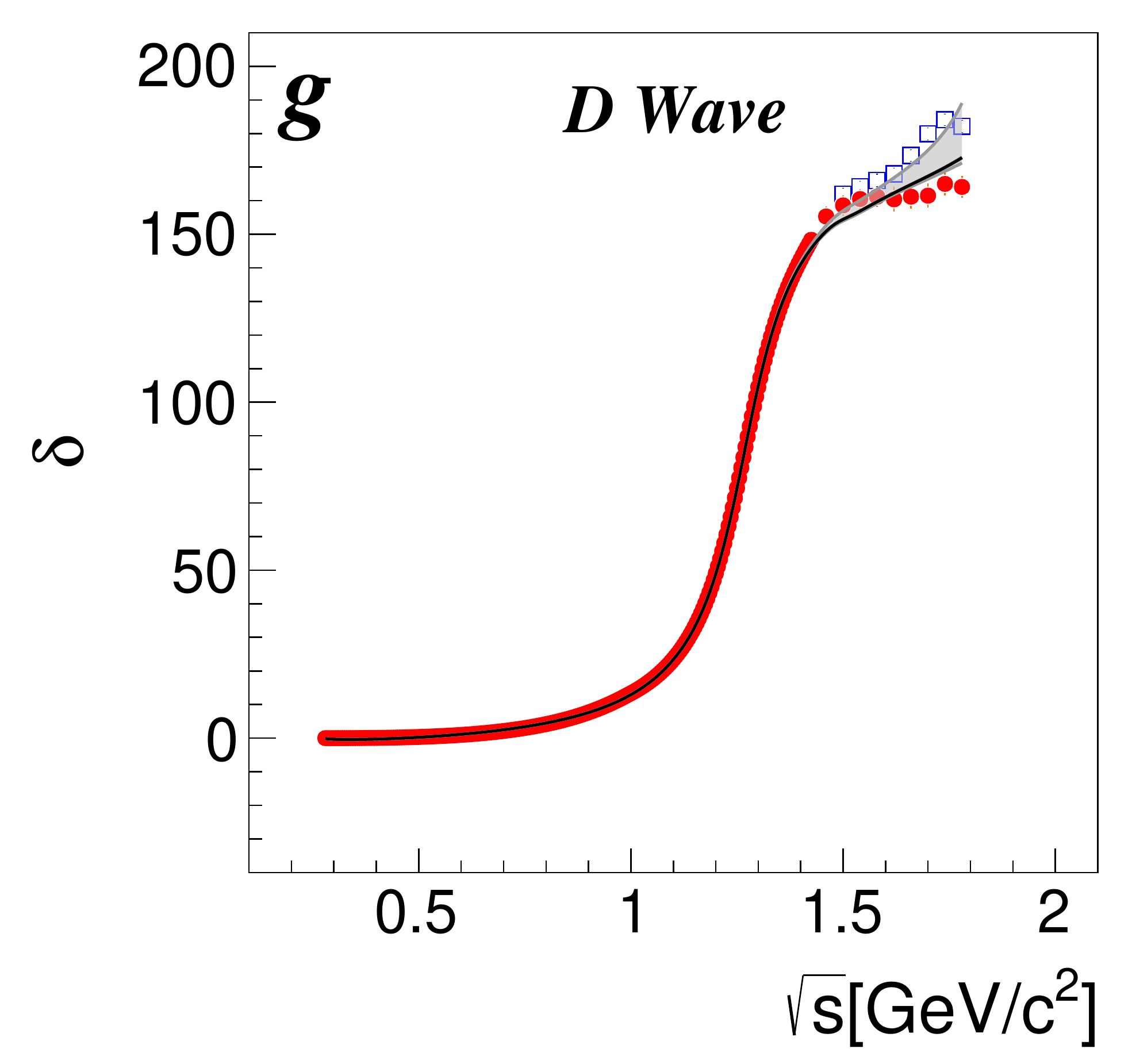} &
\hspace{-3mm}\includegraphics[width=.25\textwidth]{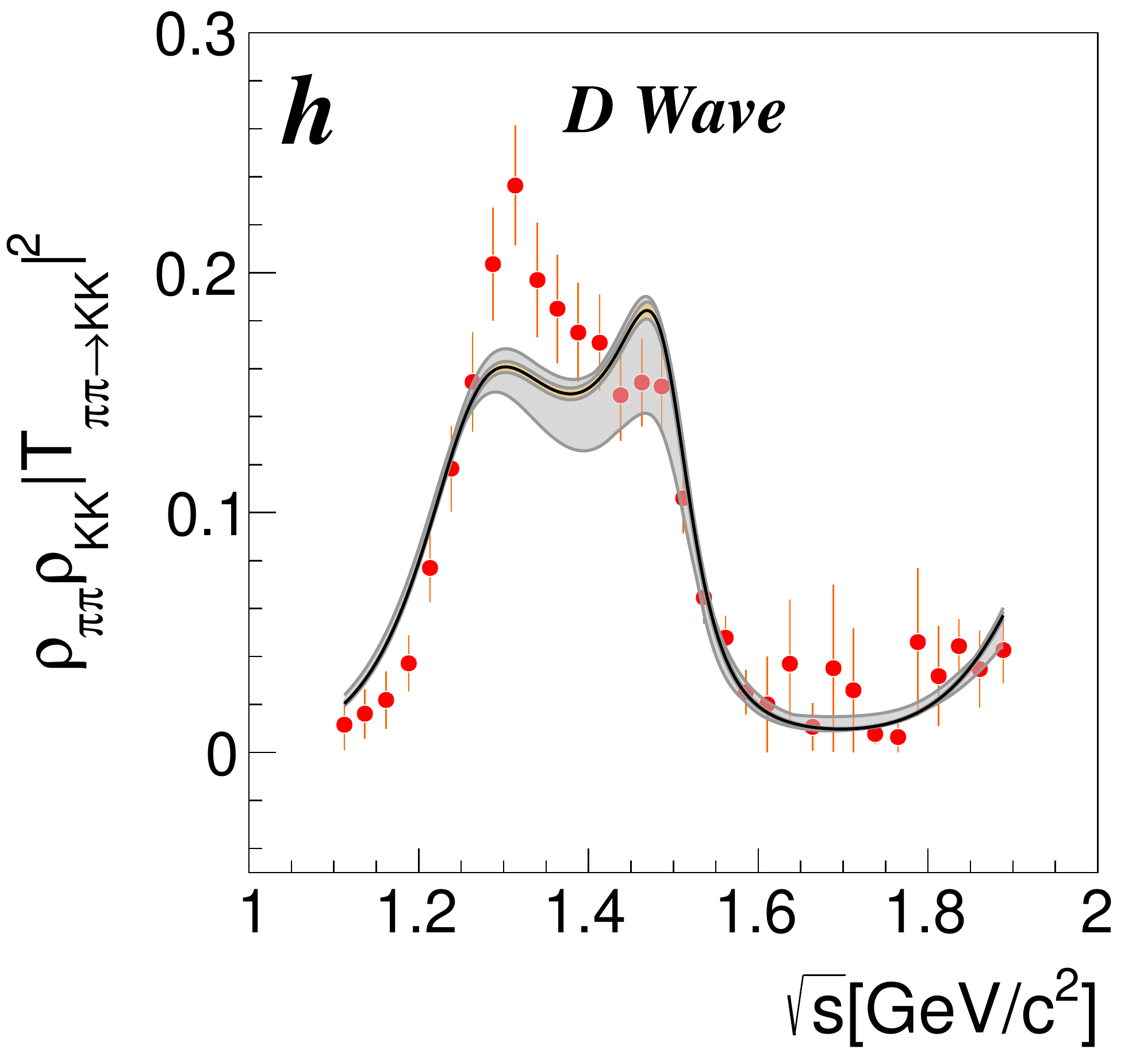}          \\
\hspace{-3mm}\includegraphics[width=.25\textwidth]{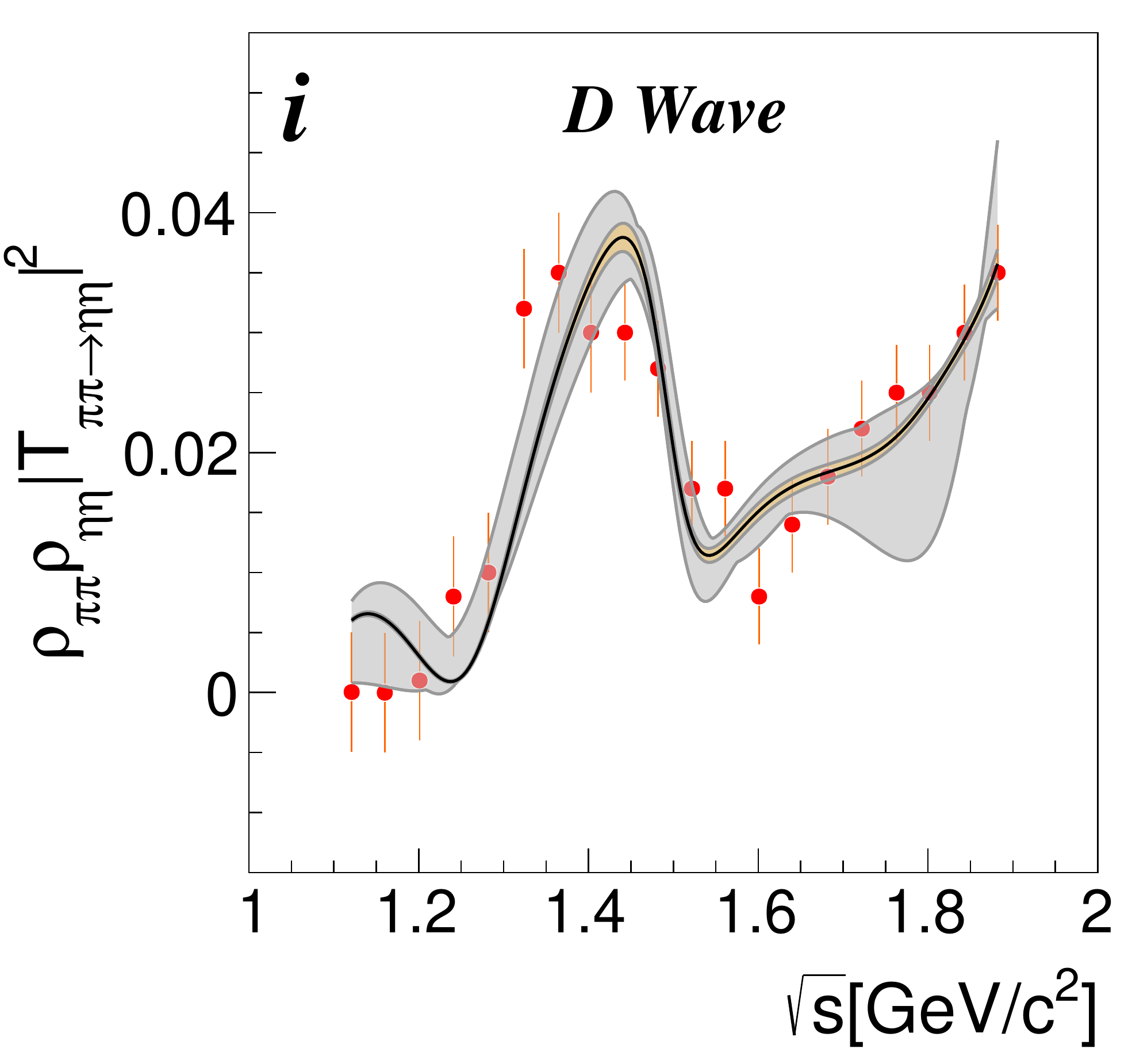}    & 
\hspace{-3mm}\includegraphics[width=.25\textwidth]{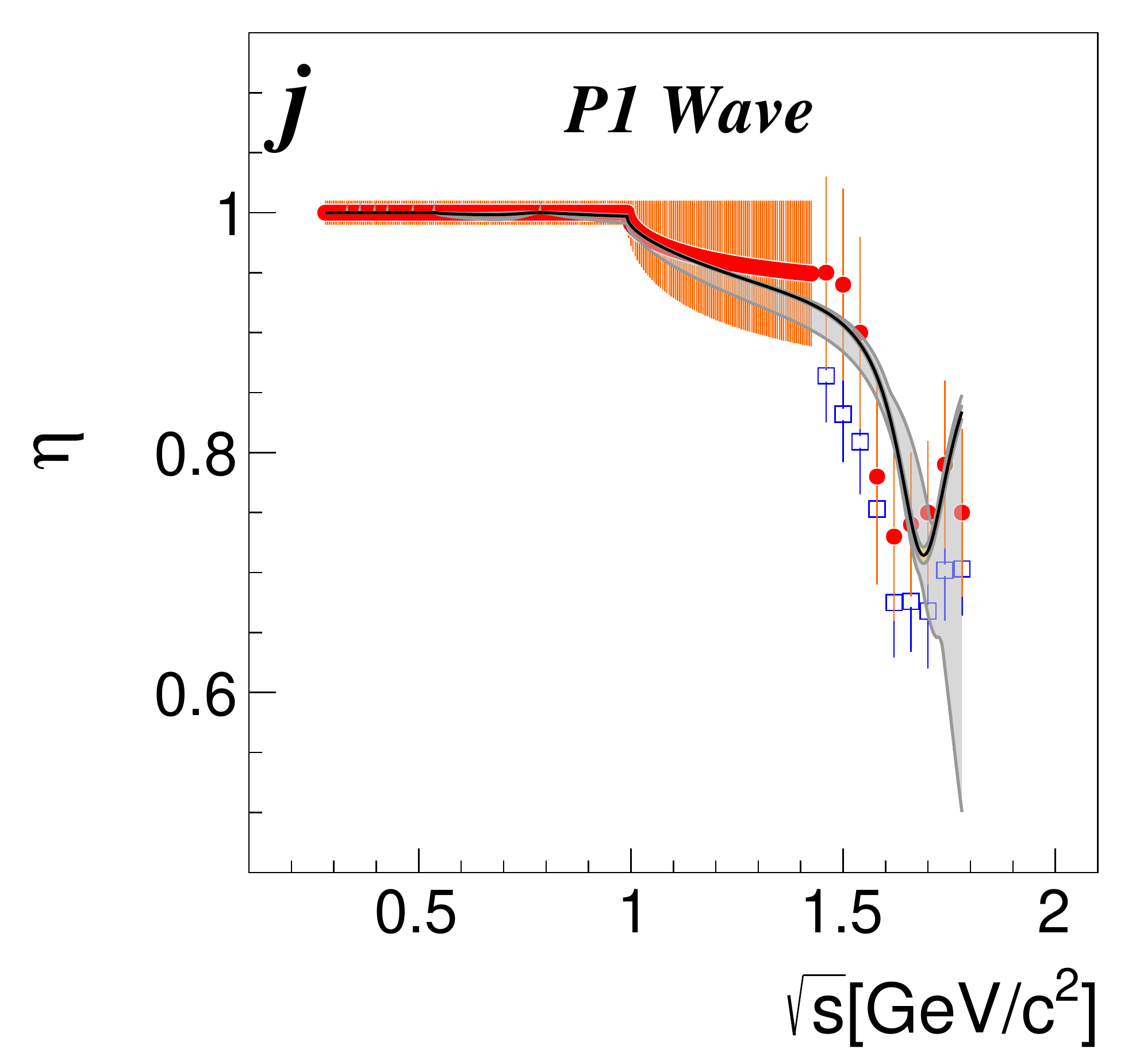}  &
\hspace{-3mm}\includegraphics[width=.25\textwidth]{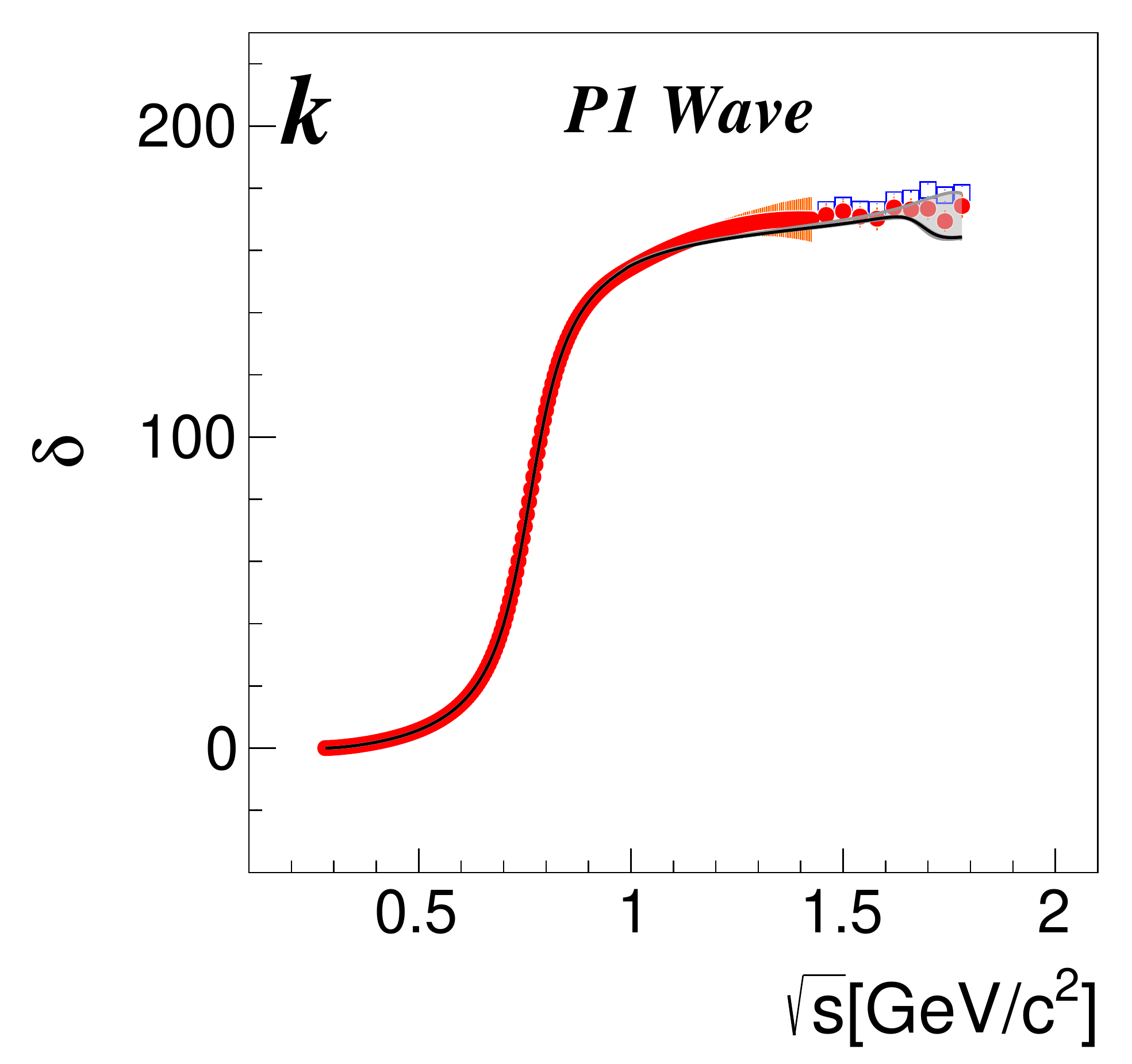}      &\\ 
\end{tabular}
\caption{Results for Scattering Data. Figs $a,b$ for S wave $\pi\pi\to\pi\pi$ inelasticity and phase shift, 
$c\,$--$\,e$ for $|T|^2$-values of processes $\pi\pi\to\eta\eta,\eta\eta',KK$ in S wave. $f,g$ for
D wave $\pi\pi\to\pi\pi$ inelasticity and phase shift. $h,i$ for
$|T|^2$-values
   of processes $\pi\pi\to KK, \eta\eta$, $j,k$ for P1 wave
$\pi\pi\to\pi\pi$ inelasticity and phase shift. The data considered in the best fit are given by
red points with error bars which includes the solution (-~-~-) from
~\cite{Ochs:2013gi} (Figs $a,b$), and solution (-~-~-) from ~\cite{Hyams:1975mc} (Figs $f, g, j, k$). 
Data sets from multiple solutions are taken into account as alternative fits. 
Data points for solution (-~+~-) from ~\cite{Ochs:2013gi} are labeled
with magenta stars (Figs $a, b$), and data points for the
solution (-~-~-) and solution (-~+~-)
from ~\cite{Hyams:1975mc}  are labeled with blue triangles (Figs $a,b$) and blue squares (Figs $a, b, f, g, j, k$). The black line represents the fit result, the tiny yellow bands 
   illustrate the statistical uncertainty and the gray
   bands stand for the systematic uncertainty obtained from the
   alternative fits. 
   The references for the individual scattering data sets are listed in section \ref{overview_scattdata}.
}
\label{fig:ResultsScattering}      
\end{figure*}

\begin{figure}[htb]
\begin{tabular}{cc}
\hspace{-3mm}\includegraphics[width=.25\textwidth]{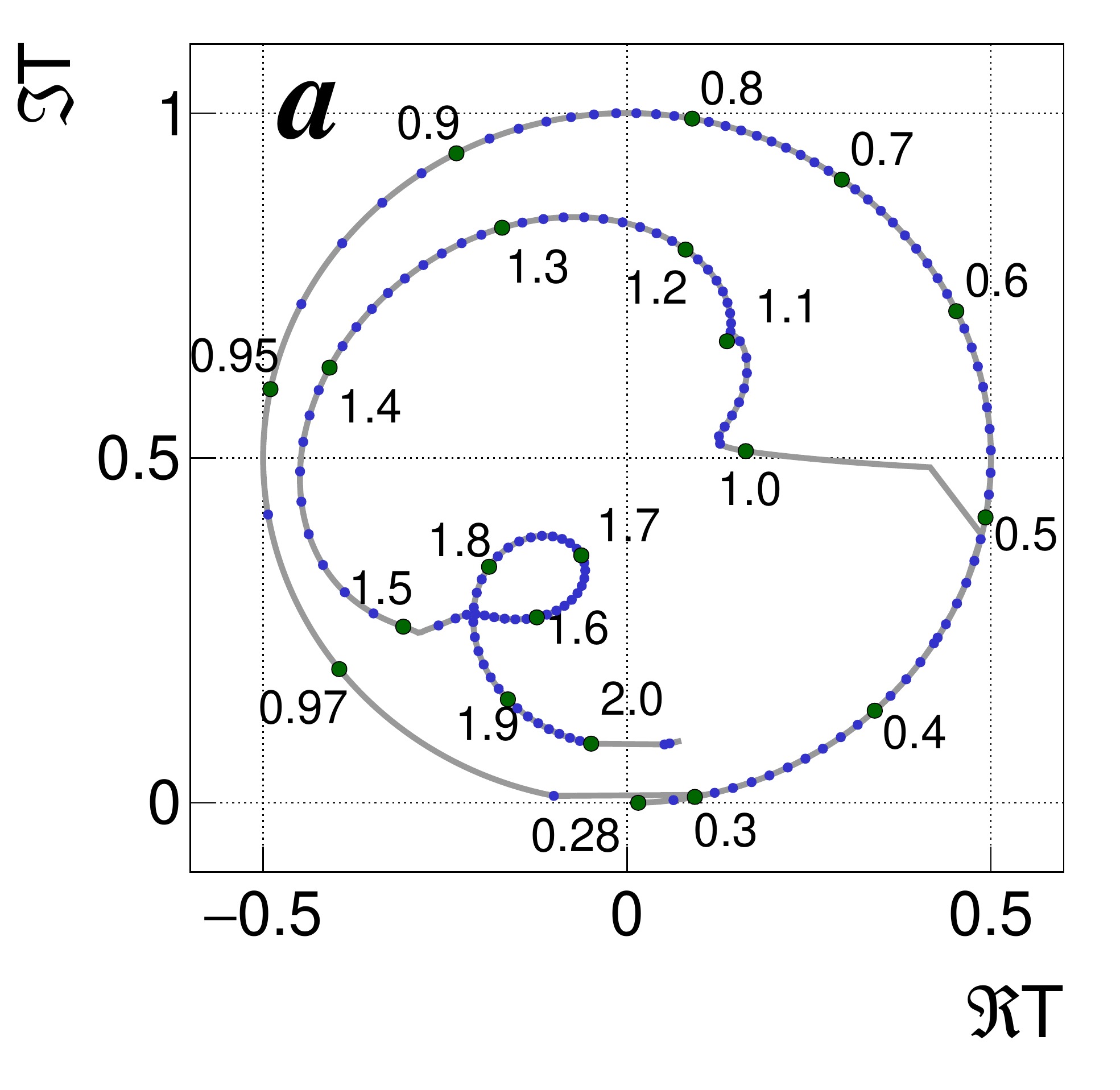} &
\hspace{-3mm}\includegraphics[width=.25\textwidth]{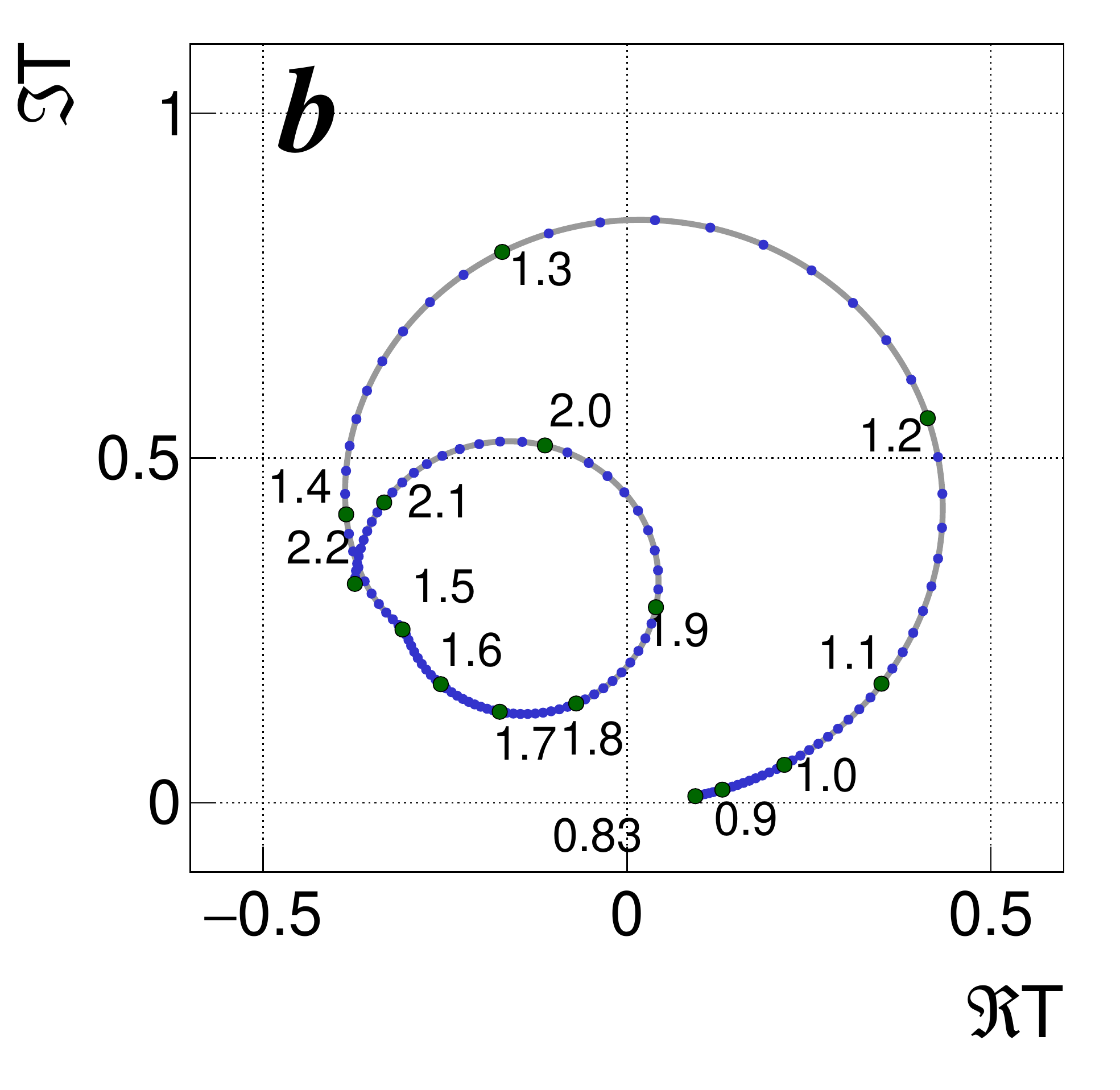} \\
\hspace{-3mm}\includegraphics[width=.25\textwidth]{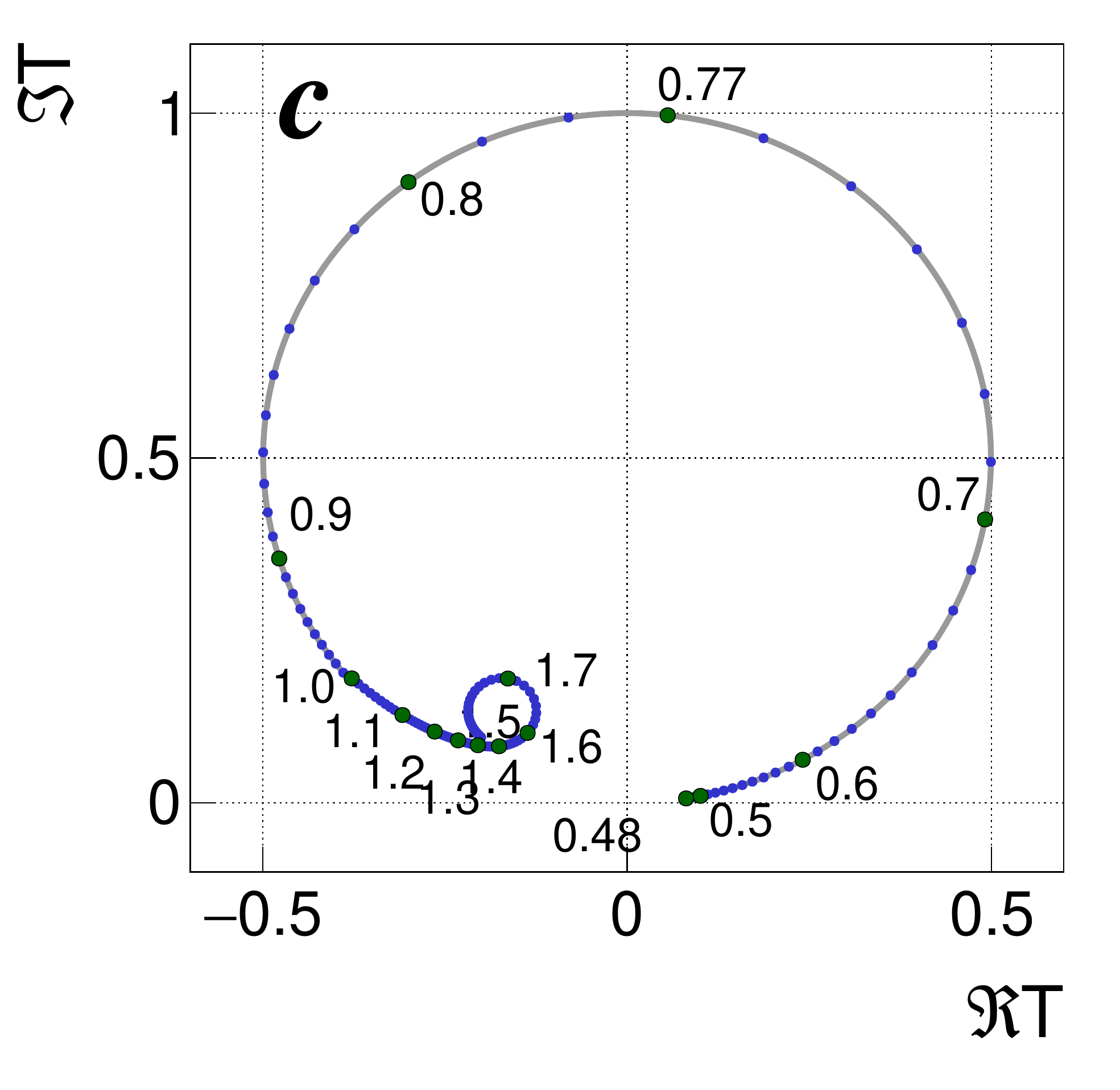} &\\
\end{tabular}
\caption{Argand plots for the projection $\pi\pi\to\pi\pi$ of the waves
  $S0  \, (a), D0 \, (b), P1 \, (c)$. The numbers
  along the lines denote the masses at these positions. Exemplarily, the results
using the scattering data for the solution (-$\;$-$\;$-) from
\cite{Ochs:2013gi} for the S0-wave and from
\cite{Ochs:2013gi} \cite{Hyams:1975mc} for D0- and P1-wave the are shown.}
\label{fig:ArgandPlot}       
\end{figure}

The outcome of the simultaneous fit for the scattering data is summarized in
Fig.~\ref{fig:ResultsScattering}. Apart from some systematic
discrepancies, appearing in Fig.~\ref{fig:ResultsScattering} (e) and (h),  the agreement is good. 
It should be emphasized that in particular a good 
consistency is achieved for the  phase shifts based on the model independent
parameterizations with small uncertainties below $\sqrt{s} <
1.425 \; $\gev2c. Fig.~\ref{fig:ArgandPlot} shows the resulting Argand diagrams
 for the individual waves. 

\subsection{Comparison to Other Measurements}
\begin{figure}[ht]
  \hspace{-2mm} \includegraphics[width=.5\textwidth]{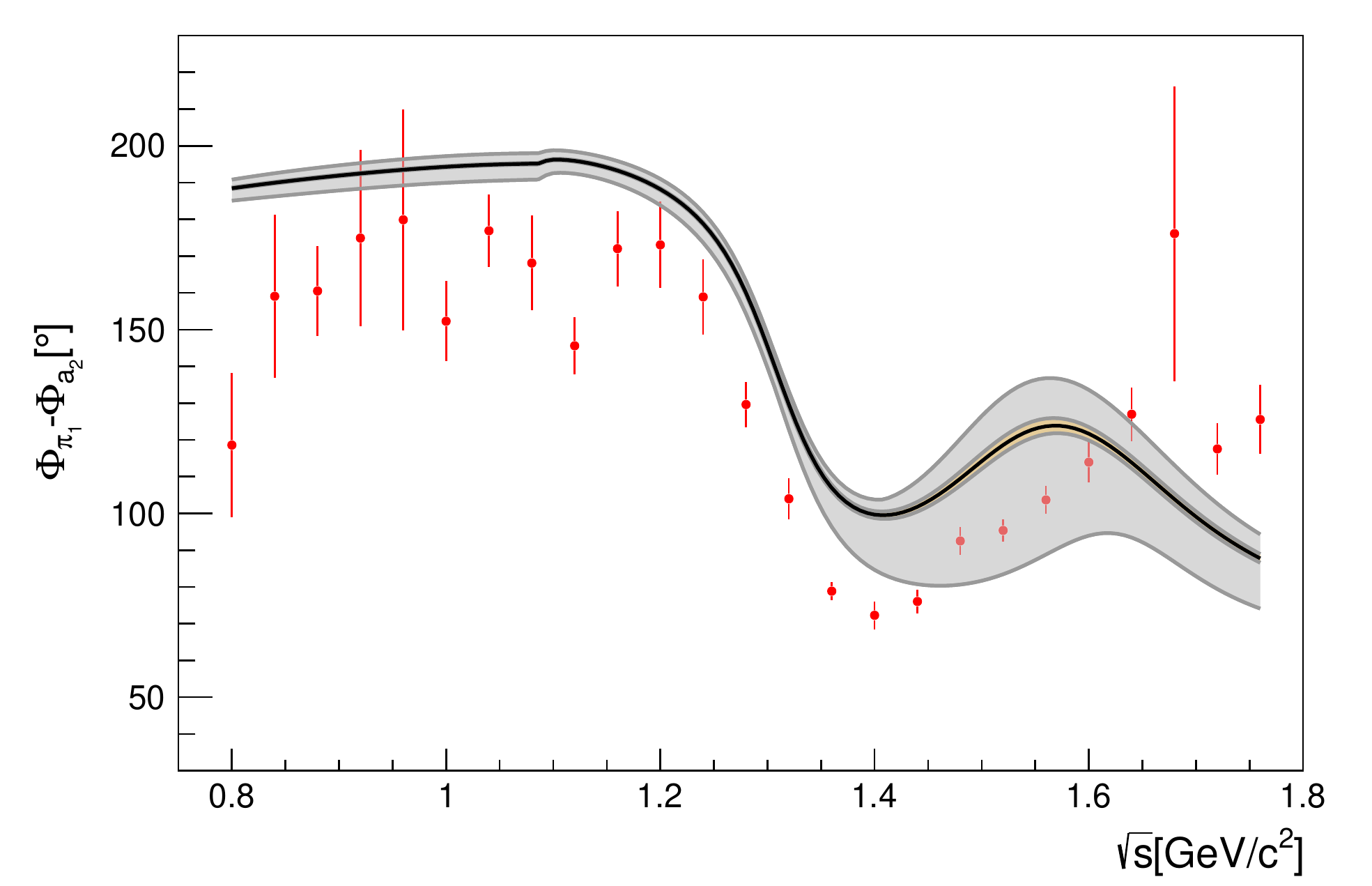}
   \caption{Phase difference between the $a_2$ and the $\pi_1$ wave in $\pi \eta$. The red dots with error bars 
show the COMPASS data for the P- and D-wave of the $\eta \pi$ system in diffractive $\pi^- p$ 
scattering~\cite{Adolph:2014rpp}. The corresponding result for the $\pi \eta$ channel extracted from the 
$a_2$ and the $\pi_1$ scattering matrices $T_{\pi\eta \rightarrow \pi\eta}$ obtained from the best fit in \pbarp\ annihilation
is represented by the black line. For comparative purposes the obtained
  phase difference is shifted by an offset of 180$^\circ$. The gray shaded area represents the systematic uncertainty obtained from the alternative fits listed in Tab.~\ref{tab:FitLHs} and the yellow band illustrates the statistical uncertainty based on the covariance error matrix from the MINUIT2 fit. Note that this figure does not show a fit to the COMPASS data but merely a comparison.}
  \label{fig:pi1Compass}
 \end{figure}

 One striking feature of the best selected fit is the agreement of the obtained phase 
difference between the $a_2$ and the $\pi_1$ wave
compared to previously published data for $\eta \pi$- and $\eta^\prime \pi$ production
in diffractive $\pi^- p$ scattering at 191 GeV/c, studied by the COMPASS collaboration~\cite{Adolph:2014rpp}. 
Fig. \ref{fig:pi1Compass} shows the P-wave versus the D-wave phase motion resulting from an $\eta \pi$ partial
wave analysis~\cite{Adolph:2014rpp}, overlayed with the present result extracted from the best fit. It should 
be stressed that Fig.~\ref{fig:pi1Compass} merely shows a comparison with no subsequent fitting.

\section{Extracted Properties}
\label{rpops_lab}
For the extraction of all properties the chosen $\pi\pi \rightarrow
   \pi\pi$ scattering data in the high energy range between
   $\sqrt{s} \; > \; 1.425$ \gev2c\ and $\sqrt{s}  \;  < \; 1.9$ \gev2c\
are the solution labeled with (-~-~-) from~\cite{Hyams:1975mc} for the D0- and P1-wave
and from~\cite{Ochs:2013gi}  for the S0-wave. Since these solutions might be not unambiguous, further 
alternative fits have been performed by
replacing these data by the ones based on the solutions labeled with
(-~+~-) in~\cite{Hyams:1975mc} and~\cite{Ochs:2013gi} for the S0-, D0-
and P1-wave and with  (-~-~-) in~\cite{Hyams:1975mc} for the S0-wave, respectively. All of these 
fit results and the ones marked with (*) in Tab.~\ref{tab:FitLHs} are taken into account for the
estimation of the systematic uncertainties. Since exchanging the
scattering data in the high mass region has a large influence on the
resonances in that region, asymmetric uncertainties for the parameters
of the  $f_0(1710)$, $f_2(1810)$, $f_2(1950)$ and $\rho(1700)$ are determined. 
Additionally, whenever asymmetries are observed for the deviation of resonance parameters from the respective values of the best fit, these are reflected by asymmetric systematic uncertainties in Tab.~\ref{tab:Properties}, as e.g. for the $f_0(1500)$, $a_2(1700)$ and $\pi_1$.

\subsection{Contributions of Different Waves}
\label{ContribsWave_lab}
\begin{figure}
\hspace{-1.6em} \includegraphics[width=0.5\textwidth]{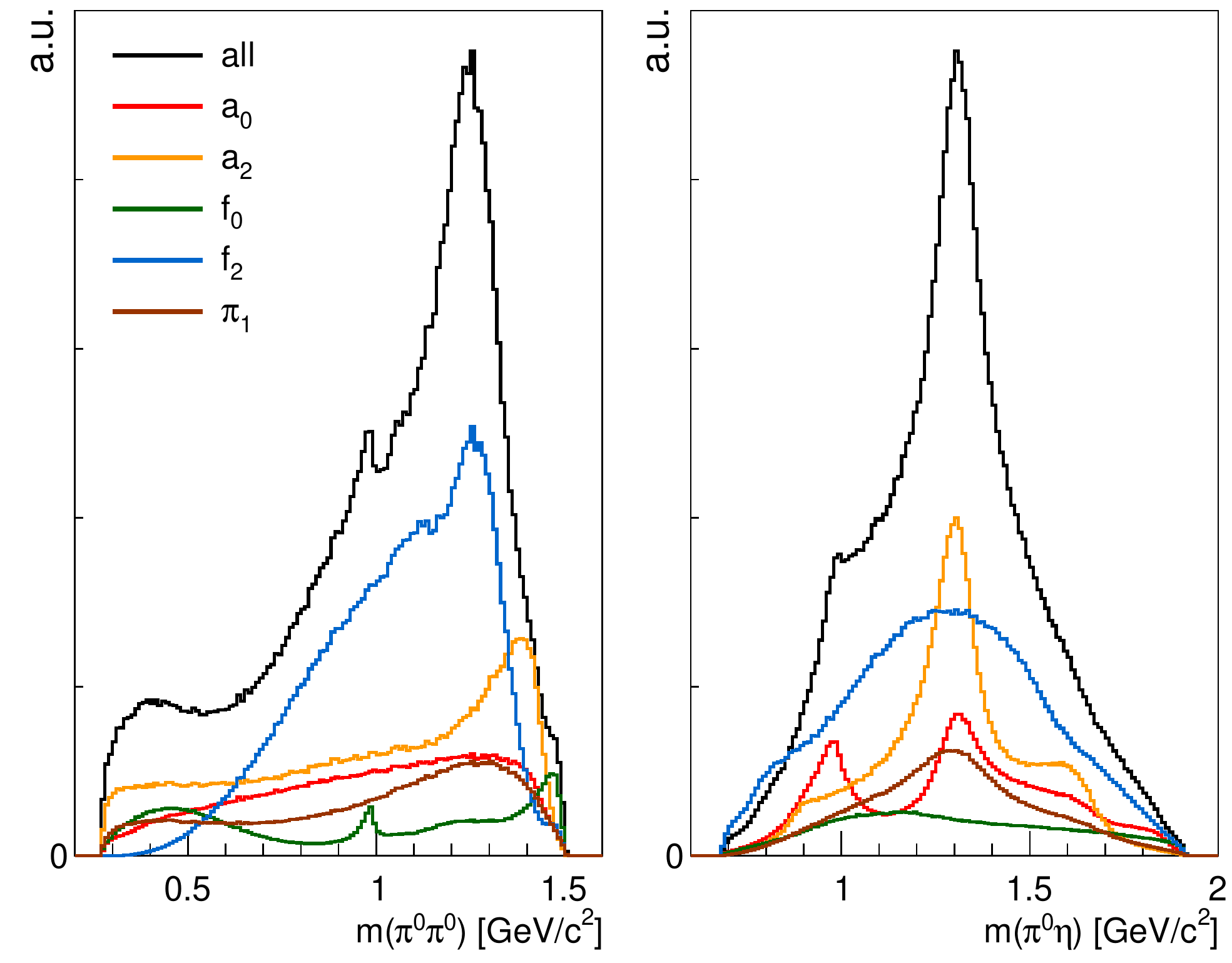}
\centering
\caption{Efficiency corrected invariant $\pi^0\pi^0$- (left) and
  $\pi^0\eta$- mass (right) for the reaction
  \pbarpToPi0Pi0Eta\ . The overall result is marked in black, while the
  individual contributions are visualized by different colors.}
\label{fig:MassFractionsPiPiEta}       
\end{figure}

\begin{figure}
\hspace{-1.6em} \includegraphics[width=0.5\textwidth]{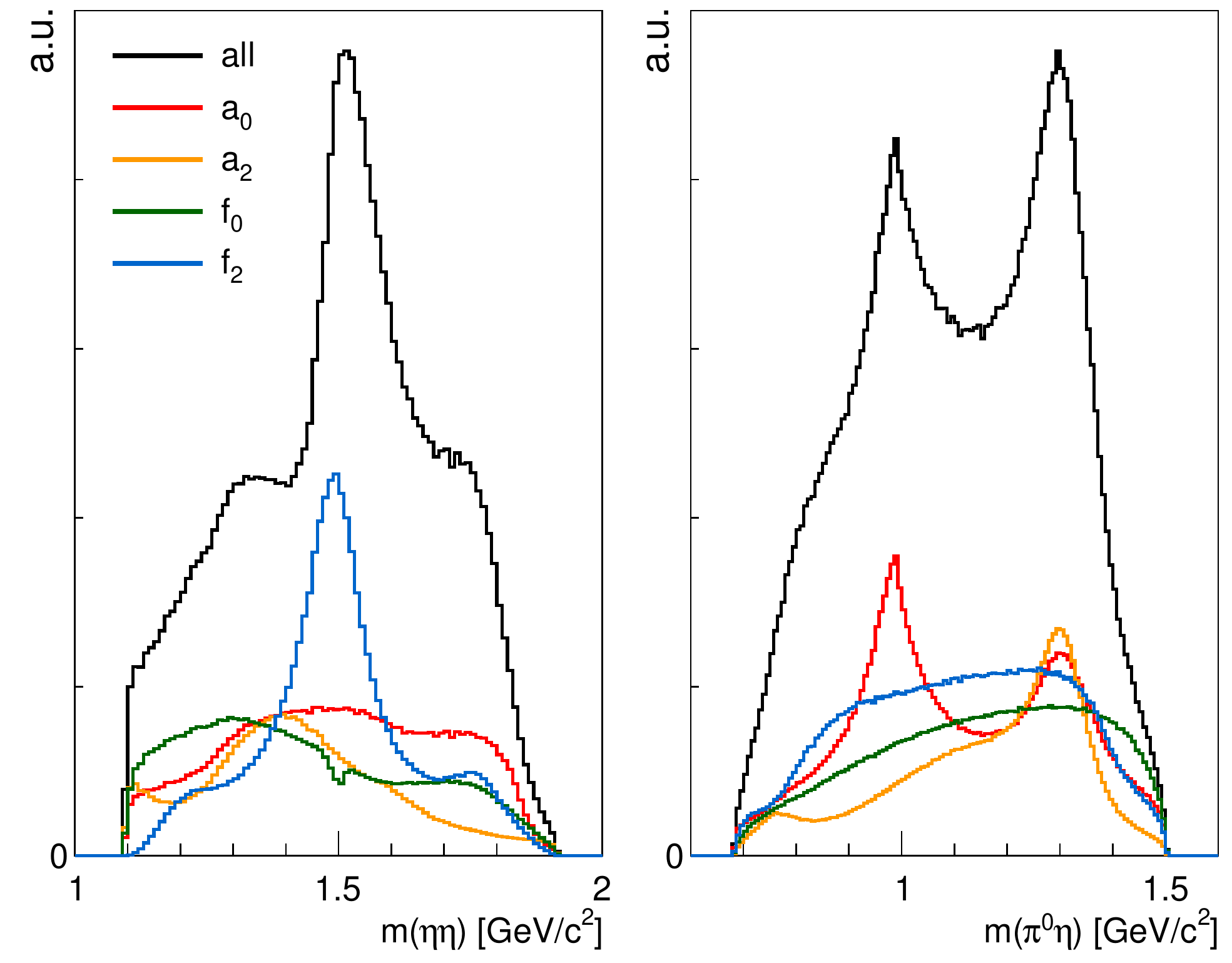}
\centering
\caption{Efficiency corrected invariant $\eta\eta$- (left) and
  $\pi^0\eta$- mass (right) for the reaction
  \pbarpToPiEtaEta\ . The overall result is marked in black, while the
  individual contributions are visualized by different colors.}
\label{fig:MassFractionsPiEtaEta}       
\end{figure}

\begin{figure}
\hspace{-1.6em} \includegraphics[width=0.5\textwidth]{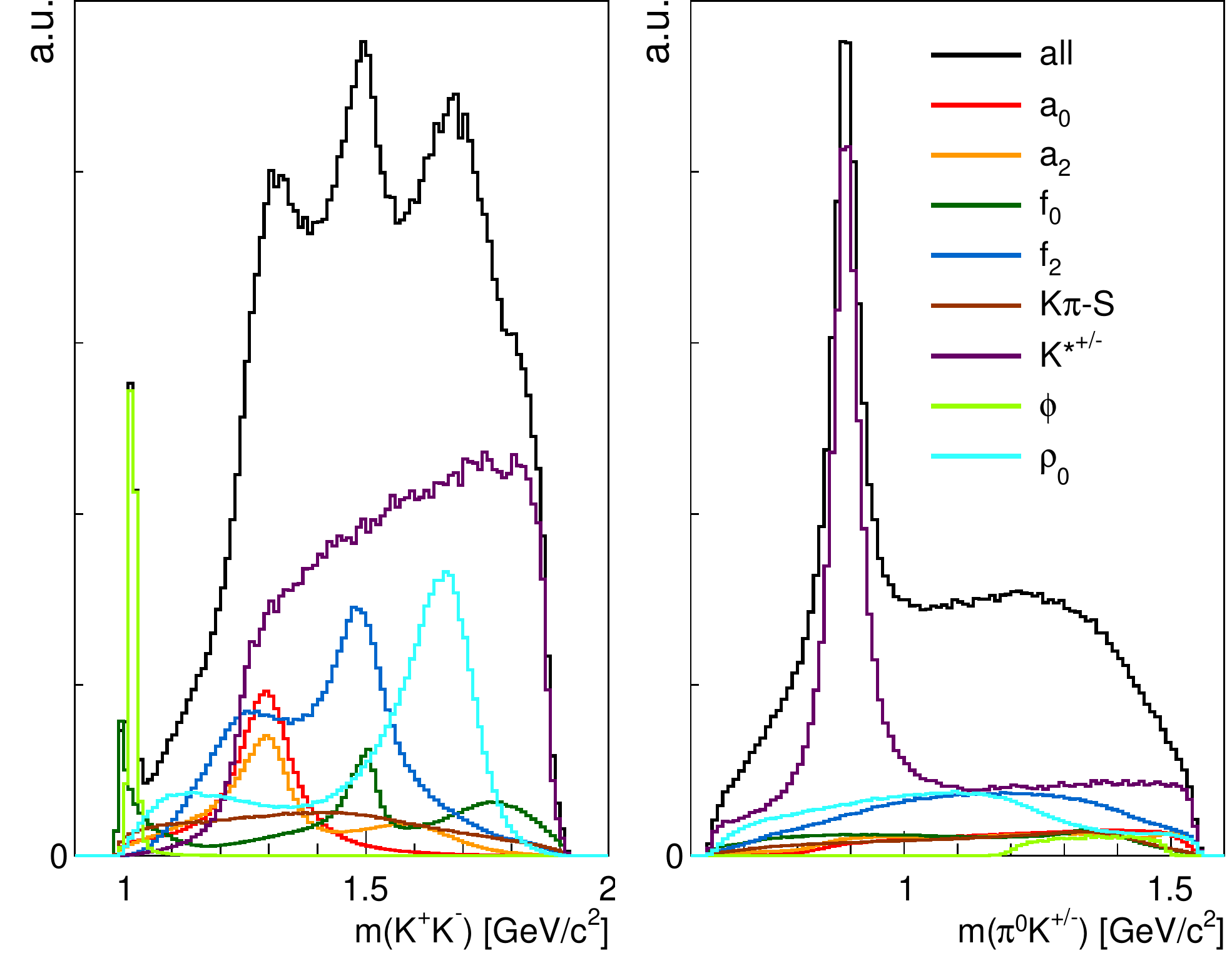}
\centering
\caption{Efficiency corrected invariant $K^{+}K^{-}$- (left) and
  $K^\pm \pi^0$- mass (right) for the reaction
  \pbarpToKpKmPi0\ . The overall result is marked in black, while the
  individual contributions are visualized by different colors.}
\label{fig:MassFractionsKKPi}       
\end{figure}

The contributions of the individual waves have been determined
according to the prescription in~\cite{Beringer:1900zz} by calculating
the absolute square of the amplitudes of the relevant wave only, and
dividing it by the absolute square of the incoherent and 
coherent sums of all amplitudes as defined in eq.~(\ref{eq:crossSection1}).
As the waves can interfere with each other the sum
of all fractions is not necessarily unity. In
Tab.~\ref{tab:FitFractions} the fractions are listed
for each wave and each annihilation channel individually. The total sum of
135.0~$\pm$ 1.2 (stat.)~$\pm$ 8.7 (sys.) $\%$
for the \PiPiEta,  101.2~$\pm$ 2.4 (stat.)~$\pm$ 11.7 (sys.) $\%$ for the \PiEtaEta\ and 107.8~$\pm$ 1.9 (stat.)~$\pm$ 12.5 (sys.) $\%$ for
the \KpKmPi0\ channel show interference effects which are
small enough to believe that the fit result relies on a reasonable
physics description. For the resonances described with the K-matrices it
is not straightforward to disentangle the individual
contributions of different resonances and background terms~\cite{Beringer:1900zz}. 
Therefore, only the contributions of the partial waves 
described by
the K-matrix are summarized in~Tab.~\ref{tab:FitFractions}. 
The dominant contributions with more than 20~$\%$
 are the  $a_0 \, \pi^0$,  $a_2 \, \pi^0$ and $f_2 \, \eta$ components for  the \PiPiEta, 
the  $f_0 \, \pi^0$, $f_2 \, \pi^0$ and $a_0 \, \pi^0$ waves for the
 \PiEtaEta\ and the $K^*(892)^\pm \, K^\mp$ reaction for the  \KpKmPi0\
 channel. It is worth mentioning that the spin exotic component $\pi_1$
 exhibits a fraction of almost 20~$\%$ in \PiPiEta. Fig.~\ref{fig:MassFractionsPiPiEta},
\ref{fig:MassFractionsPiEtaEta}  and \ref{fig:MassFractionsKKPi}
show the invariant mass spectra with the contributions of the individual waves for
the channels \PiPiEta, \PiEtaEta\ and \KpKmPi0, respectively.   
\begin{table}[htb]
  \caption{Contributions in $\%$ of the individual waves for the three
    channels \pbarpToPi0Pi0Eta\
    , \PiEtaEta\
    and \KpKmPi0\ .
  }

\label{tab:FitFractions}      
\setlength{\tabcolsep}{3pt}
\footnotesize
\centering
\adjustbox{max width=\linewidth}{
\begin{tabular}{l r r r} 
  \hline\noalign{\smallskip}
         & \multicolumn{3}{c}{contribution (in $\%$) for channel} \\
         & \PiPiEta\ & \PiEtaEta\ & \KpKmPi0\ \\
  \noalign{\smallskip}
  \hline
  \noalign{\smallskip} 
                $f_0 \, \pi^0$  &                           & 23.7~$\pm$~1.2$\pm$~2.3    & 7.4~$\pm$~0.3$\pm$~4.1   \\ 
                 $f_0 \, \eta$  & 10.7~$\pm$~0.4$\pm$~1.8   &                            &                          \\ 
   \noalign{\smallskip}                                                                                                %
                $f_2 \, \pi^0$  &                           & 30.1~$\pm$~1.3$\pm$~2.7    & 17.1~$\pm$~0.7$\pm$~10.0  \\ 
                 $f_2 \, \eta$  & 52.3~$\pm$~0.8$\pm$~5.0   &                            &                          \\ 
   \noalign{\smallskip}                                                                                                %
               $\rho \, \pi^0$  &                           &                            & 17.2~$\pm$~1.0$\pm$~4.0  \\ 
   \noalign{\smallskip}                                                                                                %
                $a_0 \, \pi^0$  & 22.4~$\pm$~0.4$\pm$~1.0   &                            & 6.1~$\pm$~0.2$\pm$~2.8   \\ 
                 $a_0 \, \eta$  &                           & 28.6~$\pm$~1.1$\pm$~7.5    &                          \\ 
   \noalign{\smallskip}                                                                                                %
                $a_2 \, \pi^0$  & 33.0~$\pm$~0.6$\pm$~2.9   &                            & 6.4~$\pm$~0.2$\pm$~2.9   \\ 
                 $a_2 \, \eta$  &                           & 18.8~$\pm$~1.1$\pm$~5.6    &                          \\ 
   \noalign{\smallskip}                                                                                                %
     $K^*(892)^\pm \, K^{\mp}$  &                           &                            & 45.0~$\pm$~1.3$\pm$~11.0  \\ 
     $(K\pi)_S^\pm \, K^{\mp}$  &                           &                            & 6.1~$\pm$~0.4$\pm$~4.9   \\ 
         $\phi(1020) \, \pi^0$  &                           &                            & 2.5~$\pm$~0.3$\pm$~0.3   \\ 
              $\pi_1 \, \pi^0$  & 16.7~$\pm$~0.5$\pm$~3.0   &                            &                          \\ 
   \noalign{\smallskip}
  \hline
  \noalign{\smallskip}  
$\Sigma$ &  135.0~$\pm$~1.2$\pm$~8.7  & 101.2~$\pm$~2.4$\pm$~11.7  & 107.8~$\pm$~1.9$\pm$~12.5 \\ 
  \hline
\end{tabular}
}
\end{table}

 For the reaction \pbarpToPi0Pi0Eta\ the contributions are comparable with 
the outcome of the former Crystal Barrel
analysis~\cite{Amsler:2002qq}. An obvious difference is that the spin-exotic wave has not
been seen in the old analysis. However, an evident contribution of this
wave has been found in $\pbarp$ annihilation data at rest
with liquid deuterium and gaseous hydrogen targets~\cite{Abele:1998gn,
  Abele:1999tf}. 
The individual fractions for the channels \PiEtaEta\ and \KpKmPi0\
show slight differences compared to former
Crystal Barrel analyses~\cite{Amsler:2002qq, Amsler:2006du}.
However, the two contributions  $\phi(1020) \pi^0$ and
  $K^*(892)^\pm K^\mp$  which are of special interest here and where
  the intermediate resonances are isolated and parameterized by
Breit-Wigner functions, are in good agreement with~\cite{Amsler:2006du}.
For most of the remaining resonances, which are described by
K-matrices in this work, it is not straightforward to compare the
contributions to those obtained from Breit-Wigner based fits. 

\subsection{Pole and Breit-Wigner Parameters}
The dynamics of the isolated resonances $K^*(892)^\pm$
and\linebreak
$\phi(1020)$ are described by relativistic Breit-Wigner
approximations. The correspon\-ding
masses and widths are treated
as free parameters so that these properties including their
statistical uncertainties can directly be obtained from the
outcome of the fit. However, the parameters of the resonances described by the
K-matrices must be determined from the pole positions in the complex
energy plane of the T-matrix on the Rieman sheet located next to the
physical sheet. A detailed description of the classification of poles
and their occurrence on the different sheets can be found in \cite{Badalian:1981xj}, for
example. Therefore the pole position properties are not direct fit
parameters. The scan of the complex energy plane is realized here by a minimization
procedure where the real and imaginary parts of the pole in the complex
T-matrix plane are the free parameters. The extraction of the statistical errors is based
on this fit method as well and makes use of a numerical approach by taking into account the
covariance error matrix obtained by the coupled channel fit. Due to 
the fact that the $f_0(980)$ and $a_0(980)$ resonances 
are located very close to the $K\bar{K}$ threshold 
their pole positions have been extracted from the two relevant sheets below and 
above this threshold. The masses, widths and pole
positions obtained for both sheets are listed in Tab.~\ref{tab:Properties}. 
The systematic uncertainties were derived as described before in~\ref{ContribsWave_lab}.
It turned out that the statistical errors in particular for the 
masses and widths provided by the
minimization tool MINUIT2 are systematically too small. Therefore the likelihood 
profiling method~\cite{Rolke:2004mj} has been used for some specific resonances. The obtained uncertainties 
based on this procedure are larger by a factor of between 2 and 5 compared to the outcome from MINUIT2. However, the 
uncertainties for the positions of all poles with a significant contribution to the $\bar{p} p$ channels
are dominated by the systematics and thus the statistical errors are negligible.\\     
 Most of the obtained quantities are in good agreement with
other measurements~\cite{Tanabashi:2018oca}. It should be noted that the $f_0(500)$
exhibits a larger mass and width. This is probably related to the chosen 
K-matrix approach. The description chosen here does not take into account properly crossing symmetries and 
other sophisticated constraints for the low $\pi \pi$ mass region, as for example used in~\cite{GarciaMartin:2011cn}. 
The pole mass of the $a_0(1450)$ meson is measured to be 1302.1~\mev2c\ , a significantly lower value compared
to the PDG average~\cite{Tanabashi:2018oca}. It is, however, comparable with the old Crystal Barrel 
analysis~\cite{Amsler:2002qq} and several other measurements collected
in~\cite{Tanabashi:2018oca}.
It should be noted that the obtained pole parameters of the 
$f_0$, $f_2$ and $\rho$ resonances with a non negligible coupling to \PiPi\ are mainly driven by the used scattering data. 
Since old analyses 
are also based on these data and their measurements already contribute to the PDG world average 
the quantities obtained here are not completely independent. However, the $\pi_1$-, $a_0$- and $a_2$-waves
are exclusively contributing in the \pbarp\ data samples. Therefore the measurements of the pole positions 
related to these waves can be claimed to be independent.

\begin{table*}[htb]
\caption{Masses and widths for the individual
  resonances. The values of the Breit-Wigner parameterization are
  listed  for the isolated resonances $K^*$(892) and $\phi(1680)$. 
  The $\phi(1020)$ has been described by a Voigtian.
  For the $\pi_1$ and all $f_0$, $f_2$, $a_0$, $a_2$ and $\rho$-states the pole
  positions are extracted from the T-matrix. Absolute and relative branching fractions are 
  extracted for the $f_2(1270)$, $f_{2}^{\prime}(1525)$, $\rho(770)$, $\rho(1570)$ and for the
  $a_0$ and $a_2$ resonances, respectively. 
 The quantities for the $a_0(980)$ and $f_0(980)$ mesons are
  determined on the two 
  relevant sheets below and above the $K\bar{K}$ threshold
  individually. The sheets of the complex plane are characterized and
  also defined by
  the signs (+ and -) of the imaginary part of the channel
  momenta. The order of the channels for the $f_0(980)$ resonance is
  $\pi \; \pi$, $2\pi \; 2\pi$, $K\;\bar{K}$, $\eta \; \eta$ and $\eta
\; \eta^\prime$ and for the $a_0(980)$ resonance is $\pi \; \eta$ and $K\;\bar{K}$. 
  The statistical uncertainty is given by the first and the 
  systematic uncertainty is provided by the second error.
  Each  resonance is assigned to the data samples which
   are contributing significantly to the extraction of the resonance properties \pbarp\ data
  (\pbarp)  or scattering data (scat) or both (\pbarp\ + scat)).
}

\label{tab:Properties}      
\footnotesize
\centering
\begin{tabular}{l l r r r r r }
  \cline{1-4}\noalign{\smallskip}
 name           & relevant data & Breit-Wigner mass   & Breit-Wigner width & & &\\ 
                &         &   [\mev2c] &   $\Gamma$ [MeV] &  & & \\
\cline{1-4}\noalign{\smallskip} 
$K^*(892)$       & \pbarp & 892.2~$\pm$~0.5$\pm$~1.7 & 54.4~$\pm$~0.9$\pm$~1.7    & & & \\
$\phi(1020)$    & \pbarp & 1018.4~$\pm$~0.5$\pm$~0.1 & 4.2~(fixed)      & & & \\
  \noalign{\smallskip} \noalign{\smallskip}
  \cline{1-4}
  \noalign{\smallskip}
  name     & relevant data & pole mass  & pole width  &  & & \\
           &             &   [\mev2c] & $\Gamma$ [MeV] &  & & \\
  \noalign{\smallskip}
  \cline{1-4}\noalign{\smallskip}
$\pi_{1}$     & \pbarp  & 1404.7~$\pm$~3.5~$^{+15.1}_{-17.7}$ & 628.3~$\pm$~27.1~$^{+35.8}_{-138.2}$   & & & \\
  \noalign{\smallskip}
$f_{0}(500)$    &  scat  & 857.0~$\pm$~5.7$\pm$~366.4 & 771.6~$\pm$~8.3$\pm$~291.1    & & & \\
$f_{0}(980)^{---++}$    &  scat  & 977.8~$\pm$~0.6$\pm$~1.6 & 97.8~$\pm$~1.2$\pm$~5.4    & & & \\
$f_{0}(980)^{--+++}$    &  scat  & 992.8~$\pm$~0.8$\pm$~1.0 & 61.3~$\pm$~1.3$\pm$~4.4    & & & \\
$f_{0}(1370)$  &   scat  & 1280.6~$\pm$~1.6$\pm$~47.4 & 410.5~$\pm$~3.5$\pm$~41.5  & & & \\
	$f_{0}(1500)$  & \pbarp\ + scat  & 1496.0~$\pm$~1.2~$^{+4.4}_{-26.4}$ & 80.8~$\pm$~0.6~$^{+20.0}_{-5.0}$    & & & \\
$f_{0}(1710)$   & \pbarp\ + scat  & 1803.5~$\pm$~3.5~$^{+45.5}_{-10.4}$ & 289.7~$\pm$~5.0~$^{+32.6}_{-19.3}$    & & & \\
 \noalign{\smallskip}
$f_{2}(1810)$   & scat   & 1845.0~$\pm$~2.2~$^{+1.6}_{-7.2}$ & 260.9~$\pm$~3.9~$^{+199.9}_{-38.2}$ & & & \\
$f_{2}(1950)$   & scat   & 1978.2~$\pm$~1.8~$^{+28.4}_{-16.9}$ & 237.6~$\pm$~1.6~$^{+41.6}_{-15.5}$ & & & \\
 \noalign{\smallskip} \noalign{\smallskip}
  \cline{1-5}
  \noalign{\smallskip}
  name   & relevant data & pole mass  & pole width  & $\Gamma_{KK}/\Gamma_{\eta\pi^{0}}$  & & \\
             &                   &   [\mev2c] & $\Gamma$ [MeV] & $[\%]$ & & \\
  \noalign{\smallskip}
  \cline{1-5}\noalign{\smallskip}
$a_{0}(980)^{--}$  & \pbarp & 1002.4~$\pm$~1.4$\pm$~6.6 & 127.0~$\pm$~2.3$\pm$~6.7    & 14.9~$\pm$~0.1$\pm$~3.9   & & \\
$a_{0}(980)^{-+}$  & \pbarp & 1004.1~$\pm$~1.5$\pm$~6.5 & 97.2~$\pm$~1.9$\pm$~5.7    & 13.8~$\pm$~0.1$\pm$~3.5   & & \\
$a_{0}(1450)$  & \pbarp & 1302.1~$\pm$~1.1$\pm$~3.9 & 112.4~$\pm$~1.4$\pm$~3.4   & 188.7~$\pm$~4.1$\pm$~97.0    & & \\
  \noalign{\smallskip}
$a_{2}(1320)$  & \pbarp  & 1312.5~$\pm$~0.7$\pm$~2.6 & 106.9~$\pm$~1.2$\pm$3.7    & 35.2~$\pm$~1.1$\pm$~17.5 & & \\
	$a_{2}(1700)$   & \pbarp  & 1638.9 ~$\pm$~2.3~$^{+57.4}_{-0.1}$ & 224.0~$\pm$~2.5~$^{+1.8}_{-48.3}$   & 413.4~$\pm$~10.6~$^{+490.9}_{-298.8}$   & & \\
 \noalign{\smallskip} \noalign{\smallskip}
  \hline
 \noalign{\smallskip} 
  name & relevant data  & pole mass & pole width  & $\Gamma_{\pi\pi}/\Gamma $ & $\Gamma_{KK}/\Gamma$ & $\Gamma_{\eta\eta}/\Gamma $ \\
           &            &   [\mev2c]        & $\Gamma$ [MeV]  & $ [\%]$ & $[\%]$ & $ [\%]$ \\ 
\noalign{\smallskip}\hline\noalign{\smallskip}
$f_{2}(1270)$  & \pbarp\ + scat & 1263.3~$\pm$~0.2$\pm$~1.5  & 193.7~$\pm$~0.4$\pm$~1.6  & 85.6~$\pm$~0.1$\pm$~5.0  & 3.3~$\pm$~0.1$\pm$~0.5   & 0.4~$\pm$~0.1$\pm$~0.2 \\
$f_{2}^{\prime}(1525)$ & \pbarp\ + scat  & 1495.0~$\pm$~1.1$\pm$~8.1  & 104.8~$\pm$~0.9$\pm$~9.8   & 3.4~$\pm$~1.5$\pm$~1.0  & 74.6~$\pm$~0.2$\pm$~16.6    & 5.9~$\pm$~0.3$\pm$~2.6 \\
\noalign{\smallskip}
$\rho(770)$   & scat   & 766.8~$\pm$~0.2$\pm$~0.2   & 126.2~$\pm$~0.3$\pm$~0.4   & 100.5~$\pm$~0.1$\pm$~6.7       & 0.5~$\pm$~0.1$\pm$~0.1      & -  \\
   $\rho(1700)$ & \pbarp\ + scat & 1688.7~$\pm$~3.1~$^{+141.1}_{-1.3}$ & 150.9~$\pm$~2.5~$^{+60.0}_{-10.6}$   & 10.8~$\pm$~1.7~$^{+16.2}_{-0.4}$  & 0.7~$\pm$~0.6~$^{+4.1}_{-0.2}$ & -  \\
\noalign{\smallskip}
 \hline
\end{tabular}
\end{table*}

\subsection{Partial Decay Widths}
Due to the fact that the K-matrix ansatz chosen here fulfills
unitarity and analyticity it is for some specific resonances
even possible to extract not only the pole positions but also
the coupling strengths and thus the
partial widths for the individual decay channels. The widths can be extracted via the
residues of the scattering matrix $T$ with the projection to the
relevant decay channel $k$ on the sheet closest to the
physical one. The residues are determined by calculating the integral along a
closed contour $C_{z_{\tilde{\alpha}}}$ around the pole $\tilde{\alpha}$ (cf.~\cite{Tanabashi:2018oca}):
\begin{eqnarray}
\label{equ:ClosedIntResidue}
 Res^{\tilde{\alpha}}_{k \rightarrow k} = \frac{1}{2\pi i} \; \oint_{C_{z_{\tilde{\alpha}}}} \;
  \sqrt{\rho_k} \, \cdot \, T_{k
  \rightarrow k} (z) \, \cdot \, \sqrt{\rho_k}  \; dz
\end{eqnarray}
where $z_{\tilde{\alpha}}$ denotes the pole position of the resonance $\tilde{\alpha}$
in the complex energy plane.
This integral has been numerically
estimated by making use of the Laurent expansion as described
in~\cite{Doring:2010ap}, for example. By determining the Laurent coefficient
$a^k_{-1}$ numerically via
\begin{eqnarray}
 \label{equ:LaurentCoff}
 \frac{1}{a^k_{-1}} \; = \;  \frac{\partial } {\partial z} \,
  \Big|_{z=z_{\tilde{\alpha}}} \; \frac{1}{\sqrt{\rho_k}  \cdot T_{k \rightarrow k} \cdot \sqrt{\rho_k} } \; \approx \; \frac{1}{Res^{\tilde{\alpha}}_{k \rightarrow k} },
\end{eqnarray}
one approximates the partial width $\Gamma_k$ for the decay channel $k$ by:
\begin{eqnarray}
 \label{equ:PartialWidth}
 \Gamma_k \; \approx \; 2\, \cdot \, \Big| \, a^k_{-1}  \,  \Big|.
\end{eqnarray}
This method is numerically stable because the inverse T-matrix exhibits a value of zero at the pole position. In addition no integration is 
needed and thus the calculation is fast. This procedure simplifies the extraction of the statistical 
uncertainties where the calculation must be redone several times.\\
 Tab.~\ref{tab:Properties} lists the partial widths
for the $f_2(1270)$, $f_{2}^{\prime}(1525)$, $\rho(770)$ and $\rho(1700)$ resonances.
These quantities are not extracted for the
decay channel to $2\pi \, 2\pi$ which is not directly
accessible and is only treated as an effective
channel to consider unitarity. The partial widths are in good agreement with
all other measurements~\cite{Tanabashi:2018oca}. It should be noted that the obtained
quantities for the $\rho(770)$ are only based on the fit to the scattering data. This 
vector meson does not couple to the \pbarp\ channels analyzed here.  
The absolute coupling strengths for the $a_0$ and $a_2$ resonances have not been
determined because the K-matrices are only described by the two channels
$\pi\eta$ and $K\bar{K}$. Since further relevant decay modes like 3$\pi$ or
$\omega\pi\pi$ are not considered, the
extraction would result in
unreliable values for these coupling strengths. Instead, the ratios
$\Gamma_{\pi\eta} / \Gamma_{K\bar{K}}$ have been determined for these 
resonances which should
deliver more reasonable results.\\
 Due to the fact that the numerical method
based on the Laurent expansion is only a trustful approximation for
resonances located not too far from the real axis and not too close to
thresholds the coupling strengths have not been extracted for the $f_2(1810)$, 
$f_2(1950)$ and for all $f_0$ resonances. The relevant K-matrix of the $(\pi\pi)_S$-wave 
is very complex and is characterized by  5 channels and 5 poles. These poles 
are in fact located far from the real axis or close to specific thresholds.

\subsection{Production Cross Sections and Spin Density Matrix Elements for
  $\phi(1020)$, $K^*(892)^\pm$ and the $\pi^0_1$-Wave}
The differential production cross sections and the SDM
elements for the resonances $\phi(1020)$, $K^*(892)^\pm$ in the
reaction \pbarpToKpKmPi0\ and for
the spin-exotic wave  $\pi_1$ in the reaction
\pbarpToPi0Pi0Eta\ derived from the final fit result are discussed in
the following. Due to the fact that the information on the beam luminosity is
not accessible anymore, only the relative cross sections could be
extracted. Absolute values were determined by normalization to the
measured total cross sections at the beam momentum of 900\,MeV, 347~$\pm$37~$\mu$b  for 
the channel \pbarpToKpKmPi0\ \cite{Amsler:2006du} and \linebreak 
83.3~$\pm$4.9~$\mu$b for
the reaction \pbarpToPi0Pi0Eta\ with $\eta \rightarrow \gamma \gamma$ \cite{Anisovich:1999pt}.\\
 The SDM elements for the three vector mesons have
been extracted in a similarly way as already done for the $\omega$ in the
reaction \pbarpToOmegaPi0\ \cite{Amsler:2014xta}. Since interference effects
are not negligible in particular for the contributions $K^*(892)^\pm
K^\mp$ and $\pi^0_1 \pi^0$, the
extraction of these quantities is more challenging compared to
the $\omega \pi^0$ case which is characterized by an isolated narrow resonance. The
traditional way, also called Schilling method~\cite{Schilling:1969um}, cannot be utilized
for the decay topology here. This method uses
only the decay angles and it is therefore mandatory that no
interference effects rise up in connection with the resonance of
interest.
Instead, the SDM elements have been determined here by
using the relevant production amplitudes obtained by the fit which
contain the full information on these quantities. This method has already been applied successfully for the reaction $\gamma p \;
\rightarrow \; \omega p$~\cite{Williams:2009aa} and later on for the
\pbarp\ annihilation process
\pbarpToOmegaPi0\ \cite{Amsler:2014xta}.
The individual $\rho$-elements can be extracted 
from the initial \pbarp\ and production amplitudes via~\cite{Kutschke:1996}:
\begin{eqnarray}
  \label{eq:sdmPwa}
 \rho_{\lambda_i\lambda_j}=
  \frac{1}{N}\sum_{\lambda_{\pbar},\lambda_{p}}
  \Big( A^{\pbarp \rightarrow J^{PC}}_{\lambda_p \; \lambda_{\pbar}}
  B^{J^{PC} \rightarrow X s_r}_{\lambda_i \; 0} \Big)^*
  \cdot
 \Big( A^{\pbarp \rightarrow J^{PC}}_{\lambda_p \; \lambda_{\pbar}}
  B^{J^{PC} \rightarrow X s_r}_{\lambda_j \; 0} \Big),
 \nonumber \\  
  \end{eqnarray}
where $\lambda_i$ denotes the helicity of the vector meson  and $N$ is the normalization factor:
\begin{eqnarray}
\label{eq:sdmPwaNorm}
N =\sum_{\lambda_{\pbar},\lambda_{p},\lambda_X} 
\Big|A^{\pbarp \rightarrow J^{PC}}_{\lambda_p \; \lambda_{\pbar}}
  B^{J^{PC} \rightarrow X s_r}_{\lambda_X \; 0}\Big|^2 
\end{eqnarray}
According to eq.~(\ref{eq:XProdAmp})
the SDM elements are slightly depending on the invariant mass of
the two-body subsystem X which is caused by the production barrier
factor $B^{L_{X s_r}}(\sqrt{s}, m_X, m_{s_r})$. In oder to suppress the
impact of this model dependent factor the elements have been extracted
within the range of $\pm20$ \mev2c\ around the obtained mass values for all
3 vector mesons. This limitation ensures that the fluctuations related to the 
invariant mass values are small and thus negligible compared with other uncertainties.\\ 
 The spin density matrix elements averaged over the complete production angle
have also been calculated via:
\begin{eqnarray}
\label{eq:AveragedSDM}
\overline{\rho}_{ij} \; = \; \frac{\int\limits_{-1}^{1}
  \frac{d\sigma}{dcos \; \theta^\pbarp} \; \rho_{ij}(cos \; \theta^\pbarp)
  \; dcos \; \theta^\pbarp}{  \int\limits_{-1}^{1}
  \frac{d\sigma}{dcos \; \theta^\pbarp} \;  dcos \; \theta^\pbarp}
\end{eqnarray}

\subsubsection{\pbarpToPhiPi0}

\begin{figure}[htb]
\begin{tabular}{cc}
\hspace{-3mm}\includegraphics[width=.5\textwidth]{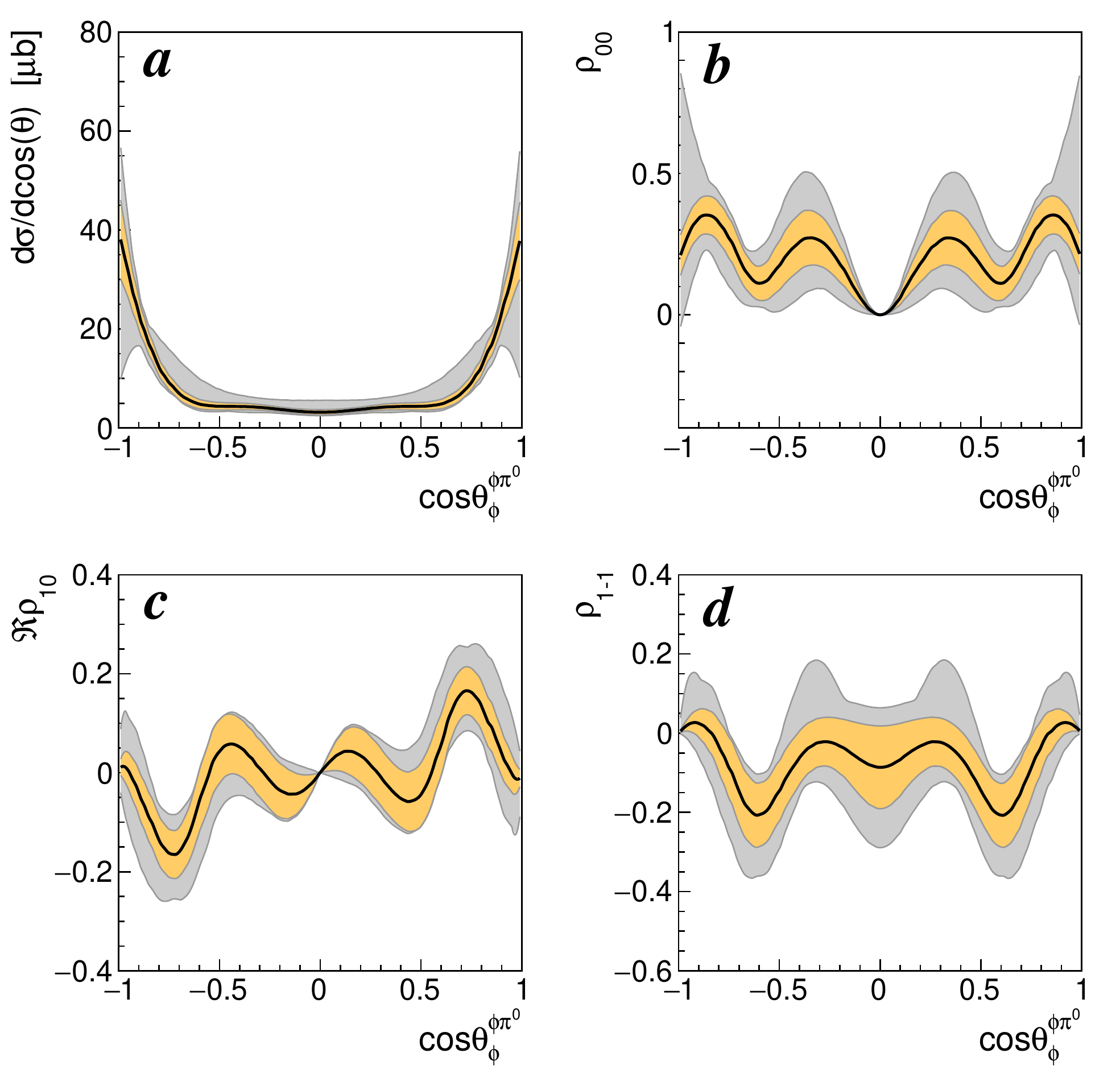} 
\end{tabular}
\caption{Differential production cross section (a) and spin density
  matrix elements $\rho_{00}$ (b), $\Re \rho_{10}$ (c) and  $\rho_{1
    -1}$ (d) of the $\phi(1020)$ in $\pbarpToPhiPi0$. All elements are
shown in the helicity system of the $\phi(1020)$. The yellow and the gray 
bands stand for the statistical and systematic uncertainty, respectively.}
\label{fig:sdmPhi}     
\end{figure}

The differential cross section for the produced $\phi(1020)$ is
shown in Fig.~\ref{fig:sdmPhi} (a). It is clearly visible that this vector meson is
produced strongly in the forward and backward direction and is
symmetric in $cos\theta^{\pbarp}_{\phi}$, as expected according to the
underlying strong interaction process. Based on this outcome and the one from~\cite{Amsler:2006du}
the total cross section for the
reaction \pbarpToPhiPi0\ at a beam momentum of 900~MeV/c is determined
to be $\sigma (\pbarpToPhiPi0 )$ = 17.5~$\pm$~1.9 (stat.)~$\pm$~2.1~
(exp.)~$\pm$~1.9~(ext.) $\mu$b. The first error is the
statistical and the second one the systematic uncertainty from this
analysis. The third error represents the uncertainty for the total
cross section of the reaction \pbarpToKpKmPi0\ extracted from \cite{Amsler:2006du}.\\
 The SDM elements for $\phi(1020)$  in its respective helicity system
are shown in Fig.~\ref{fig:sdmPhi} (b-d). All matrix elements exhibit a strong oscillatory dependence on the production angle
$cos \; \theta^\pbarp_{\phi}$. This oscillatory behavior was already
observed in the SDM elements of the $\omega(782)$~\cite{Amsler:2014xta}. The integrated elements averaged over the complete production
angle are consistent with no spin
alignment (Tab.~\ref{tab:AveragedSDMs}) which means that
all diagonal elements are in agreement with  $\overline{\rho}_{ii} = 1/3$.

\subsubsection{\pbarpToKstarK}
\begin{figure}[htb]
\begin{tabular}{cc}
\hspace{-3mm}\includegraphics[width=.5\textwidth]{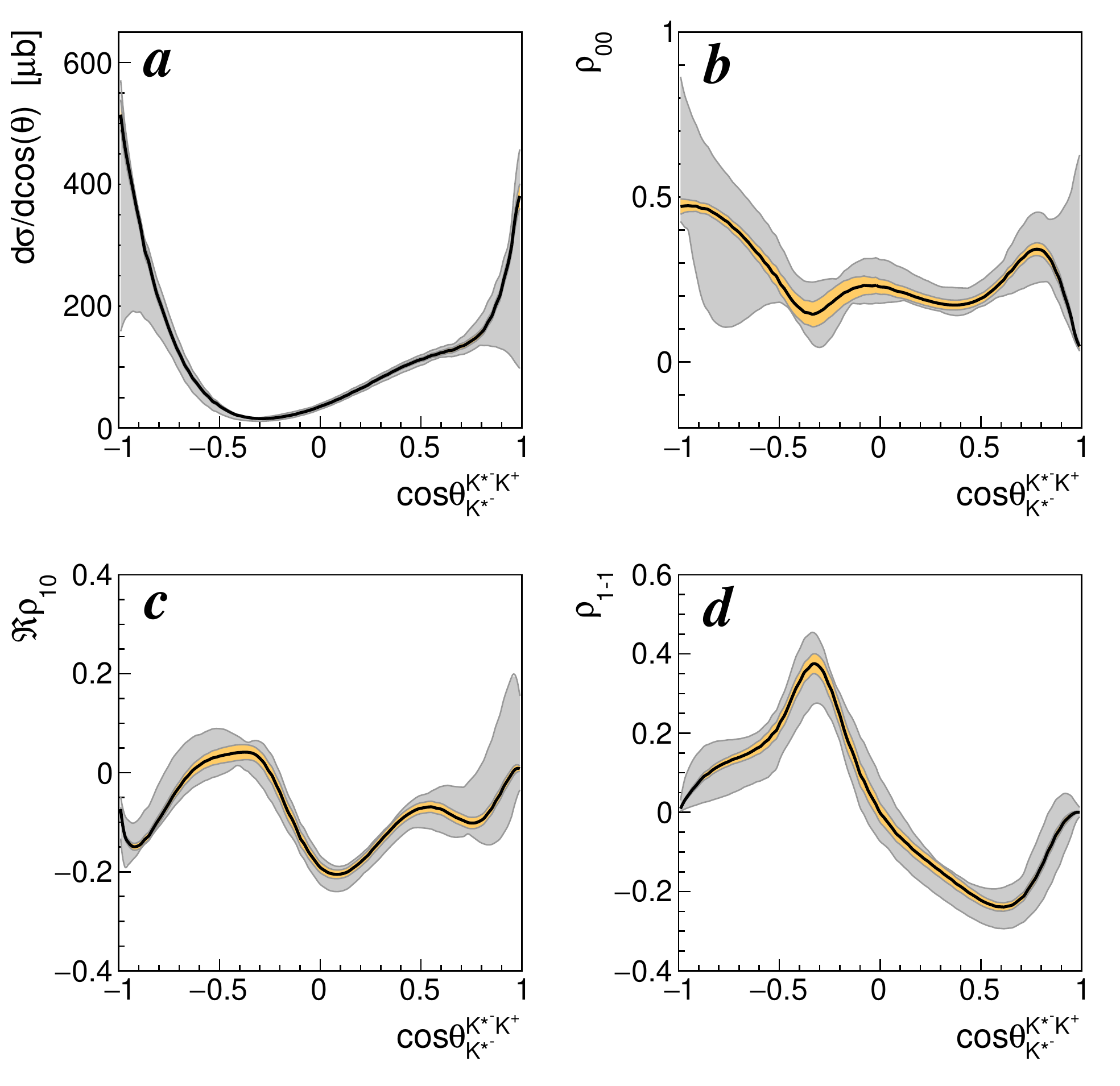} 
\end{tabular}
\caption{Differential production cross section (a) and spin density
  matrix elements $\rho_{00}$ (b), $\Re \rho_{10}$ (c) and  $\rho_{1
    -1}$ (d) of the $K^{*}(892)^-$ in $\pbarpToKstarK$. All elements are
shown in the helicity system of the $K^{*-}$. The yellow and the gray 
bands stand for the statistical and systematic uncertainty, respectively.}
\label{fig:sdmKstar}     
\end{figure}
In contrast to the $\phi(1020)$ the cross section of the  $K^*(892)^-$ is characterized by  a very significant
asymmetric dependence on the production angle
(Fig.~\ref{fig:sdmKstar} (a)) . The production of the
$K^*(892)^+$ is directly related to the one
of the $K^*(892)^-$ by
\begin{equation}
 \label{equ:KpmSymRelation}
\frac{d\sigma}{dcos\theta^\pbarp_{K^{*-}}} (cos\theta^\pbarp_{K^{*-}})\; = \;
\frac{d\sigma}{dcos\theta^\pbarp_{K^{*+}}}(-cos\theta^\pbarp_{K^{*+}})
\end{equation}
and thus the corresponding histograms are not explicitly
shown here. The reaction
\pbarpToKstarmKp\ exhibits very similar characteristics like \pbarpToKmKp\ measured by a spark chamber
experiment for 20 incident \pbar\ momenta between 0.8 and 2.4 \gevc\
\cite{Eisenhandler:1974cn}. There the forward peak
becomes stronger by increasing beam momenta and it has been suggested that
this observed s-de\-pen\-dence might be caused by Regge
exchange effects~\cite{Eisenhandler:1974cm,
  Eisenhandler:1974xp}. It might be possible that similar underlying
processes are also relevant for the charged $K^*(892)^\pm$ production in this
energy region. The total cross section for the
reaction \pbarpToKstarmKp\ at a beam momentum of 900~MeV/c is determined
to be
$\sigma (\bar{p}p \, \rightarrow \, K^*(892)^{\pm}K^{\mp})$ = 474.5~$\pm$~14.1 (stat.)~
$\pm$~116.0~(exp.)~$\pm$~50.6~(ext.) $\mu$b. Also here the third error
is due to the uncertainty of the total cross section for the reaction \pbarpToKpKmPi0\ \cite{Amsler:2006du}.\\
 The SDM elements for $K^{*}(892)^-$  in its respective helicity system
are shown in Fig.~\ref{fig:sdmKstar} (b-d). Similar to the
$\phi(1020)$ and $\omega(782)$ case all matrix
elements exhibit a strong oscillatory dependence on the production
angle
$cos \; \theta^\pbarp_{K^{*-}}$. Also here the integrated elements averaged over the complete production
angle are consistent with no spin
alignment (Tab.~\ref{tab:AveragedSDMs}).

\subsubsection{$\bar{p}p\,\rightarrow\,\pi_1^0 \pi^0$}
\begin{figure}[htb]
\begin{tabular}{cc}
\hspace{-3mm}\includegraphics[width=.5\textwidth]{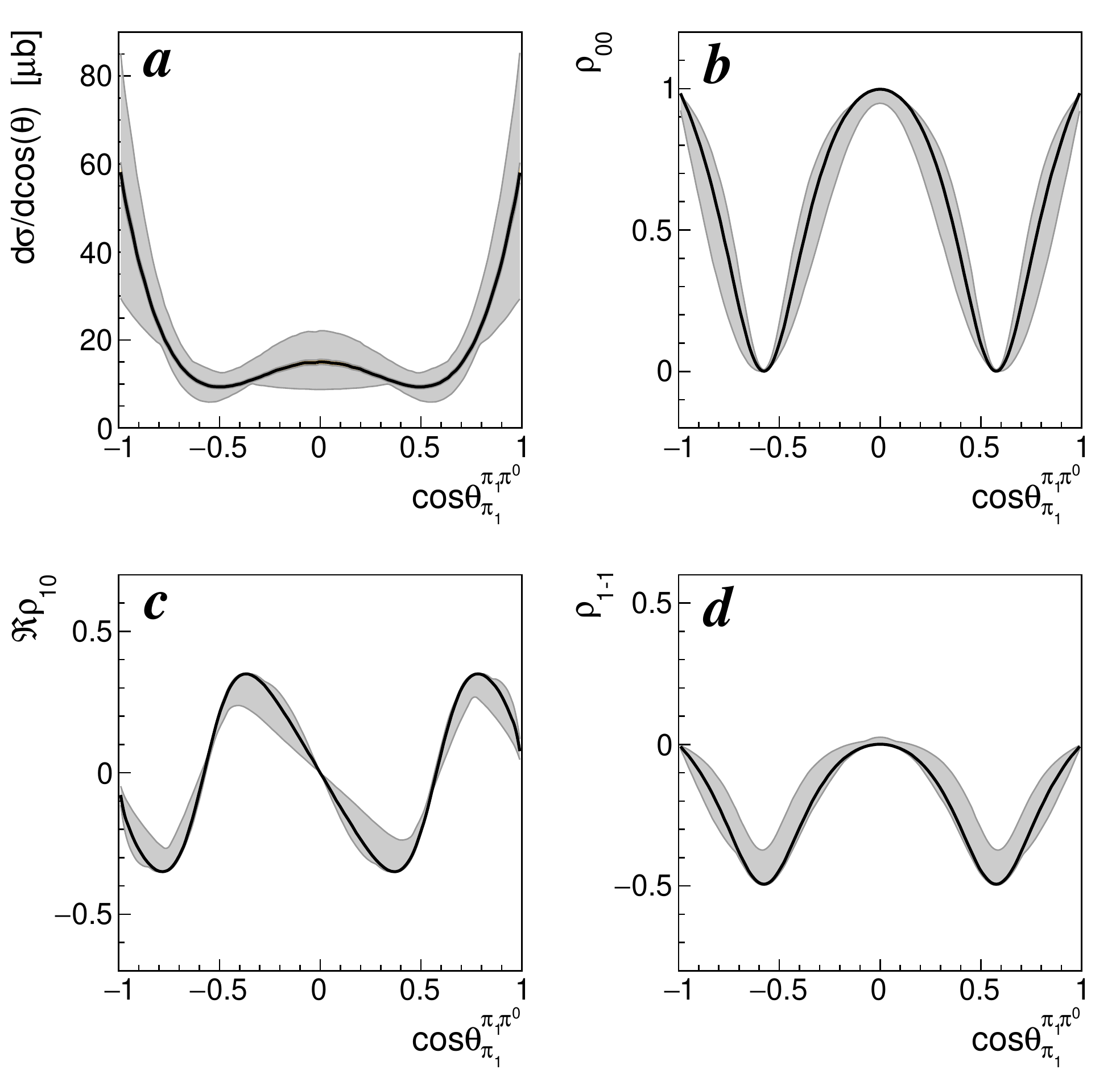} 
\end{tabular}
\caption{Differential production cross section of
  $\bar{p}p\to\pi_1^0 \pi^{0}$ with $\pi_1^0 \to
  \pi^{0}\eta$ (a) and spin density
  matrix elements $\rho_{00}$ (b), $\Re \rho_{10}$ (c) and  $\rho_{1
    -1}$ (d) of the $\pi_1^0$ in $\bar{p}p\to\pi_1^0 \pi^0$. All elements are
  shown in the helicity system of the $\pi_1^0$. The yellow and the gray 
bands stand for the statistical and systematic uncertainty, respectively.}
\label{fig:sdmPi1}     

\end{figure}

The differential production cross sections and the SDM
elements for the $\pi_1^0$-wave are summarized in
Fig.~\ref{fig:sdmPi1}. In comparison to the $\phi(1020)$ case the
forward and backward peak is similarly pronounced and the SDM element $\overline{\rho}_{00}$
averaged over the production angle exhibit a value of $66.7\%$. The total cross section for the
reaction $\pbarp \rightarrow \pi_1^0 \pi^0$  with the decay
$\pi_1^0 \rightarrow \eta \pi^0$ at a beam momentum  
of 900~MeV/c is calculated to be
$\sigma (\pbarp  \rightarrow \pi_1^0 \pi^0, \pi_1^0 \rightarrow \eta \pi^0)$ = 36.1~$\pm$~1.0 (stat.)
~$\pm$~6.5~(exp.)~$\pm$~2.1~(ref.) $\mu$b. The third error
represents the uncertainty of the total cross section for the
reaction \pbarpToPi0Pi0Eta\ extracted from \cite{Anisovich:1999pt}.

\begin{table}[htb]
  \caption{Spin density matrix elements averaged over the whole
  production cross section for the
  $\phi(1020)$,  $K^*(892)^+$ and $\pi_1^0$ vector mesons in their respective helicity system. The
 errors give the statistical and the systematic uncertainties. Due to symmetry reasons the elements $\Re\,\overline{\rho}_{1\;0}$  
for $\phi(1020)$ and $\pi_1^0$ are exactly 0 and therefore not listed here.}
\label{tab:AveragedSDMs}      
\setlength{\tabcolsep}{3pt}
\footnotesize
\centering
\begin{tabular}{l  r r r} 
  \hline\noalign{\smallskip}
         & $\phi (1020)$ & $K^*(892)^+$ & $\pi_1$\\
\noalign{\smallskip}
  \hline \noalign{\smallskip}
                 $\overline{\rho}_{0\;0} \; [\%]$ & 
$25.2~\pm~6.9^{+~8.3}_{-~8.0}$ & $31.1~\pm~1.8^{+~13.2}_{-~2.5}$ & $66.7~\pm~0.1^{+~10.7}_{-~22.8}$ \\ 
\noalign{\smallskip}
                    $\overline{\rho}_{1\;-1} \; [\%]$ & 
$-3.5~\pm~5.0^{+~1.7}_{-~5.3}$ & $-1.8~\pm~1.1^{+~0.1}_{-~3.4}$ & $-16.3~\pm~0.1^{+~8.8}_{-~6.2}$ \\ 
\noalign{\smallskip}
                   $\Re\,\overline{\rho}_{1\;0} \; [\%]$ & 
  & $-8.4~\pm~1.0^{+~3.1}_{-~0.3}$ & \\
  \noalign{\smallskip}
  \hline
\end{tabular}
\end{table}

\section{Summary}
\label{conclusion_lab}
A coupled channel analysis of \pbarp\ annihilation to \PiPiEta ,\linebreak
\PiEtaEta\ and \KpKmPi0\ at a beam momentum of 900~MeV$/c$ has been
performed by considering
 $\pi\pi$-scattering data for the S0-, D0- and P1-waves. 
The usage of the K-matrix approach for the description of the dynamics ensures 
a sufficient fulfillment of unitarity and analyticity conditions. It was demonstrated, that it is possible to extract the properties of all contributing resonances in a simultaneous fit to all data. In lots of analyses in the past, part of the properties were taken from previous fits to a sub-set of data and are not treated as free parameters. All
data are reproduced reasonably well by the simultaneous fit. The dominant contributions in the
three \pbarp\ channels are the $a_2 \, \pi^0$,
and $f_2 \, \eta$ components for \PiPiEta, the
$f_0 \, \pi^0$ and  $a_2 \, \eta$ waves for
 \PiEtaEta\ and the 
$K^*(892)^\pm \, K^\mp$ reaction for  \KpKmPi0. Masses
 and widths obtained from the Breit-Wigner pa\-ra\-merte\-ri\-za\-tions for isolated
 resonances and pole positions extracted from the K-matrix descriptions
 have been determined and are within the ballpark of other individual measurements. In the channel  \PiPiEta\  a significant
 contribution of the spin
 exotic $I^G=1^-$ $J^{PC}=1^{-+}$ wave  decaying to $\pi^0 \eta$ has been observed. By choosing the K-matrix approach with one
pole and two decay channels ($\pi \eta$, $\pi \eta^\prime$) 
 for the description of the dynamics, a mass of (1404.7~$\pm$ 3.5 (stat.)~$^{+9.0}_{-17.3}$ (sys.)) \mev2c\ and a width
 of (628.3~$\pm$ 27.1 (stat.)~$^{+35.8}_{-138.2}$~(sys.)) MeV are obtained. An
 analysis explicitly focused on this spin-exotic wave will be presented
 in a forthcoming paper. Partial decay widths for the $f_2(1270)$, $f_{2}^{\prime}(1525)$, $\rho(770)$ 
and $\rho(1700)$ states and ratios of
 these properties for the $a_0$ and $a_2$ resonances have been retrieved via the residues
 of the pole positions. 
The differential production cross section and
the spin-density-matrix elements for the $\phi(1020)$ and the $K^*(892)^\pm$
have been extracted. While the $\phi(1020)$ vector meson
is produced
strongly in the forward and backward direction, the $K^*(892)^-$ instead, is characterized by  
a very significant asymmetric dependence on the production angle. No spin-alignment
effects are observed for both vector mesons but the individual spin-density matrix
elements exhibit an oscillatory dependence on the production angle.
The SDM elements have also been determined for the spin-exotic wave
$\pi_1^0$ with an averaged value of $\overline{\rho}_{00} =66.7\%$.

\begin{acknowledgements}
 The study was funded by the Collaborative Research Center under the
 project CRC 110: {\it Symmetries and the Emergence of Structure in
   QCD}. W.~D\"unnweber and M.~Faessler are supported
 by the {\it DFG Cluster of Excellence Origins}.
 The authors wish to thank R.~Kaminski, W.~Ochs, J.~Ruiz de Elvira and A.~Rodas for
 providing the scattering data and related information 
 and to C.~Hanhart, D.~R\"onchen and S.~Ropertz for the helpful
 hints related to unitarity and analyticity conditions and to the
 extraction of resonance properties. We also gratefully acknowledge U.~Thoma 
 for the constructive suggestions and advices to improve this paper. Most 
 of the time-consuming fits
 have been performed on the Green Cube at GSI in Darmstadt.
\end{acknowledgements}

\appendix


\end{document}